\documentclass[aps,prb,superscriptaddress,twocolumn,floatfix]{revtex4}
\usepackage{color,graphicx,pstricks}
\usepackage{psfrag}
\usepackage{textcomp}
\usepackage{hyperref}
\hypersetup{
  colorlinks=true,
  citecolor=blue,
  linkcolor=red,
  }
\usepackage{amsmath}
\begin{document}

\title{Transport and Phase competition in three-dimensional Hubbard-Holstein model at half-filling}

\author{Sandip Halder}
\email{halder@post.bgu.ac.il}
\affiliation{Department of Physics, Ben-Gurion University of the Negev,
  Beer-Sheva 84105, Israel}
\author{Moshe Schechter}
\email{smoshe@bgu.ac.il}
\affiliation{Department of Physics, Ben-Gurion University of the Negev,
  Beer-Sheva 84105, Israel}

\begin{abstract}
We investigate the magnetotransport properties of the one-band Hubbard-Holstein model
at half-filling in three dimensions (3D) using exact diagonalization based semi-classical Monte Carlo simulations with
phonons treated in the adiabatic limit. The low-temperature electronic correlation $U$ vs
electron-phonon coupling $V$ phase diagram reveals two insulating phases--antiferromagnetic
(AF-I) and charge-ordered (CO-I)--separated by a first-order transition, with no metallic phase
observed at their intersection, indicating robustness of these phases in 3D. For $U \sim bandwidth$,
the $V$ vs temperature $T$ phase diagram exhibits multiple phases including AF-I, CO-I, Mott-Hubbard
insulator, bipolaronic insulator, and two bipolaronic metallic states. Several first-order transitions
occur near $V\sim 3.75$. Above the ordering temperature, the density of states shows universal behavior
dominated by electronic contribution, while susceptibility and DOS analyses reveal pseudogap features.
Magnetic and transport properties along the phase boundary highlight strong proximity effects between
competing phases, suggesting routes for tuning correlated materials and emergent electronic states.

\end{abstract}

\maketitle

\section{Introduction}
The Hubbard interaction ($U$) and electron-phonon coupling ($g$) are two fundamental and competing
interactions, which are found widely in nature~\cite{Dagotto,Imada,Damascelli} and govern the
electronic, magnetic, and structural properties of strongly correlated materials~\cite{Kulic,Capone1,Mannhart,Dagotto1}.
The onsite Hubbard $U$ is amenable for electronic localization, local moment formation, Mott insulating
behavior, and robust magnetic and orbital orders~\cite{Choi,Imada,Dagotto}. The modulation of $U$
through bandwidth control, dimensional confinement, epitaxial strain, chemical substitutions,
interfacial engineering is leveraged in a vast range of technological applications~\cite{Choi,Mannhart},
such as Mott transistors~\cite{Ahn,Scheiderer}, oxide-based neuromorphic devices~\cite{YZhou,Chang,Goteti},
and antiferromagnetic spintronics~\cite{Wadley,Marti}. Here, correlation-driven phase transitions (metal-insulator
transitions) make possible low-power electronic switching and solid-state sensor devices~\cite{Yang}.
Also strong electronic correlation-driven mechanisms, such as history-dependent resistance and
multistability, are exploited to create memristive and neuromorphic elements~\cite{Zhou,Kwon,YZhou},
which are important for neuromorphic computing~\cite{Zhou,YLi,SDHa,Boybat}. On the other hand, electron-phonon
coupling ($g$) connects the electronic degrees of freedom with lattice distortions,
leading to polaron formation~\cite{Franchini}, charge density waves (CDW)~\cite{Peng,Peierls},
structural phase transitions~\cite{Alonso,Kim}, and, at strong coupling, bipolaronic states~\cite{Prokofev,Kornilovitch,Bonca}.
Materials whose properties are dominated by electron-phonon coupling are technologically relevant for
optoelectronic devices~\cite{Bai,Ulbricht}, thermoelectrics~\cite{Bai}, and ultrafast switching applications~\cite{Chen}.
Strong electron-phonon coupling helps improve electromechanical responses, which play a key role
in sensors~\cite{Sayers,Benyamini,Yan}, actuators~\cite{Sayers,Benyamini}, and strain-controlled electronic devices~\cite{Niehues,Tang}.

In a large class of correlated materials, such as high $T_{C}$ cuprates~\cite{Lanzara,Damascelli},
fullerides~\cite{Gunnarsson}, nickelates~\cite{Haule,Stepanov,Alonso,Tranquada}, manganites~\cite{Millis,Millis1},
vanadates~\cite{Imada}, and organic salts~\cite{Powell}, the Hubbard $U$ and electron-phonon coupling $g$
coexist and, in some cases, compete, giving rise to emergent phases~\cite{Zhan} that are not
observed when the individual interactions $U$ or $g$ act alone.  For instance, correlation-assisted
charge ordering~\cite{RWang,QWang,Imada}, lattice-driven metal-insulator transitions~\cite{Georgescu,Imada,Ge},
and coupled magnetic-structural phases~\cite{Duan,Kim}, are realized due to the subtle interplay of
electronic correlation and electron-phonon coupling. From a technological standpoint, this synergy
enables multifunctional materials in which electronic, magnetic, and structural properties can be
simultaneously tuned by electric field, strain, or temperature~\cite{Ngai,Bark,Watanabe,Mirjolet}.
Balanced manipulation of the interplay between U and g stands as an important asset toward adaptive electronics~\cite{ZOu,Zhan},
ultrafast switching devices~\cite{Baldini}, and quantum and oxide technologies~\cite{Singh,ZOu,Zhan,Baldini},
where phase competition and tunability are not limitations but key functional advantages. Overall, in
addition to the individual roles of $U$ and $g$, their cooperation and antagonistic competition are central
to many technologically relevant functionalities.

All these practical features urge for a model Hamiltonian that includes both Hubbard $U$ and electron-phonon coupling $g$,  
comprehensively delineates their interplay, and sheds light on the understanding of these systems. In this context, the
Hubbard-Holstein (HH) model is widely investigated, as it explores the interplay among charge order, antiferromagnetism, 
and superconductivity~\cite{Karakuzu,Han,Koller,Koller1,Koller2,Costa,Nowadnick,Johnston,Mendl,Ohgoe,Huang}. Most of the studies
have been reported on one-dimensional ($1D$) systems, with well-established phase diagrams that include spin-density waves, bond-ordered waves,
CDW,  metallic phases, and phase separation behavior~\cite{Nocera,SLi,Xiao,Hebert,Lavanya,Hohenadler1,Matsueda,Ning}. Density matrix renormalization 
group (DMRG) studies and quantum Monte Carlo (QMC) simulations on $1D$ also revealed several intriguing general features~\cite{Tezuka,Tezuka1,Clay,Fehske}
at finite phonon frequencies, as well as in the adiabatic ($\omega \rightarrow 0$) and anti-adiabatic limits ($\omega \rightarrow \infty$)~\cite{Clay,Fradkin}. Some of them 
emphasized the stability of pairing correlations when the system is doped or when the electronic band structure is modified, i.e., 
in the particle-hole symmetry broken scenario~\cite{Tezuka,Tezuka1}. A few other reports were dedicated to establishing an intermediate
metallic state between the Peierls insulator and the Mott insulator, where the metallic regime depends on the phonon frequency~\cite{Clay,Fehske,Lavanya,Hohenadler1}.

Apart from the $1D$ studies, a myriad of striking phases have also been identified in the two-dimensional Hubbard-Holstein
model~\cite{Nowadnick,Karakuzu,Brink,Mendl,Johnston,Berger,Hotta,Nowadnick1}. The phases include, naturally, the AF phase,
CO phase, superconducting phases, and other phases such as polaronic, bipolaronic, quasiparticles~\cite{Mendl},
and pseudogap~\cite{Dupuis,Pai} phases. Moreover, phase separation was reported upon doping the CDW insulator whereas
a uniform superconducting ground state was evidenced in case of doping the superconducting phase~\cite{Karakuzu}. A recently
developed variational non-Gaussian wave function approach within exact diagonalization predicts two intervening
phases between the spin-ordered and charge-ordered phases~\cite{Wang}. One of them may exhibit superconductivity and the other displays
large charge fluctuations and dominant binding energy, whereas both the phases reside within the superconducting regime
according to the variational Monte Carlo (VMC) studies~\cite{Karakuzu,Ohgoe}.      

In a nutshell, extensive studies of the Hubbard-Holstein model had been performed using a plethora of methods, such as exact
diagonalization (ED) in $1D$ and $2D$~\cite{Wang,Marsiglio}, density matrix renormalization group (DMRG) studies in $1D$~\cite{Fehske,Hohenadler1},
different variants of quantum Monte Carlo (QMC) simulations [such as exact QMC, contineous-time QMC, determinant QMC]
in both $1D$ and $2D$~\cite{Clay,Hohenadler1,Weber,Murakami,Costa}, exact diagonalization based semi-classical Monte Carlo (s-MC) technique
in $2D$~\cite{Pai}, dynamical mean field theory (DMFT)~\cite{Bauer,Bauer1} and some times in conjecture with the numerical
renormalization group (NRG) in infinite dimension~\cite{Bauer}, Hartree-Fock mean-field theory in $2D$~\cite{Brink}, and perturbative
approaches~\cite{Berger,Brink}. However, there are no reports in real-space three dimensions till now to the best of
our knowledge. This lacking might be a result of the enhanced numerical complexities and computational cost.

Specifically, the family of nickelates is important in the context of the interplay between electron-phonon
coupling and strong electronic correlations, which drive various intriguing phenomena~\cite{Hayashida,Talantsev,Ouyang,Stepanov,Zhang,Munoz,Alonso,Johnston1}.
The $RNiO_{3}$ ($R$ is a rare-earth lanthanide element) compounds exhibit staggered-type phononic modes, i.e., alternate $NiO_{6}$
octahedra are expanded and compressed in a rocksalt-type distortion~\cite{Haule,Munoz,Alonso}. This contraction (expansion) of the $NiO_{6}$ octahedra
leads to charge disproportionation into $Ni^{3+\delta}O_{6}$ ($Ni^{3-\delta}O_{6}$), stabilizing the insulating
state via energy gain~\cite{Munoz}.
The charge segregation ($\delta$) increases with decreasing lanthanide ionic radius~\cite{Alonso1,Staub}, accompanied by an enhanced
disparity in Ni--O bond lengths between electron-rich and electron-depleted Ni sites~\cite{Munoz}.  An exact diagonalization
based Hartree-Fock calculation has shown that rocksalt-type lattice  distortions (contraction/expansion of O octahedra) are energetically
favorable in these nickelates ($RNiO_{3}$), and that a strong electron-phonon coupling drives these rocksalt-like distortions in nickelates~\cite{Johnston1}.

Considering the absence of theoretical studies in three dimensions and the experimental relevance of systems where
both $U$ and $g$ are important, we investigate the magnetotransport properties of the Hubbard-Holstein model in $3D$
using an exact diagonalization based semiclassical Monte Carlo (s-MC) method. To access large system sizes ($8^{3}$),
we employ the traveling cluster approximation (TCA)~\cite{Kumar} with a TCA cluster size of $4^{3}$. The phonon variables are treated
in the adiabatic limit with two modes: expansion and contraction~\cite{Haule,Munoz,Alonso,Johnston}.
The low-temperature $U$-$V$ ($V\sim g^{2}/K$, scaled in terms of energy unit $t$) phase diagram at half-filling reveals two phases: an antiferromagnetic insulator and a charge-ordered
insulator, separated by a first-order transition line. In contrast to previous reports~\cite{Hardikar,Pai}, no metallic phase is observed at their
intersection. For $U = 8$, away from both the perturbative and strong correlation limits, the $V$-$T$ phase diagram exhibits
several phases, including antiferromagnetic insulator (AF-I), charge-ordered insulator (CO-I), Mott-Hubbard insulator (MH-I), bipolaronic insulator (BP-I), and two bipolaronic metals
(BP-M and BP-M*). The distinction between BP-M and BP-M* is identified from the temperature evolution of the bipolaronic order parameter.
The transitions AF-I$\rightarrow$CO-I, MH-I$\rightarrow$BP-I, and BP-M*$\rightarrow$BP-M are all first order and occur near $V\sim 3.75$
at different temperatures. Near this transition line, the density of states (DOS) shows universal behavior above the ordering temperature,
indicating that high-temperature properties are dominated by electronic contributions. Pseudogap features are also captured in our
calculations from susceptibility and DOS analyses. Finally, we examine magnetic and transport properties along the transition line
of the $U$-$V$ phase diagram, where proximity to competing phases and external factors such as frustration, strain, doping, long-range
hopping, and dimensional confinement may lead to tunable material properties and emergent phases.

The paper is organized as follows: In Sec. \textbf{II}, we briefly describe our model
Hamiltonian and numerical method in order to study the ground state and finite temperature properties 
of the system. In Sec. \textbf{III}, we outline various physical observables which will be used to calculate the
magnetic, transport, and electronic properties of the system. The details of the ground-state $U$-$V$ phase
diagram at half-filling are delineated in Sec. \textbf{IV}. The $V$-$T$ phase diagram for a fixed  $U = 8$ is
illustrated in Sec. \textbf{V}. Sec. \textbf{VI} is devoted to explaining the universality near the
transition line and the physics of pseudogap features. In Sec. \textbf{VII}, we describe the magnetotransport
properties for small $U$ and $V$ combinations. The features along the transition (boundary) line of the $U$-$V$ phase
diagram are represented in Sec. \textbf{VIII}. Finally, we summarize our results in Sec. \textbf{IX}.

\section{Model Hamiltonian and Method}
We explore the magnetotransport properties of the one band Hubbard-Holstein model in three dimensions, which is
represented as follows
\begin{align}
H=&-t\sum_{<i,j>,\sigma}(c_{i,\sigma}^{\dagger}c_{j,\sigma}+H.c.)+U\sum_{i}n_{i,\uparrow}n_{i,\downarrow} \nonumber\\
&+g\sum_{i}n_{i}.{\bf{Q'}}_{i}+\frac{K}{2}\sum_{i}{\bf{Q'}}_{i}^{2}-\mu\sum_{i}n_{i},\label{eq:1}
\end{align}

\noindent
where $c_{i,\sigma}^{\dagger}$ and $c_{i,\sigma}$ indicate the electron creation and annihilation 
operator at site $i$ with spin $\sigma\;(=\uparrow or, \downarrow)$. $<i,j>$  represents the nearest 
neighbor sites on a $3D$ cubic lattice and $t$ is the hopping amplitude among these nearest neighbor
sites. The strength of onsite Hubbard repulsive interaction is denoted by $U$. 
$n_{i}=\sum_{\sigma}n_{i,\sigma}=\sum_{\sigma}c_{i,\sigma}^{\dagger}c_{i,\sigma}$, is the charge density
operator at site $i$. The average filling (density) of the system is controlled 
by the chemical potential $\mu$.
Here we concentrate on the case of half-filling where $\mu = 0$.
$g$ is the coupling strength of phonon $Q'_{i}$ with the electronic density 
$n_{i}$ at arbitrary lattice site $i$.  The stiffness constant of the distortion is denoted by $K$. In our 
investigation, two possible distortions (expansion and contraction modes) are considered and the phonon
modes are treated in the adiabatic limit, which are justified in some experiments~\cite{Haule,Munoz,Alonso,Alonso1}
and theoretical~\cite{Johnston1,Brink} investigations.  

The pure Holstein and Hubbard models are recovered in the limits $U=0$ and $g=0$, respectively. At half-filling in pure 
classical Holstein model, the single site energy is (for $t=0$)
\begin{align}
E=g{Q'}_{i}+\frac{K}{2}{Q'}_{i}^{2}.\label{eq:2}
\end{align}
\noindent
The energy is minimum with respect to the distortion $Q'_{min}=-\frac{g}{K}=\rho$ and the corresponding polaronic minimum energy 
(polaronic binding energy) is $E_{pol}=-\frac{g^{2}}{2K}$. Thereafter, we rescale the phonon coordinate
$Q'$ by $\rho$ such as $Q = Q'/\rho$ and the phonon energies by $|E_{pol}|$ which leads the phononic part of the Hamiltonian in Eq.~\eqref{eq:1}
in terms of a single dimensionless parameter $V$ (scaled in term of energy unit $t$) to be presented as 

\begin{align}
H=&-t\sum_{<i,j>,\sigma}(c_{i,\sigma}^{\dagger}c_{j,\sigma}+H.c.)+U\sum_{i}n_{i,\uparrow}n_{i,\downarrow} \nonumber\\
&+V\sum_{i}n_{i}.{\bf{Q}}_{i}+\frac{V}{2}\sum_{i}{\bf{Q}}_{i}^{2}-\mu\sum_{i}n_{i}\nonumber \\
&=H_{p}+H_{int}.\label{eq:3}
\end{align}

Here, $H_{p}$ contains all other terms except the Hubbard interaction term. Therefore, $H_{p}$
is bilinear (quadratic) in electronic creation and annihilation operators and holds the phononic terms.
$H_{int}$ contains the interacting Hubbard term which is quartic in nature. This quartic interaction
term is then decomposed into two different quadratic terms in order to solve the Hamiltonian in the
following way~\cite{Schulz,Mukherjee,S1,S2}:

\begin{align}
Un_{i,\uparrow}n_{i,\downarrow}=U\left[\frac{1}{4}n_{i}^{2}-({\bf{S}}_{i}\cdot{\hat{\Omega}}_{i})^{2})\right]\label{eq:4}
\end{align}

\noindent
with ${\bf{S}}_{i}$ and  ${\hat\Omega}_{i}$ are the spin vector and the arbitrary unit vector at the $ith$ site, respectively.
The spin vector ${\bf{S}}_{i}$ is defined as ${\bf{S}}_{i}=\frac{\hbar}{2}\sum_{\alpha,\beta}c_{i,\alpha}^{\dagger}\sigma_{\alpha,\beta}c_{i,\beta}$
where $\hbar=1$,  and $\sigma$ is the Pauli matrices. In Eq.~\eqref{eq:4}, the applied decoupling is rotationally invariant.
Next the partition function of the total Hamiltonian in Eq.~\eqref{eq:3} is written as $Z=Tre^{-\beta H}$, where $\beta=\frac{1}{T}$
is the inverse temperature and the Boltzmann constant $k_{B}$ is set to $1$. The interval $[0,\beta]$ is split into $M$ number 
of equally spaced slices of width $\Delta\tau$ ($\beta=M\Delta\tau$). Following this, for very large $M$ (i.e., in the limit $\Delta\tau\rightarrow0$),
we express  $e^{-\beta(H_{p}+H_{int})}=(e^{-\Delta\tau H_{p}}e^{-\Delta\tau H_{int}})^{M}$ up to first order in $\Delta\tau$ using
Suzuki-Trotter transformation. Thereafter, implementing Hubbard-Stratonovich (H-S) transformation, the interacting 
part of the partition function $e^{-\Delta\tau U\sum_{i}[\frac{1}{4}n_{i}^{2}-({\bf{S}}_{i}\cdot{\hat{\Omega}}_{i})^{2}]}$ 
takes a form which is shown to be proportional to 
\begin{align}
\sim\int &d\phi_{i}(l) d\Delta_{i}(l)d^{2}\Omega_{i}(l) \nonumber\\
&\times e^{-\Delta\tau [\sum_{i}\{\frac{{\phi_{i}(l)}^{2}}{U}
+i\phi_{i}(l)n_{i}+\frac{{\Delta_{i}(l)}^{2}}{U}
-i\Delta_{i}(l){\hat{\Omega}}_{i}(l)\cdot{\bf{S}}_{i}\}],}\label{eq:5}
\end{align}

\noindent
for a generic time slice `$l$'. Here, the H-S auxiliary fields $\phi_{i}(l)$ and $\Delta_{i}(l)$ at arbitrary site i, are coupled
with the charge density $n_{i}$ and with the spin vector ${\bf{S}_{i}}$, respectively. Then by introducing a new vector auxiliary
field ${\bf{m'}}_{i}(l)=\Delta_{i}(l)\cdot{\hat\Omega}_{i}(l)$, the total partition function~\cite{Santos,Mukherjee,Pai,S1,S2} is written as

\begin{align}
Z=const. & \times  Tr \prod_{l=M}^{1} \int d\phi_{i}(l)d^{3}{\bf{m'}}_{i}(l)dQ_{i}\times \nonumber\\
& e^{-\Delta\tau[H_{p}+\sum_{i}\{\frac{{\phi_{i}(l)}^{2}}{U}
+i\phi_{i}(l)n_{i}+\frac{{{\bf{m'}}_{i}(l)}^{2}}{U}
-2{\bf{m'}}_{i}\cdot{\bf{S}}_{i}\}],}\label{eq:6}
\end{align}

\noindent
where the product follows the time order product, i.e., former time appears at the right ($l$ runs from $M$ to $1$).

An effective model Hamiltonian is derived from this partition function. At this step, we
neglect the $\tau$ dependence of the classical auxiliary fields and retain the spatial
fluctuations of the auxiliary fields which capture the physics of inhomogeneities in real
space system. Then, using the saddle point value of the auxiliary field $i\phi_{i}(l)=\frac{U}{2}<n_{i}>$ and
redefining ${\bf{m}}_{i} = \frac{U}{2}{\bf{m'}}_{i}$, we express the effective Hamiltonian~\cite{Pai}
in the following form

\begin{align}
H_{eff}=& -t\sum_{<i,j>,\sigma}(c_{i,\sigma}^{\dagger}c_{j,\sigma}+H.c.)\nonumber\\
&+\frac{U}{2}\sum_{i}(<n_{i}>n_{i}-{\bf{m}}_{i}\cdot\sigma_{i})+\frac{U}{4}\sum_{i}({{\bf{m}}_{i}}^{2}-{<n_{i}>}^{2})\nonumber \\
&+V\sum_{i}n_{i}.{\bf{Q}}_{i}+\frac{V}{2}\sum_{i}{\bf{Q}}_{i}^{2}-\mu\sum_{i}n_{i}.\label{eq:7}
\end{align}

We solve the final Hamiltonian in Eq.~\eqref{eq:7} using the semiclassical Monte Carlo (s-MC) technique via the exact
diagonalization of the fermions in the fixed background of classical fields $\{{\bf{m}}_{i}\}$ and $\{{\bf{Q}}_{i}\}$,
and $\{<n_{i}>\}$ configurations. The classical fields (${\bf{m}}_{i}$, ${\bf{Q}}_{i}$) are annealed from a very high
temperature, $T = 10$, with random initial configurations using the standard Metropolis algorithm~\cite{Metropolis,Halder,S3}
by visiting every lattice site sequentially. In this approach, we first propose an update of the ${\bf{m}}_{i}$
and ${\bf{Q}}_{i}$ fields at the $i$th site. The proposed update is accepted/rejected according to the Boltzmann probability,
which is estimated from the energy calculated by diagonalization before and after the proposed update at that temperature.
This process is executed over all the lattice sites and constitutes a Monte Carlo system sweep. We perform 2000 MC system
sweeps at each temperature, where the first $1000$ MC sweeps are used to thermalize the system and the remaining $1000$ MC
sweeps are utilized to measure the physical observables. We estimate $\{<n_{i}>\}$ self-consistently at every $10$th step
of the MC sweep, which are used in the next $10$ MC sweeps. We discard $10$ MC sweeps between two consecutive measurements
to avoid spurious self-correlation in the data. Then, we decrease the temperature in a small steps and repeat the above
mentioned updates and measurements. The procedure is repeated while lowering the temperature, adopting the updated classical-field
configurations from the last MC step at the previous temperature as the initial classical-field configurations for the next
temperature. We also average the data over $10$ different independent runs starting with different random initial configurations,
in addition to the Monte Carlo averages. A travelling cluster approximation (TCA) based Monte Carlo scheme~\cite{Kumar,Chakraborty,Halder}
is adopted to access a large system size of $8^3$, with a TCA cluster size of $4^3$.

We simulate the model Hamiltonian in Eq.~\eqref{eq:7} on a $3D$ real-space lattice with $N = L^3$ sites  ($L=8$).
The computationally expensive step is diagonalizing the Hamiltonian to calculate the energy change associated with
updating the classical variables ${\bf{m}}_{i}$ and ${\bf{Q}}_{i}$ at site $i$. For a single site update, this costs
$O(N^3)$. Then a complete MC sweep costs $O(N^4)$. Consequently, exact diagonalization based Monte Carlo is limited
to small system sizes ($L \le 6$). To overcome this limitation, we employ the well-established travelling cluster
approximation (TCA), in which a smaller cluster of $N_{C}=L_{C}^{3}$ sites ($L_{C}=4$) surrounding the  updated site
is diagonalized instead of the full lattice Hamiltonian. This approximation is justified by the observation that the
effect of a single-site update is confined to a limited region. The computational cost is thereby reduced to $O(N_{C}^{3}$)
per site update and $O(NN_{C}^{3})$ per MC sweep, enabling simulations of larger systems. The convergence test of our results
with respect to TCA cluster sizes is discussed in section \textbf{IV}.

\section{Physical Observables}

For exploring the magnetotransport properties and various ground state phases of the one-band Hubbard-Holstein
model in three dimensions, we calculate various physical quantities associated with magnetic, transport,
electronic, and charge profiles. We first evaluate the structure factor of the quantum spin correlation
as follows 

\begin{align}
S(\boldsymbol{q})=\frac{1}{N^{2}}\sum_{i,j}\langle {\boldsymbol{S}}_{i}.{\boldsymbol{S}}_{j} \rangle
e^{-i{\boldsymbol{q}}.({\boldsymbol{r}}_{i}-{\boldsymbol{r}}_{j})}, \nonumber
\end{align} 
with $\mathbf{S_{i}}$ being the quantum spin vector at site i, determined from the eigenvalues and eigenvectors
obtained through the diagonalization of the effective Hamiltonian in Equation~\eqref{eq:7}. Here, $\bf{q}$ is the 
wave vector, and $N$ is the total number of lattice sites in the system. The thermal and quantum mechanical averages
of the observables are represented by the angular brackets. The indices $i$ and $j$ run over all lattice sites.

Then, we determine the charge structure factor [CO(q)] in order to measure the charge order (CO) of the system as follows,

\begin{align}
CO(\boldsymbol{q})=\frac{1}{N^{2}}\sum_{i,j}\langle ({n}_{i}-<n>).({n}_{j}-<n>) \rangle
e^{-i{\boldsymbol{q}}.({\boldsymbol{r}}_{i}-{\boldsymbol{r}}_{j})}, \nonumber
\end{align}
\noindent
where $<n>$ is the average density of the system. We basically estimate the charge structure factor at 
$\bf{q}=(\pi, \pi, \pi)$, i.e., CO($\pi, \pi, \pi$), which measures long-range staggered-type (G-type) 
charge order. 

The average local momemt ($M$), which is a measure of the system averaged magnetization squared, is evaluated
using the expression  $M = \langle (n_{\uparrow}-n_{\downarrow})^{2}\rangle = \langle n \rangle -2\langle n_{\uparrow}n_{\downarrow}\rangle$, 
where $\langle n \rangle = \langle n_{\uparrow}+n_{\downarrow}\rangle$. Subsequently, the distribution of the 
local moment is represented by $P(M)$, defined as $P(M)=\sum_{i}\delta(M-M_{i})$.

Numerically, we calculate the specific heat $C_{v}(U,T)$ of the system by differentiating the 
total energy with respect to temperature [$C_{v}(U,T) = \frac{dE(U,T)}{dT}$] by 
implementing the central difference formula. We then examine the temperature evolution of this 
thermodynamic observable.

We compute the electronic properties of the system by calculating the density of state [DOS($\omega$)]
at an arbitrary frequency $\omega$, which is defined as $DOS(\omega) = \frac{1}{N}\sum_{\alpha}\delta (\omega - \epsilon_{\alpha})$. 
Here, $\epsilon_{\alpha}$ represents the single-particle eigenvalues, and $\alpha$ runs over the total 
number ($= 2N$) of eigenvalues of the system. The delta function is calculated using its Lorentzian representation 
with a broadening $\sim BW/2N$ ($BW$ is the bare bandwidth) to evaluate the DOS.

To analyze the transport properties, we estimate conductivity of the system in the 
dc limit using the Kubo-Greenwood formalism~\cite{Mahan,Kumar1,Bulanchuk} in the following way,

\[\sigma(\omega)=\frac{A}{N}\sum_{\alpha, \beta}(n_{\alpha}-n_{\beta})
\frac{{|f_{\alpha \beta}|}^{2}}{\epsilon_{\beta}-\epsilon_{\alpha}}
\delta[\omega-(\epsilon_{\beta}-\epsilon_{\alpha})]\]

\noindent
where $A=\pi e^{2}/\hbar a$ ($a$ is the lattice parameter). $f_{\alpha \beta}$
represents the matrix elements of the paramagnetic current operator
${\hat{j}}_{\gamma}=it\sum_{i,\sigma}
(c_{i,\sigma}^{\dagger}c_{i+\hat{\gamma},\sigma}-c_{i+\hat{\gamma},\sigma}^{\dagger}c_{i,\sigma})$
between the eigenstates $|\psi_{\alpha}>$ and $|\psi_{\beta}>$ with corresponding
eigenenergies $\epsilon_{\alpha}$ and $\epsilon_{\beta}$, respectively. $\gamma$ can be $x$, or $y$, or $z$.
Here, $n_{\alpha} = \mathcal{F}(\mu - \epsilon_{\alpha})$ is the Fermi function for the 
single-particle energy level $\epsilon_{\alpha}$.
Then, we determine the averaged dc-conductivity by averaging over a small low-frequency interval
($\Delta \omega$) in the following way 
\[ \sigma_{av}(\Delta\omega)=\frac{1}{\Delta\omega}\int_{0}^{\Delta\omega}\sigma(\omega) d\omega,\]
where $\Delta\omega$ is chosen to be three to five times larger than the mean finite-size gap (average eigenvalue
separation) of the system, which is actually the ratio of the bare bandwidth to the total number
of eigenvalues.

The carriers localization/delocalization is determined by the the effective hopping parameter (a measure 
of the gain in kinetic energy)~\cite{White,Mondaini,S2,Araujo} as follows:
\[t_{eff}\equiv \left(\frac{t^{sys}}{t}\right)_{\gamma}
=\frac{{\Big\langle\sum_{i,\sigma}(c_{i+\hat{\gamma},\sigma}^{\dagger}c_{i,\sigma}
+c_{i,\sigma}^{\dagger}c_{i+\hat{\gamma},\sigma})\Big\rangle}_{U,V}}{{\Big\langle\sum_{i,\sigma}(c_{i+\hat{\gamma},\sigma}^{\dagger}c_{i,\sigma}
+c_{i,\sigma}^{\dagger}c_{i+\hat{\gamma},\sigma})\Big\rangle}_{0}},\]
where the angular brackets represent the expectation value in the system. 

Additionally, inverse participation ratio is defined as IPR = $\sum_{i}(\psi_{l}^{i})^{4}$, where $\psi_{l}^{i}$ is the normalized
single-particle wave function associated with the $l$th eigenvalue at site $i$~\cite{Chakraborty}. Here, IPR($\epsilon_{F}$) characterizes
the localized or extended nature of the wave function at the Fermi level $\epsilon_{F}$. All the physical parameters, such as $U$, $T$,
and $\omega$ are measured in units of $t$.
In our calculations of physical observables, the observed error bars are smaller than the size of the data points in the plots.
Therefore, we ommit the error bars from the figures.

%*****************************************************************************************
\begin{figure}[!t]
\centerline{
\includegraphics[width=8.5cm,height=6.30cm,clip=true]{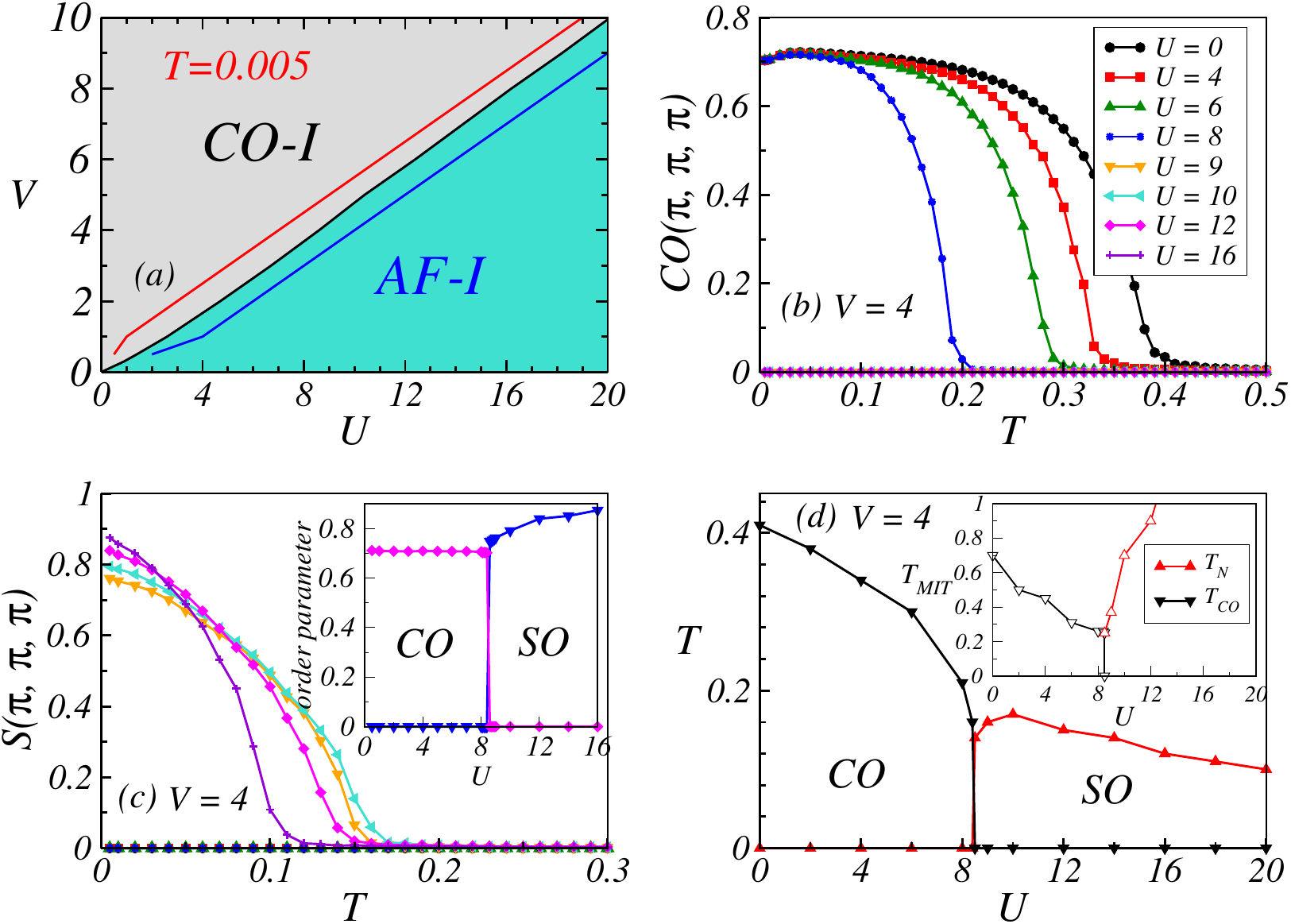}}
\caption{
(a) The ground-state $U$-$V$ phase diagram at electron density $n = 1$ (half-filling) is shown at
low temperature $T = 0.005$. The charge-ordered insulating (CO-I) and antiferromagnetic insulating
(AF-I) phases are separated by a phase boundary (see main text for details). For a fixed $V = 4$,
the temperature dependence of $CO(\pi, \pi, \pi)$ and $S(\pi, \pi, \pi)$ with varying $U$ ($U = 0$--$16$)
is shown in panels (b) and (c), respectively, with identical legends. The results indicate a transition
from the CO phase (at $U = 8$) to the AF phase (at $U = 9$). The precise transition point is determined
to be $U = 8.4$ from the order parameters at $T = 0.005$, as shown in the inset of (c). The inset reveals
a discontinuous transition from the charge-ordered (CO) phase to the spin-ordered (SO) phase, confirming
its first-order nature. (d) The variation of $T_{CO}$ and $T_{N}$ with $U$ is shown for fixed $V = 4$.
While $T_{CO}$ decreases with increasing $U$ and vanishes abruptly above $U \approx 8.4$, $T_{N}$ emerges
in the AF phase, further supporting the first-order CO--AF transition. The corresponding $T_{MIT}$ values
for $U < 8.5$ (open black symbols) and $U \ge 8.5$ (open red symbols) are displayed in the inset.
Temperature $T$ is measured in units of $t$. The blue (red) line in AF (CO) phase along the phase boundary in (a)
is utilized in \textbf{Sec. VIII}.
} 
\label{uv_01}
\end{figure}
%*****************************************************************************************

\section{Exploring U-V phase diagram at half-filling in real-space three dimensions}
To analyze the interplay between $U$ and $V$ in strongly correlated systems, we investigate the
magnetotransport properties of the three-dimensional Hubbard-Holstein model. We first construct
the ground-state $U$-$V$ phase diagram at half-filling ($n = 1$) at a very low temperature $T = 0.005$,
using Monte Carlo simulations for an $8^{3}$ lattice [Fig.~\ref{uv_01}(a)]. To the best of our knowledge,
there has not been any study of the ground-state phase diagram in $3D$ system.
For small $V$ and any finite $U$, the system stabilizes in an antiferromagnetic insulating (AF-I) phase,
whereas for small $U$ and finite $V$, it exhibits a charge-ordered insulating (CO-I) phase due to gap
opening at the Fermi level~\cite{Peierls}. In the weak-coupling regime, these ordered phases originate
from Fermi-surface instabilities (Slater or Peierls)~\cite{Peierls,Slater,Gruner,Jeckelmann,Hohenadler}.
Notably, no metallic phase is observed in our calculations, indicating that the observed AF-I and CO-I phases
are more robust in three dimensions than in lower dimensions.
The transition between AF-I and CO-I occurs abruptly upon increasing $V$ (for fixed $U$) or $U$ (for fixed
$V$), indicating a first-order phase transition. Furthermore, the critical $V$ for the AF--CO transition increases
with increasing $U$ (as $J \sim 1/U$ decreases), consistent with earlier studies of the Holstein $t$-$J$ model~\cite{Prelov,Greco}.
This first-order nature is confirmed by the discontinuous change of the order parameters $S(\pi, \pi, \pi)$ and
$CO(\pi, \pi, \pi)$ with $U$ at fixed $V = 4$ [inset of Fig.~\ref{uv_01}(c)], where the system abruptly switches
from charge order at $U = 8.4$ to spin order at $U = 8.5$. 

%*****************************************************************************************
\begin{figure}[!t]
\centerline{
\includegraphics[width=8.5cm,height=6.30cm,clip=true]{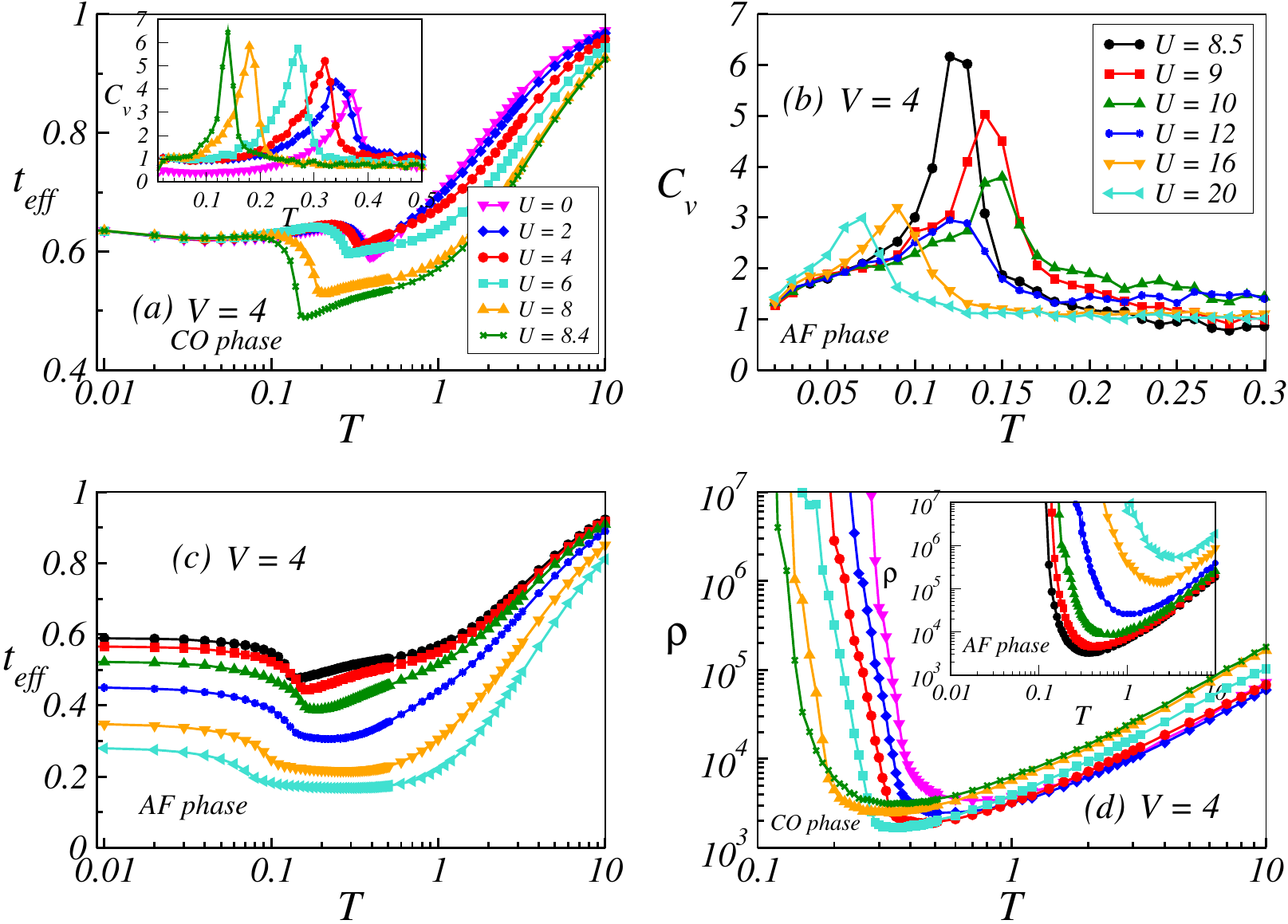}}
\caption{
(a) Temperature dependence of $t_{eff}$ for fixed $V = 4$ and varying $U = 0, 2, 4, 6, 8$, and $8.4$.
In this range ($U < 8.5$), the ground state remains in the CO phase. The dip in $t_{eff}$, associated
with the charge-ordering temperature, shifts to lower $T$ with increasing $U$. Inset: The peak in $C_{v}$,
corresponding to CO formation, also shifts to lower $T$ as $U$ increases.
(b) For $U \ge 8.5$ and $V = 4$, the peak in $C_{v}$, associated with the antiferromagnetic transition,
initially shifts to higher $T$ with increasing $U$ up to $U = 10$, and then decreases.
(c) The temperature corresponding to the dip in $t_{eff}$ vs $T$ follows the same trend as in (b),
increasing with $U$ up to $U = 10$ and decreasing thereafter. Panels (a)--(c) collectively support
the behavior of $T_{CO}$ and $T_{N}$ shown in Fig.~\ref{uv_01}(d).
(d) For $U < 8.5$ (CO phase), the metal-insulator transition temperature $T_{MIT}$ decreases with increasing $U$,
whereas for $U \ge 8.5$ (AF phase) $T_{MIT}$ increases with $U$ (inset), consistent with the inset of Fig.~\ref{uv_01}(d).
Temperature $T$ is measured in units of $t$.
} 
\label{uv_02}
\end{figure}
%*****************************************************************************************

We now present the charge structure factor [$CO(\pi, \pi, \pi)$] and spin structure factor [$S(\pi, \pi, \pi)$] 
in Fig.~\ref{uv_01}(b) and (c), respectively, for a fixed $V = 4$ and varying U, that were used to prepare the
$U$-$V$ phase diagram in Fig.~\ref{uv_01}(a). The charge ordering temperature ($T_{CO}$) decreases with increasing
$U$ up to $U = 8.4$, then drops to zero at $U = 8.5$ and above [see Fig.~\ref{uv_01}(b) and (d)]. For larger V,
charge carriers form bipolarons that organize in a staggered pattern, establishing a charge-ordered (CO) phase below $T_{CO}$. 
The electronic correlation $U$ affects bipolaron formation and their effective interactions~\cite{Nowadnick,Sangiovanni,Sangiovanni1},
leading to a decrease in $T_{CO}$ with increasing $U$, as the onsite $U$ weakens charge ordering. For $U \ge 8.5$, the system
converts to an antiferromagnetic phase, confirmed by the antiferromagnetic structure factor [S($\pi, \pi, \pi$)] in
Fig.~\ref{uv_01}(c) and $T_{N}$ in Fig.~\ref{uv_01}(d). In this AF phase, the Neel temperature $T_{N}$ increases with
decreasing $U$ in the larger $U$ range ($U \ge 10$) but begins to decrease near the phase boundary for $U < 10$
due to the proximity effect of the charge-ordered phase. Moreover, the metal-insulator transition temperature $T_{MIT}$ 
[estimated from $\rho$ ($\rho \sim \frac{1}{\sigma}$) vs $T$ plot by the temperature with minimum resistivity in Fig.~\ref{uv_02}(d)] decreases 
in both CO and AF regimes as one approaches the transition line, as illustrated in the inset of Fig.~\ref{uv_01}(d).

%*****************************************************************************************
\begin{figure}[!t]
\centerline{
\includegraphics[width=8.5cm,height=5.00cm,clip=true]{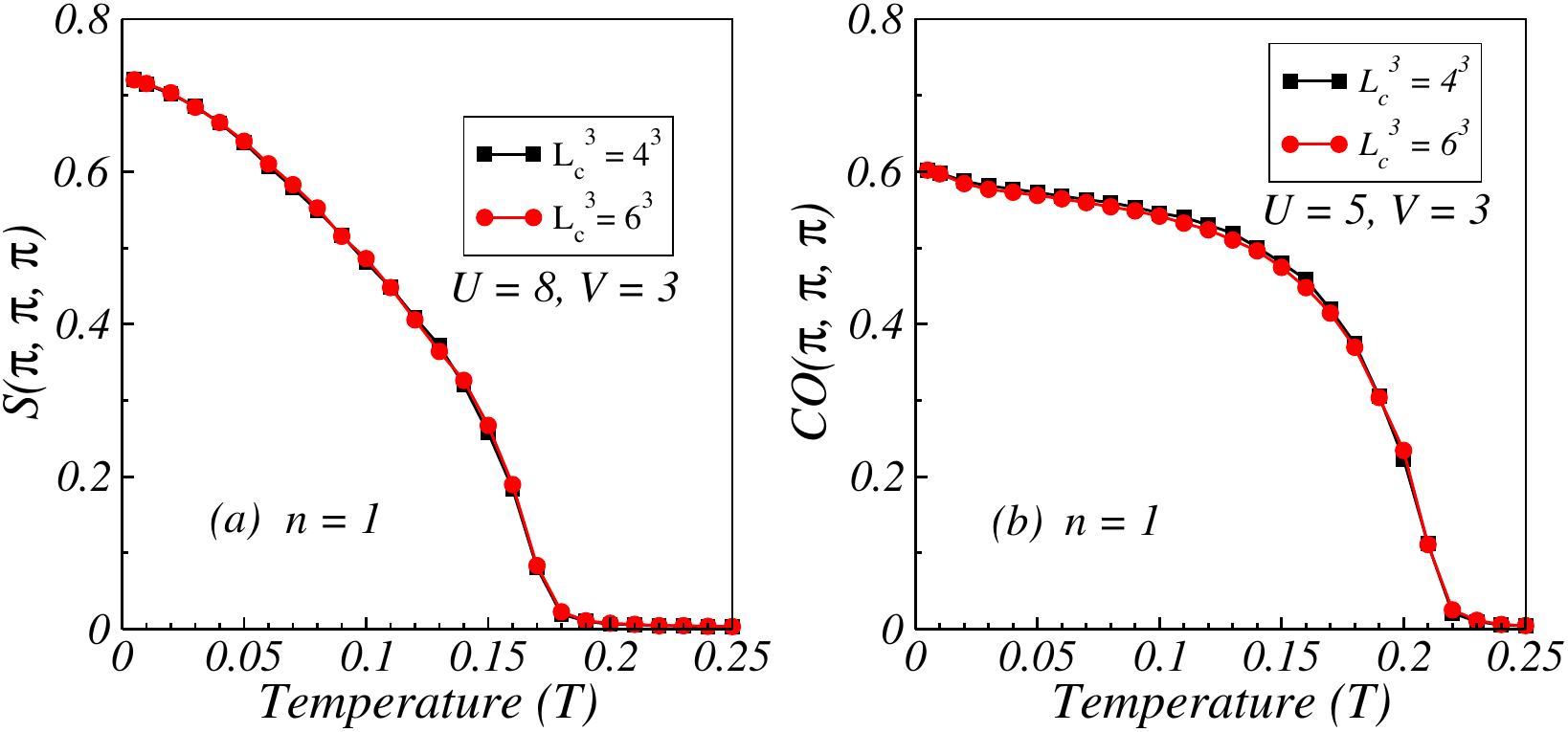}}
\caption{
(a) $S(\pi, \pi, \pi)$ vs $T$ is plotted for the parameter set $U = 8, V= 3$ for two different TCA cluster sizes, $4^3$ and $6^3$.
The plots overlap with each other. (b) $CO(\pi, \pi, \pi)$ vs $T$ is plotted for the parameter set $U = 5, V = 3$ using two different
TCA cluster sizes, $4^3$ and $6^3$. Here, too, the two plots are indistinguishable. Both (a) and (b) are prepared at half-filling $n = 1$.
Temperature $T$ is measured in units of $t$.
} 
\label{tca_size}
\end{figure}
%*****************************************************************************************

For characterization of the typical behavior of $T_{CO}$, $T_{N}$, and $T_{MIT}$ with $U$ in Fig.~\ref{uv_01}(d), we evaluate 
the effective hopping ($t_{eff}$), specific heat ($C_{v}$), resistivity ($\rho$) (see Fig.~\ref{uv_02}). The dip in 
$t_{eff}$ marks the ordering temperature (AF or CO), below which $t_{eff}$ increases due to kinetic energy gain for 
stabilizing the order. The peak in $C_{v}$ also corresponds to the ordering temperature. In the CO regime ($U < 8.5$), 
the dip in $t_{eff}$ and the peak in $C_{v}$ both shift to lower temperatures with increasing $U$, indicating
a suppression of $T_{CO}$ (see Fig.~\ref{uv_02}(a) and Fig.~\ref{uv_01}(d)). In the AF regime ($U \ge 8.5$), the peak position of
$C_{v}$ exhibits nonmonotonic behavior. It shifts to higher temperature as $U$ increases from $8.5$ to $10$, then moves to lower
temperature for larger $U$ (Fig.~\ref{uv_02}(b)). The dip in $t_{eff}$(Fig.~\ref{uv_02}(c)) follows the same trend, corroborating
the behavior of $T_{N}$ in Fig.~\ref{uv_01}(d). The resistivity $\rho$ (Fig.~\ref{uv_02}(d)) further shows that the metal-insulator
transition temperature ($T_{MIT}$) decreases (increases) with increasing $U$ in the CO (AF) regime, consistent with the inset of
Fig.~\ref{uv_01}(d).

Overall, our obtained $U$-$V$ phase diagram is consistent with the phase diagram at half-filling in $2D$~\cite{Pai}, except for the metallic
phase that arises in $2D$ at low values of $U$ and $V$. In quantum Monte Carlo (QMC) study at half-filling in $1D$~\cite{Hardikar},
a metallic phase is observed between the CO and AF phases, and the region of this metallic phase contracts as approaches to the adiabatic
limit, ultimately leading to a phase diagram similar to Fig.~\ref{uv_01}(a).

To examine the convergence of our results with respect to the traveling cluster size, we calculate the temperature dependence of $S(\pi,\pi,\pi)$
and $CO(\pi,\pi,\pi)$ for two representative parameter sets corresponding to the AF and CO phases in Fig.~\ref{uv_01}(a), using TCA cluster sizes
of $4^3$ and $6^3$. As shown in Figs.~\ref{tca_size}(a) and (b), the curves obtained using the two cluster sizes are nearly indistinguishable over
the entire temperature range, including the ordering transitions. This agreement supports the adequacy of the $4^3$ traveling cluster size for
describing the magnetic and charge-ordering behavior in the parameter regimes examined.

%*****************************************************************************************
\begin{figure}[!t]
\centerline{
\includegraphics[width=8.5cm,height=5.00cm,clip=true]{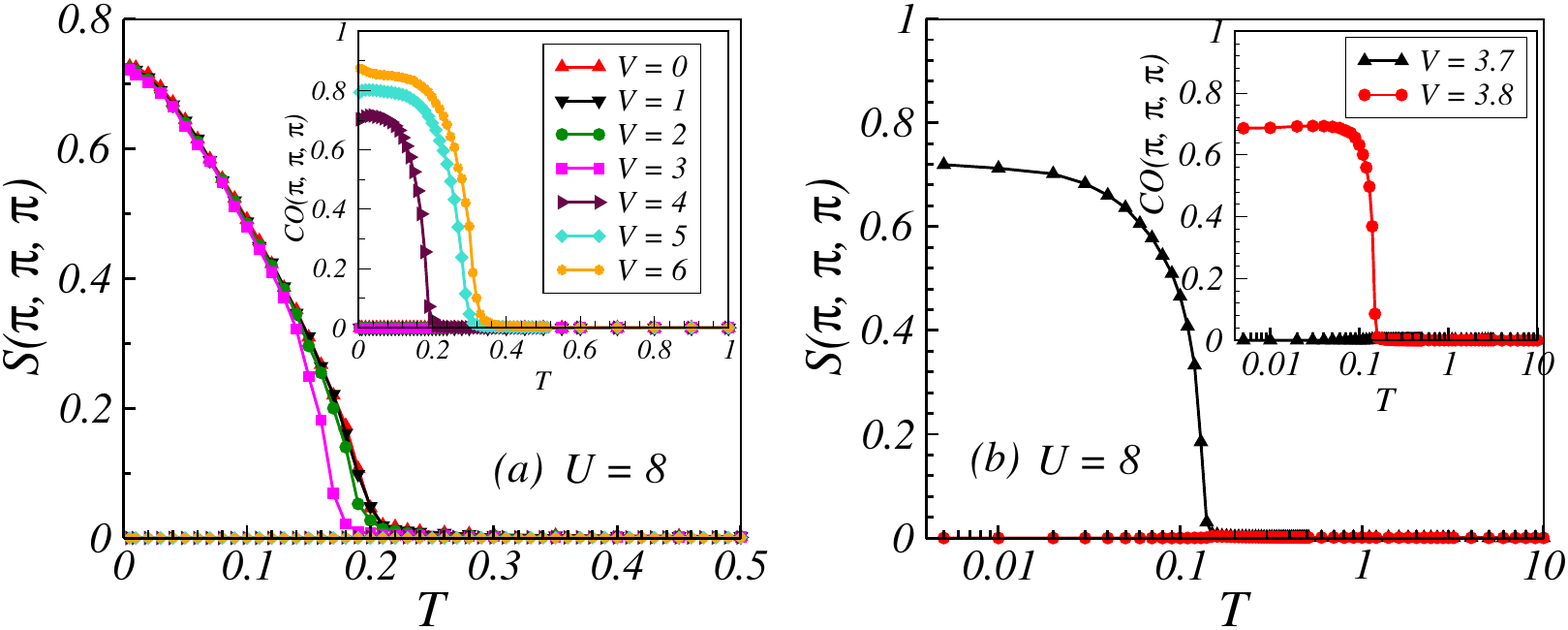}}
\caption{
(a) Antiferromagnetic structure factor $S(\pi, \pi, \pi)$ vs $T$ for fixed $U = 8$  and varying $V = 0$--$6$.
Up to $V = 3$, the system exhibits an AF transition with the Neel temperature $T_{N}$ decreasing as $V$ increases.
Inset: $CO(\pi, \pi, \pi)$ vs $T$ shows that for $V = 4, 5, 6$ the system is in the CO phase, with the
charge-ordering temperature $T_{CO}$ increasing with $V$. Legends in the main panel and inset are identical.
(b) $S(\pi, \pi, \pi)$ vs $T$ for $V = 3.7$ and $3.8$. Inset: $CO(\pi, \pi, \pi)$ vs $T$ for the same $V$ values.
The system remains in the AF phase for $V = 3.7$ and in the CO phase for $V = 3.8$. Temperature $T$ is measured
in units of $t$.
} 
\label{vt_01}
\end{figure}
%*****************************************************************************************

\section{Phase characterization across the boundary for U = 8}

In this section, we examine the effect of $V$ on the MH-I and AF-I states
and search for distinct phases from low to high temperature regimes. We explore the magnetic, charge,
and transport properties for a fixed $U = 8$ (away from perturbative and large $U$ limits) and varying $V$
and present the $V$ - $T$ phase diagram. Apparently, the system remains AF for $V \le 3$ and transitions to a CO phase
for $V \ge 4$, depicted in Fig.~\ref{vt_01}(a). An abrupt AF to CO transition between $V = 3.7$ and $3.8$,
indicating a first-order transition, is observed as illustrated in Fig.~\ref{vt_01}(b).

%*****************************************************************************************
\begin{figure}[!t]
\centerline{
\includegraphics[width=8.5cm,height=6.30cm,clip=true]{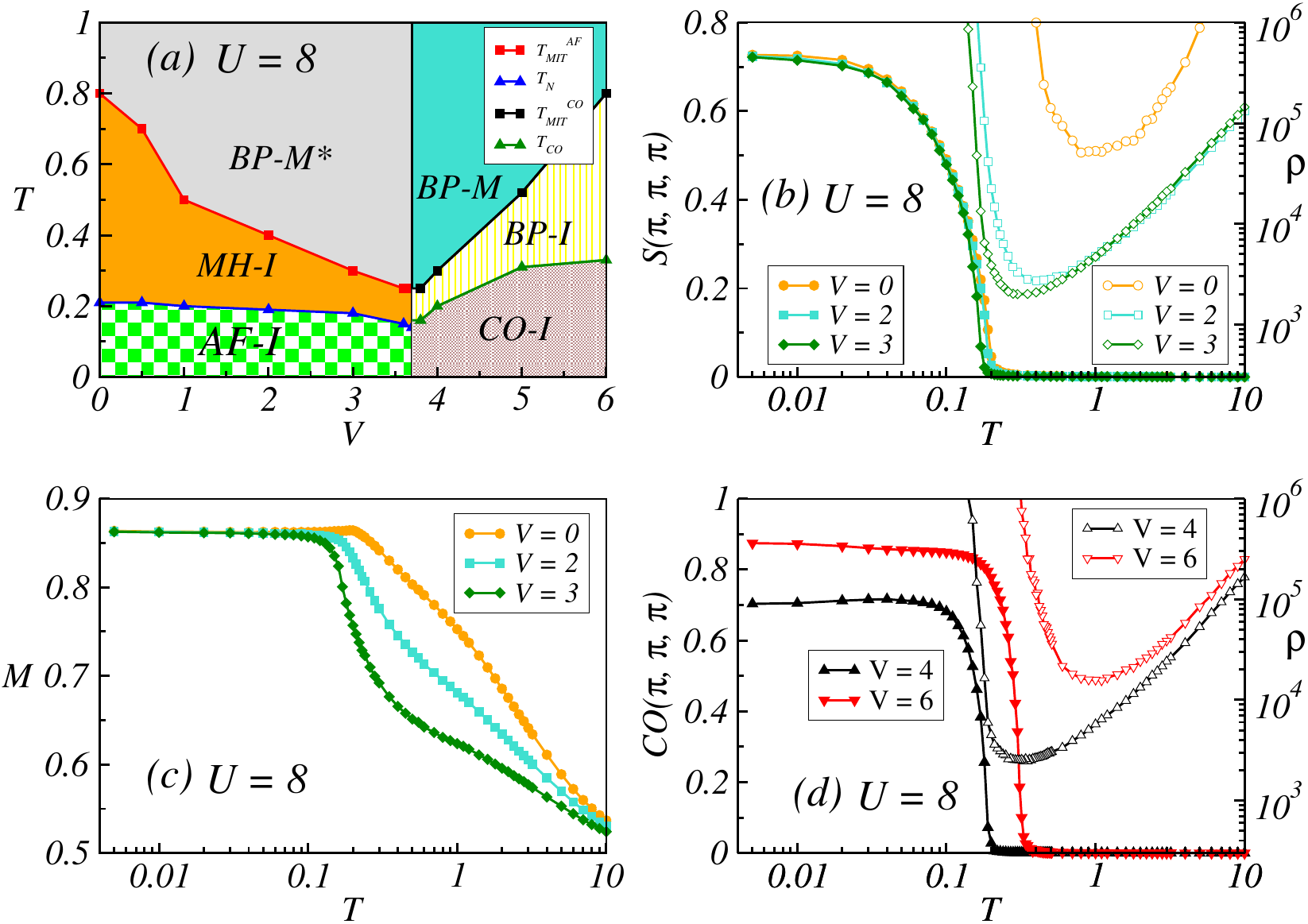}}
\caption{
(a) $V$-$T$ phase diagram for fixed $U = 8$. At low temperatures, the system undergoes a
first-order transition from the AF-I to CO-I phase near $V \approx 3.7$. Additional phases
such as the Mott-Hubbard insulator (MH-I), bipolaronic insulator (BP-I), and two bipolaronic
metallic phases (BP-M and BP-M*) are also observed at high temperatures (see text for details).
(b) $S(\pi, \pi, \pi)$ vs $T$ (left) and resistivity $\rho$ vs $T$ (right) for $V = 0, 2$,
and $3$ at $U = 8$. The Neel temperature $T_{N}$ decreases slightly with increasing $V$,
while the metal-insulator transition temperature $T_{MIT}$ decreases more significantly.
Near the phase boundary, $T_{MIT}$ remains larger than $T_{N}$.
(c) Temperature dependence of the local moment $M$ for $V = 0, 2$, and $3$. At high
temperatures, $M$ decreases with $V$.
(d) Temperature evolution of $CO(\pi, \pi, \pi)$ (left) and resistivity $\rho$ (right). Both the
charge-ordering temperature $T_{CO}$ and $T_{MIT}$ decrease as $V$ is reduced from $6$ to $4$.
Near the phase boundary, $T_{MIT} > T_{N}$, indicating the persistence of the BP-I regime.
Temperature $T$ is measured in units of $t$. 
} 
\label{vt_02}
\end{figure}
%*****************************************************************************************

%*****************************************************************************************
\begin{figure}[!t]
\centerline{
\includegraphics[width=8.5cm,height=5.00cm,clip=true]{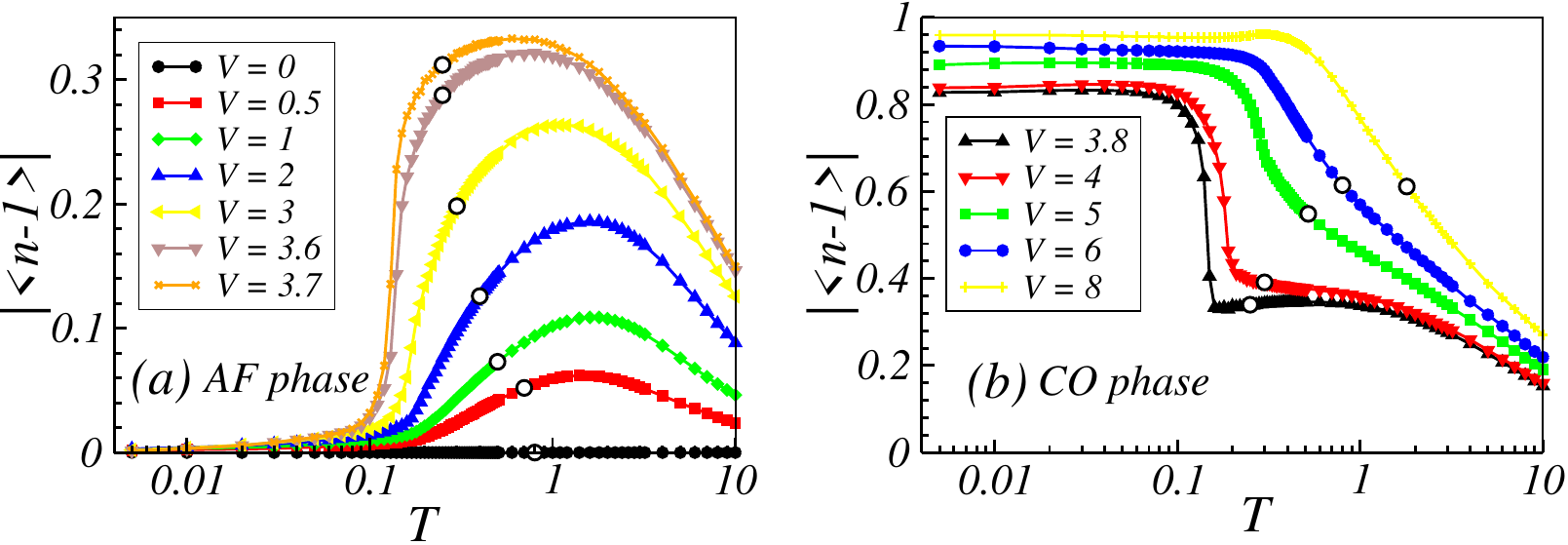}}
\caption{
(a) Temperature dependence of the bipolaronic order parameter $BPO = |<n-1>|$ for fixed $U = 8$ and
varying $V \le 3.7$ (up to the phase boundary in Fig.~\ref{vt_02}(a)). In this range, the ground
state is antiferromagnetic. BPO increases with decreasing $T$, reaches a maximum near $T \approx 1$,
and decreases at lower temperatures. Its high-temperature magnitude increases with $V$, while in the
pure Hubbard limit ($V = 0$) BPO vanishes at all temperatures.
(b) For $V > 3.7$, where the ground state is charge ordered, BPO increases with decreasing $T$ and
saturates at low temperature. The low-temperature magnitude of BPO increases monotonically with $V$.
Open circles in (a) and (b) indicate the temperatures where the system shows metal-insulator transition
for the corresponding $V$. 
} 
\label{vt_03}
\end{figure}
%*****************************************************************************************
In Fig.~\ref{vt_02}(a) we plot the $V$ - $T$ phase diagram for $U = 8$. At the pure Hubbard case, $V = 0$,
we find a successive metal-insulator (Mott) and antiferromagnetic transitions at temperatures $T_{MIT}\sim 0.8$
and $T_{N} = 0.21$, respectively, upon decreasing the temperature.
In the temperature range $T_{N} < T \le T_{MIT}$, the system remains in a MH-I state, or in other words,
paramagnetic insulating (PM-I) state. When $V$ is turned on, the ground state at low tempertures remains
an AF-I state up to $V = 3.7$ and then transitions discontinuously to a CO-I state above this point. The transition is first-order
in nature, consistent with previous studies~\cite{Karakuzu}, although those studies were reported in lower
dimensions. The MH-I region between $T_{MIT}$ and $T_{N}$ decreases with increasing
$V$ but survives in a finite region even near the transition line at $V = 3.7$. To illustrate the reduction
of the MH-I region, we plot  $S(\pi, \pi, \pi$) vs $T$ (left panel) and $\rho$ vs $T$ (right panel)
at fixed $U = 8$ for $V = 0, 2, 3$ in Fig.~\ref{vt_02}(b). The Neel temperature $T_{N}$ decreases minutely
with increasing $V$, whereas $T_{MIT}$ decreases significantly. Eventually, $T_{MIT}$ remains higher than $T_{N}$
even near the transition line, indicating the presence of the MH-I region between $T_{MIT}$ and $T_{N}$.
In the pure Hubbard limit ($V = 0$), large onsite local moment formation at high temperature ($T\sim 1$;
see Fig.~\ref{vt_02}(c)) drives the metal-insulator transition~\cite{Borejsza,Borejsza1}. But in Hubbard-Holstein
case ($V \ne 0$), the electron-phonon coupling ($V$) hinders onsite local moment formation at high temperature.
As a result, larger local moments develop only at low temperatures compared to the pure Hubbard case (Fig.~\ref{vt_02}(c)),
leading to a significant decrease in $T_{MIT}$ with increasing $V$.
For $V > 3.7$, bipolarons form at high temperature and order below $T_{CO}$ (Fig.~\ref{vt_02}(a)). Both $T_{CO}$
and $T_{MIT}$ increase with $V$ (compare Fig.~\ref{vt_02}(a) and (d)). The intermediate BP-I region between $T_{MIT}$
and $T_{CO}$ narrows as $V$ decreases but remains finite near the transition line $V \sim 3.8$.
Notably, just above $T_{N}$ and $T_{CO}$, the system exhibits a first-order transition between the Mott-Hubbard
insulator and the bipolaronic insulator, consistent with earlier DMFT+NRG studies~\cite{Jeon} at strong coupling
and zero temperature.

The system, which at low temperatures is in either the AF or CO phases, evolves into bipolaronic metallic phases (BP-M* or BP-M)
at higher temperatures. To characterize the BP-M and BP-M* phases, we compute the bipolaronic order parameter
$BPO = |<n-1>| =\frac{1}{N} \sum_{i}|n_{i}-1|$~\cite{Jeon}, where $N$ is the number of lattice sites.
This quantity measures average local charge deviations (i.e., average charge imbalance) from half-filling ($<n>=1$).
The real-space distributions of charge density, lattice distortion, and double occupancy reveal spatial correlations
among these quantities, highlighting the interplay between charge inhomogeneity, lattice distortions, and electronic
correlations, discuss in Section \textbf{VI}. These observations indicate bipolaronic tendencies
but do not, by themselves, establish the formation of bound bipolarons.
In the BP-M* phase ($V \le 3.7$), $|<n-1>|$
decreases as the temperature is lowered from $T \sim 1$  to $T_{MIT}$ (Fig.~\ref{vt_03}(a)). In contrast, in the
BP-M phase, it increases upon cooling over the same temperature range (Fig.~\ref{vt_03}(b)). Thus, the bipolaronic
order parameter distinguishes BP-M* from BP-M, although both phases are metallic. In the CO regime ($V > 3.7$), the
system undergoes successive transitions from CO-I to BP-I to BP-M with increasing temperature (Fig.~\ref{vt_02}(a)).
Similarly, in the AF regime ($V \le 3.7$), it transitions from AF-I to MH-I to the distinct bipolaronic
metallic phase (BP-M*).
Here, the transition between BP-M* and BP-M is first-order, occurring abruptly as $V$ changes from $3.7$ to $3.8$.

%*****************************************************************************************
\begin{figure}[!t]
\centerline{
\includegraphics[width=8.5cm,height=6.30cm,clip=true]{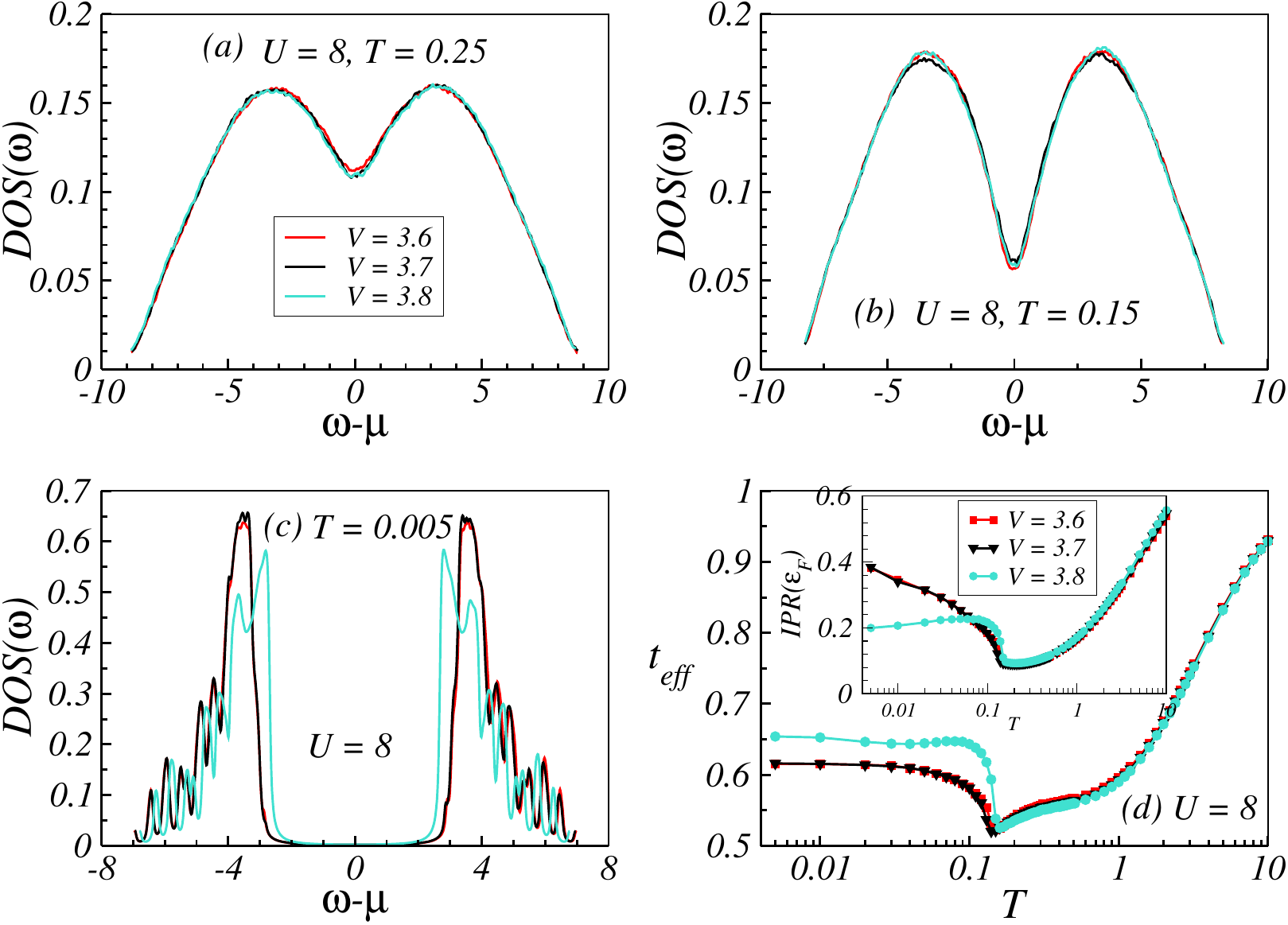}}
\caption{
Density of states (DOS) for fixed $U = 8$ and $V = 3.6, 3.7$, and $3.8$ at three temperatures:
(a) $T = 0.25$ (near $T_{MIT}$), (b) $T = 0.15$ (near $T_{N}$ and $T_{CO}$), and (c) $T = 0.005$. The DOS
for different $V$ values overlap at $T = 0.25$ and $T = 0.15$. At low temperature ($T = 0.005$),
the DOS for $V = 3.6$ and $3.7$ (AF ground state) separates from that of $V = 3.8$ (CO ground state).
(d) Temperature dependence of $t_{eff}$ for $V = 3.6, 3.7$, and $3.8$, which overlap down to the
ordering temperature and separate below it. Inset: $IPR(\epsilon_{F})$ shows a similar behavior.  
} 
\label{upg_01}
\end{figure}
%*****************************************************************************************

\section{Universality and Pseudogap features near the transition line}
In the $V$-$T$ phase diagram shown in Fig.~\ref{vt_02}(a), several phases emerge at different temperature
and $V$ scales. To understand the origin of these transitions, we focus on the region near the phase boundary
for $U = 8$ and $V = 3.6$, $3.7$, and $3.8$. As discussed earlier, the ground state at low temperature is
antiferromagnetic (AF) for $V = 3.6$ and $3.7$, while it becomes charge ordered (CO) for $V = 3.8$. The
corresponding ordering temperatures, namely the Neel temperature $T_{N}$ and the charge-ordering temperature
$T_{CO}$, are nearly identical ($T_{N} \sim T_{CO} \sim 0.15$).
To examine the origin of these ordered states, we analyze the density of states (DOS) at three representative
temperatures: $T = 0.25$ (near the metal-insulator transition temperature $T_{MIT}$), $T = 0.15$ (near $T_{N}$
and $T_{CO}$), and a low temperature $T = 0.005$, as shown in Fig.~\ref{upg_01}(a)--(c). At low temperature,
the DOS in both AF and CO phases exhibits a clear gap around the Fermi level ($\omega = 0$), indicating insulating
behavior. The opening of this gap coincides with the onset of long-range order. Notably, across $V = 3.7$, the
gap appears at nearly the same temperature for both AF and CO phases, suggesting that the electronic gap drives
the transition to the ordered phase, while the values of $U$ and $V$ determine the nature of the ordered state.
At higher temperatures, the metal-insulator transition occurs at approximately the same temperature ($T_{MIT}\sim 0.25$)
for both AF and CO ground states. The DOS curves collapse onto each other near $T_{MIT}$ (Fig.~\ref{upg_01}(a))
and also near the ordering temperature (Fig.~\ref{upg_01}(b)), indicating that the electronic spectra are indistinguishable
above the ordering temperatures.
To support this observation, we compute the effective hopping parameter $t_{eff}$ (a measure of the average kinetic
energy) and the inverse participation ratio IPR($\epsilon_{F}$) (a measure of electronic localization), shown in
Fig.~\ref{upg_01}(d). The $t_{eff}$ curves overlap down to the ordering temperature and separate below it, consistent
with the DOS results. The IPR shows a similar trend (see the inset of Fig.~\ref{upg_01}(d)). These results indicate
a universal electronic behavior above the ordering temperature, suggesting that the electronic DOS plays a dominant
role in driving the metal-insulator transition.

%*****************************************************************************************
\begin{figure}[!t]
\centerline{
\includegraphics[width=8.5cm,height=6.30cm,clip=true]{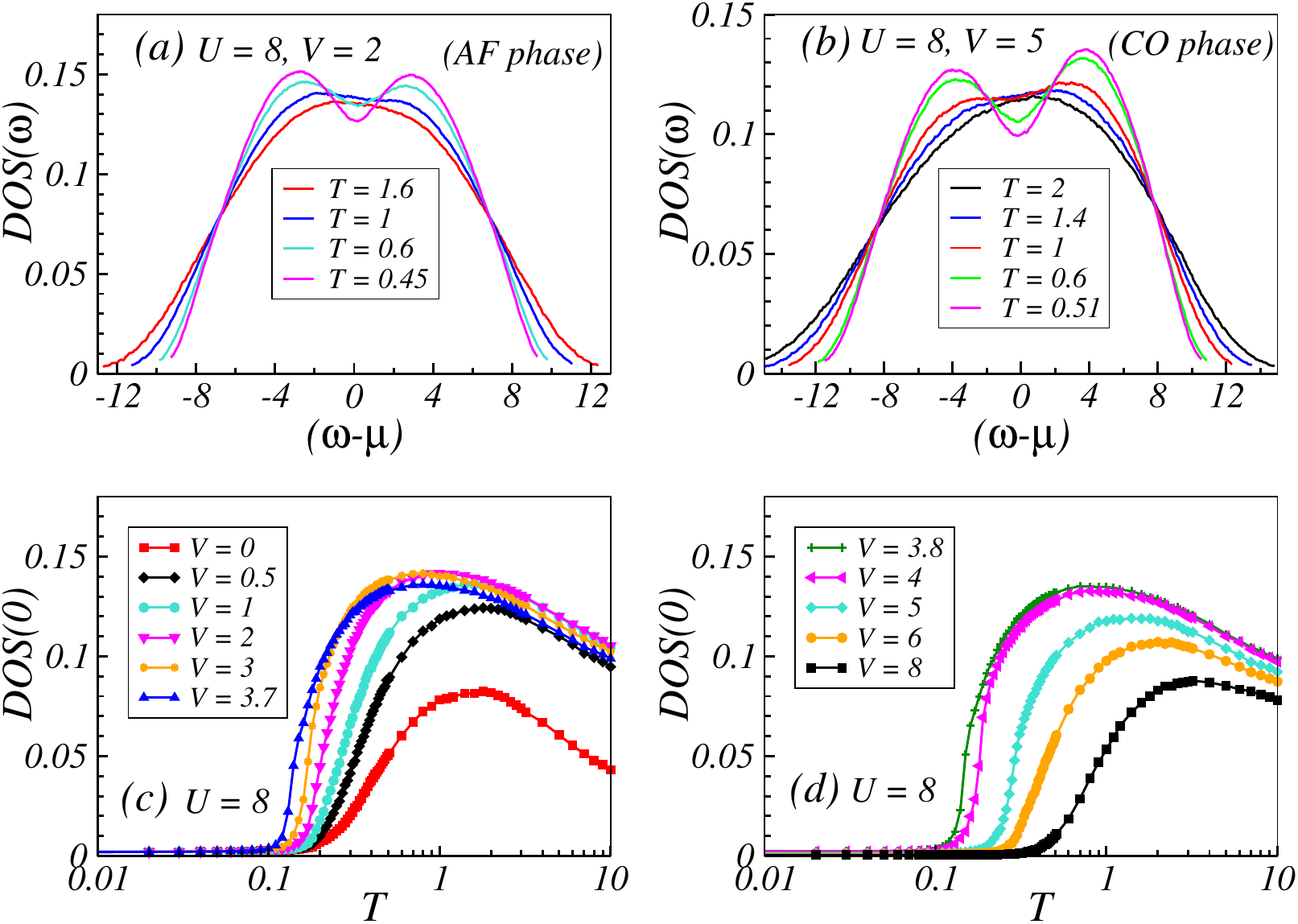}}
\caption{
(a) Density of states (DOS) for $U = 8, V = 2$,  and temperatures $T = 1.6, 1, 0.6$, and $0.45$.
In this parameter regime, the ground state is antiferromagnetic (AF) [Fig.~\ref{vt_02}(a)], and
a pseudogap develops below $T\sim 1$. (b) DOS for $U = 8$, $V = 5$, and $T = 2, 1.4, 1, 0.6$,
and $0.51$. Here the ground state is charge ordered (CO), and a pseudogap appears below $T\sim 1.4$.
Temperature dependence of the DOS at the Fermi level, $DOS(0)$, for (c) $V \le 3.7$ (AF phase)
and (d) $V > 3.7$ (CO phase), respectively. The peak in $DOS(0)$, associated with pseudogap
formation, shifts to lower (higher) temperatures with increasing $V$ in the AF (CO) regime.
The increasing magnitude of $DOS(0)$ for $V \le 3.7$ reflects enhanced charge fluctuations,
consistent with the $\chi_{n}$ results in Fig.~\ref{upg_03}(b).  
} 
\label{upg_02}
\end{figure}
%*****************************************************************************************

%*****************************************************************************************
\begin{figure}[!t]
\centerline{
\includegraphics[width=8.5cm,height=6.30cm,clip=true]{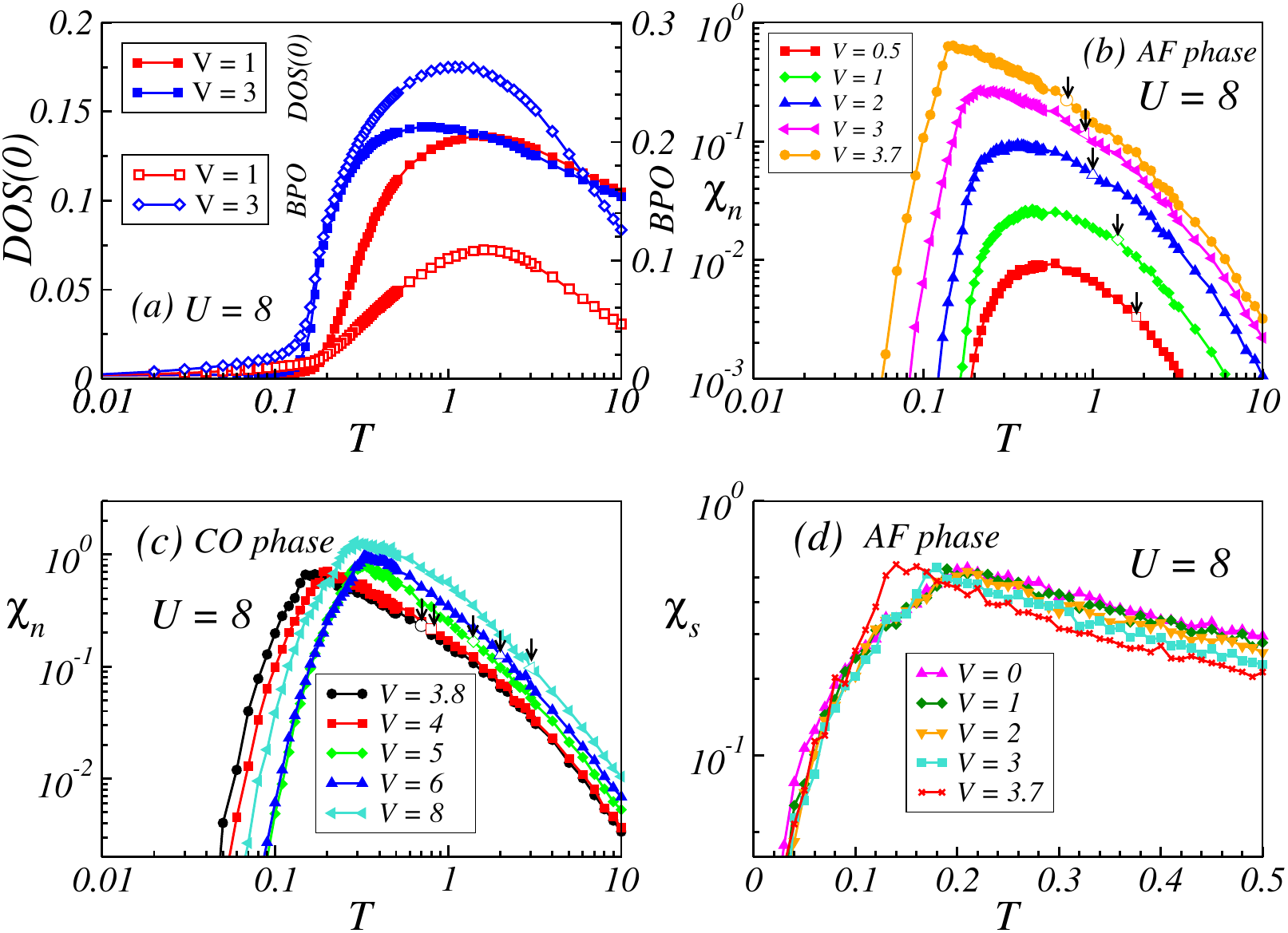}}
\caption{
(a) Temperature dependence of the bipolaronic order parameter BPO (right) and $DOS(0)$ (left) for
fixed $U = 8$ and $V = 1$ and $3$. The peaks of BPO and $DOS(0)$ occur at nearly the same temperature,
indicating a close connection between pseudogap formation and charge imbalance.
Density susceptibility $\chi_{n}$ for (b) $V \le 3.7$ (AF ground state) and (c) $V > 3.7$ (CO ground state).
Arrows mark the pseudogap formation temperature $T_{PG}$, obtained from the peak in $DOS(0)$ [Fig.~\ref{upg_02}(c) and (d)].
Significant charge fluctuations below $T_{PG}$ lead to pseudogap formation.
(d) Spin susceptibility $\chi_{s}$ for $V \le 3.7$; the peak corresponds to the antiferromagnetic Neel temperature $T_{N}$.   
} 
\label{upg_03}
\end{figure}
%*****************************************************************************************

Since the spectral properties exhibit a degree of universality near the transition line, it is important
to investigate them across the entire $V$-$T$ phase diagram in Fig.~\ref{vt_02}(a), particularly at higher
temperatures. For this purpose, we consider two representative points away from the transition line: $U = 8,
V = 2$ and $U = 8, V = 5$, corresponding to AF and CO ground states, respectively.
Figure ~\ref{upg_02}(a) shows that for $U = 8, V = 2$, the DOS begins to develop a local minimum around the
Fermi level ($\omega = 0$) below $T \sim 1$, while the corresponding metal-insulator transition occurs at
$T_{MIT} \sim 0.4$. Similarly, for $U = 8, V = 5$, a local minimum appears below $T \sim 1.4$ (Fig.~\ref{upg_02}(b)),
which is significantly higher than the transition temperature $T_{MIT}\sim 0.51$. These results indicate
that insulating behavior emerges when a sufficient gap forms in the DOS at the Fermi level, leading to a
metal-insulator transition with $T_{MIT} < T_{PG}$.
Here, $T_{PG}$ denotes the pseudogap formation temperature, defined as the highest temperature below which
the DOS develops a minimum at the Fermi level, following the conventional definition used in correlated
materials~\cite{Esterlis}.
To determine $T_{PG}$, we compute the DOS weight at the Fermi level, $DOS(0)$, for both $V \le 3.7$ and $V > 3.7$,
as shown in Fig.~\ref{upg_02}(c) and (d). In the AF regime ($V \le 3.7$), the peak position of $DOS(0)$, which signals $T_{PG}$,
shifts to lower temperatures as $V$ approaches the transition line $V = 3.7$, while its magnitude increases. This
behavior likely arises from enhanced charge fluctuations (short-range charge correlations) at higher temperatures
as the system approaches the boundary. The trend is consistent with the resistivity results in Fig.~\ref{vt_02}(b)
and indicates enhanced metallicity near the transition line.
In contrast, for $V > 3.7$, the peak position of $DOS(0)$ shifts to higher temperatures with increasing $V$, while
its magnitude decreases (Fig.~\ref{upg_02}(d)). This reduction in $DOS(0)$ is associated with the increasing tendency
toward localized bipolaron formation with increasing $V$ (see also Fig.~\ref{vt_03}(b)).

%*****************************************************************************************
\begin{figure}[!t]
\centerline{
\includegraphics[width=8.5cm,height=4.00cm,clip=true]{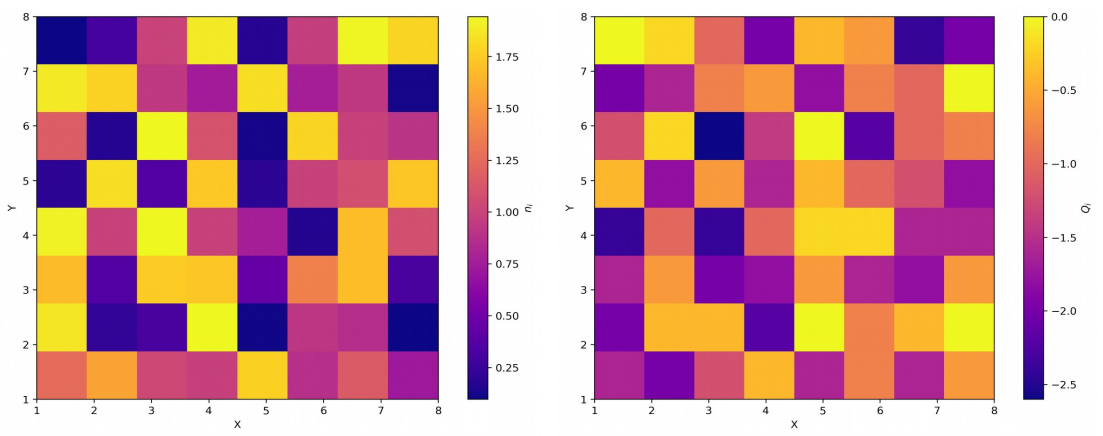}}
\caption{
Real-space snapshot of the charge density $n_{i}$ (left column) and lattice distortion $Q_{i}$ (right column) are
represented in 2D XY plane of the $8^3$ lattice. Here, the variables are shown at high temperature $T = 0.8$, which
is below the pseudogap formation temperature $T_{PG}$, for the parameter set $U = 8, V = 5$. The charge densities
are inhomogeneously distributed. The associated lattice distortions closely follow the density profile with higher (lower)
density site has larger (smaller) amplitude of lattice distortions.
} 
\label{snap_nQ}
\end{figure}
%*****************************************************************************************

To examine the correlation between pseudogap formation and charge imbalance at high temperatures for $V \le 3.7$
(where the ground state is AF), we plot the bipolaronic order parameter (BPO) and $DOS(0)$ for $V = 1$ and $3$
in Fig.~\ref{upg_03}(a). The peak positions of $DOS(0)$ coincide with those of the BPO, indicating that pseudogap
formation is closely related to charge imbalance. In other words, sufficiently strong charge fluctuations (short-range
charge correlations) drive the pseudogap formation.
To further support this observation, we calculate the density susceptibility $\chi_{n} = \frac{1}{N}(<n^{2}>-<n>^{2})$~\cite{Weber,Esterlis}
for different $V$ values in the AF regime, shown in Fig.~\ref{upg_03}(b). A pronounced enhancement of charge
fluctuations is observed below the pseudogap formation temperature $T_{PG}$ (indicated by the solid black arrow),
estimated from the peak positions in Fig.~\ref{upg_02}(c).
As $V$ decreases (moving away from the transition line),
the charge fluctuations weaken, leading to a reduction in $DOS(0)$, consistent with Fig.~\ref{upg_02}(c).
Similarly, for $V > 3.7$, a significant density susceptibility is observed below $T_{PG}$, as shown in Fig.~\ref{upg_03}(c).
In the temperature regime where the pseudogap is formed, the real-space snapshots (Fig.~\ref{snap_nQ}) reveal inhomogeneous
charge distributions accompanied by corresponding lattice distortions due to electron-phonon coupling. These observations suggest
that coupled charge and lattice inhomogeneities may contribute to pseudogap formation. However, the possible role of preformed
bipolaron pairing, as discussed in Ref.~\cite{Kovac}, can not be excluded. Distinguishing between these mechanism requires a more
detailed investigation of pairing correlations using a fully quantum treatment of the phonon degrees of freedom, which is beyond
the scope of the present study.
In this regime (CO phase), the peak in $\chi_{n}$ corresponds to the charge ordering transition, and its temperature
follows the same $V$-dependence as $T_{CO}$. Furthermore, the magnetic susceptibility
$\chi_{s}$ [$\chi_{s} = \frac{1}{N}(<(S^{z})^{2}>-<S^{z}>^{2})$, $S^{z}=\frac{1}{N}\sum_{i}{S_{i}^{z}}$] also exhibits
peaks that coincide with the antiferromagnetic transition for $V \le 3.7$, as shown in Fig.~\ref{upg_03}(d). The Neel
temperature $T_{N}$ aligns well with the peak positions of $\chi_{s}$.

\section{Magnetotransport properties for low to intermediate U and V combinations across the boundary}

\begin{figure}[!t]
\centerline{
\includegraphics[width=8.5cm,height=6.30cm,clip=true]{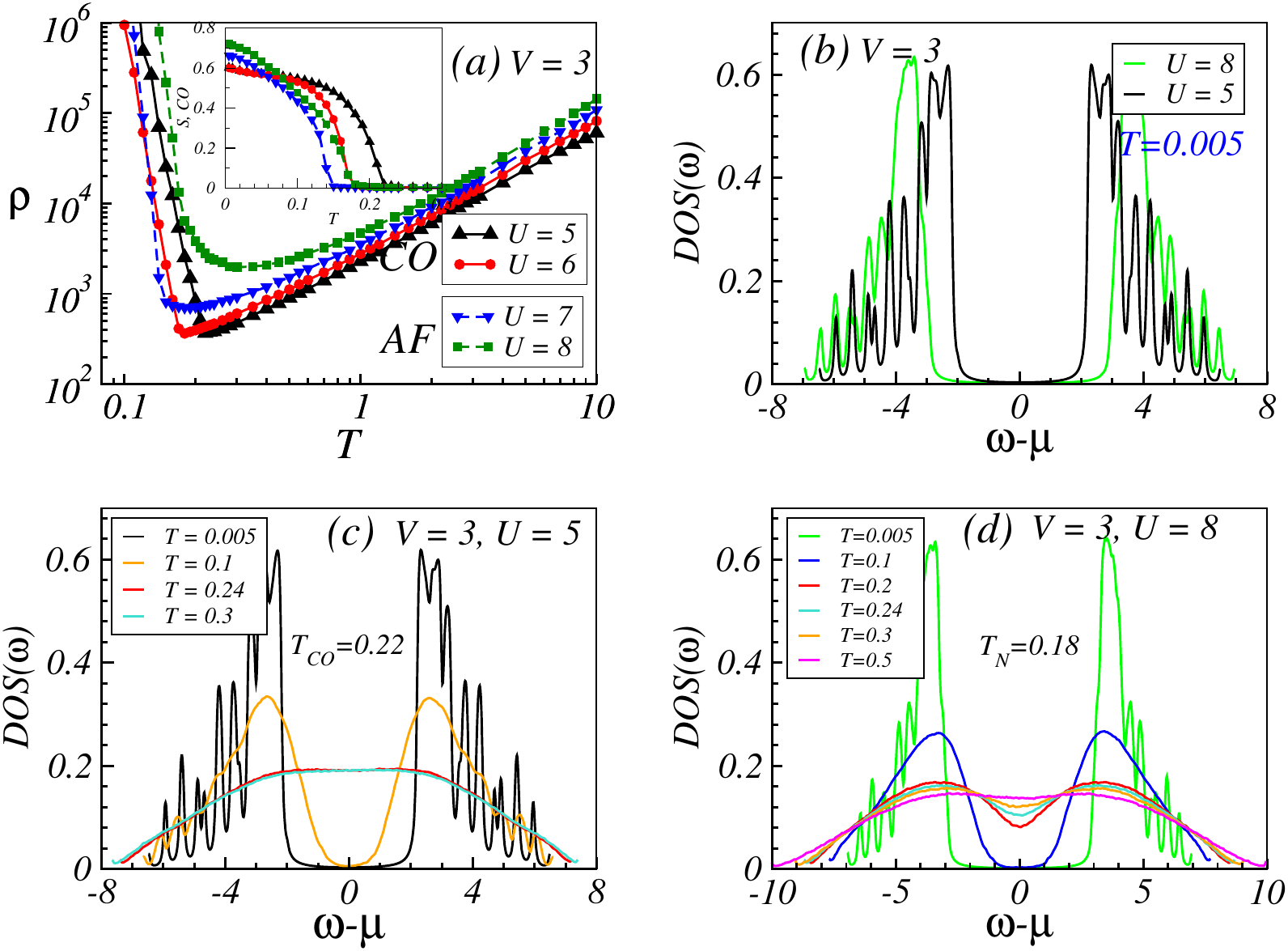}}
\caption{
(a) Resistivity $\rho$ vs $T$ for fixed $V = 3$ and varying $U = 5, 6, 7$, and $8$ across the phase
boundary in Fig.~\ref{uv_01}(a). The system exhibits a metal-insulator transition (MIT) for all $U$.
Inset: structure factors $CO(\pi, \pi, \pi)$ (solid) and $S(\pi, \pi, \pi)$ (dashed). $T_{MIT}$ coincides
with $T_{CO}$, while for $U = 8$ (AF phase) $T_{MIT} > T_{N}$.
(b) DOS near the Fermi level ($\omega = \epsilon_{F} = 0$) at $T = 0.005$, showing a clear gap in both
CO ($U = 5$) and AF ($U = 8$) phases.
(c) For $U = 5$ (CO phase), the gap weakens with increasing $T$ and closes above $T_{CO}\sim 0.22$, yielding
a finite DOS at the Fermi level.
(d) For $U = 8$ (AF phase), the gap softens with increasing $T$ and evolves into a pseudogap above $T_{N}\sim 0.18$,
with spectral weight at the Fermi level arising from local spin correlations.  
} 
\label{mtp_01}
\end{figure}

In this section, we explore the physics of low to intermediate values of $U$ and $V$ at half-filling,
which may be relevant for materials where both electronic correlations and electron-phonon coupling are
moderate. We focus on the magnetotransport and electronic spectral properties by evaluating the resistivity
($\rho$) and density of states (DOS), as shown in Fig.~\ref{mtp_01}.
Figure ~\ref{mtp_01}(a) presents the resistivity for fixed $V = 3$ and varying $U = 5, 6, 7, 8$ across
the boundary of the $U$-$V$ phase diagram in Fig.~\ref{uv_01}(a). The system is in the charge-ordered (CO) phase
for $U = 5$ and $6$, while it exhibits an antiferromagnetic (AF) phase for $U = 7$ and $8$ at low temperatures,
as shown in the inset of Fig.~\ref{mtp_01}(a). In both phases, the system undergoes a metal-insulator transition
and becomes insulating at low temperatures. In the CO phase, the charge-ordering temperature $T_{CO}$ decreases
with increasing $U$, reflecting the suppression of onsite double occupancy by the Hubbard interaction.
In this regime, the metal-insulator transition temperature approximately coincides with the charge-ordering temperature
($T_{MIT}\sim T_{CO}$). In contrast, in the AF phase for $U = 8$, the metal-insulator transition temperature
exceeds the Neel temperature ($T_{MIT} > T_{N}$), indicating the presence of a Mott insulating state.
To further understand the spectral properties, we analyze two representative points in the phase diagram:
$V = 3, U = 5$ (CO phase) and $V = 3, U = 8$ (AF phase). The $DOS(\omega)$ for both cases shows a clear gap
at low temperature (Fig.~\ref{mtp_01}(b)) around the Fermi level ($\omega = 0$), confirming the insulating
behavior observed in the transport results [Fig.~\ref{mtp_01}(a)]. The gap is larger in the AF phase due to
the stronger Hubbard interaction. In the CO phase ($U = 5$), the gap decreases with increasing temperature
(Fig.~\ref{mtp_01}(c)) and disappears just above the charge-ordering temperature ($T_{CO}\sim 0.22$), resulting
in a metallic state consistent with the resistivity results.
In contrast, in the AF phase ($U = 8$), a pseudogap feature appears in the DOS above the Neel temperature
($T_{N}\sim 0.18$), as shown in Fig.~\ref{mtp_01}(d). This pseudogap originates from local fluctuations of
spin, charge, and phonon degrees of freedom above $T_{N}$, which introduce finite spectral weight inside
the gap region.
Such fluctuation-driven pseudogap behavior also results in well-known non-Drude features in the optical conductivity~\cite{Pai}.

%*****************************************************************************************
%*****************************************************************************************

%*****************************************************************************************
\begin{figure}[!t]
\centerline{
\includegraphics[width=8.5cm,height=6.30cm,clip=true]{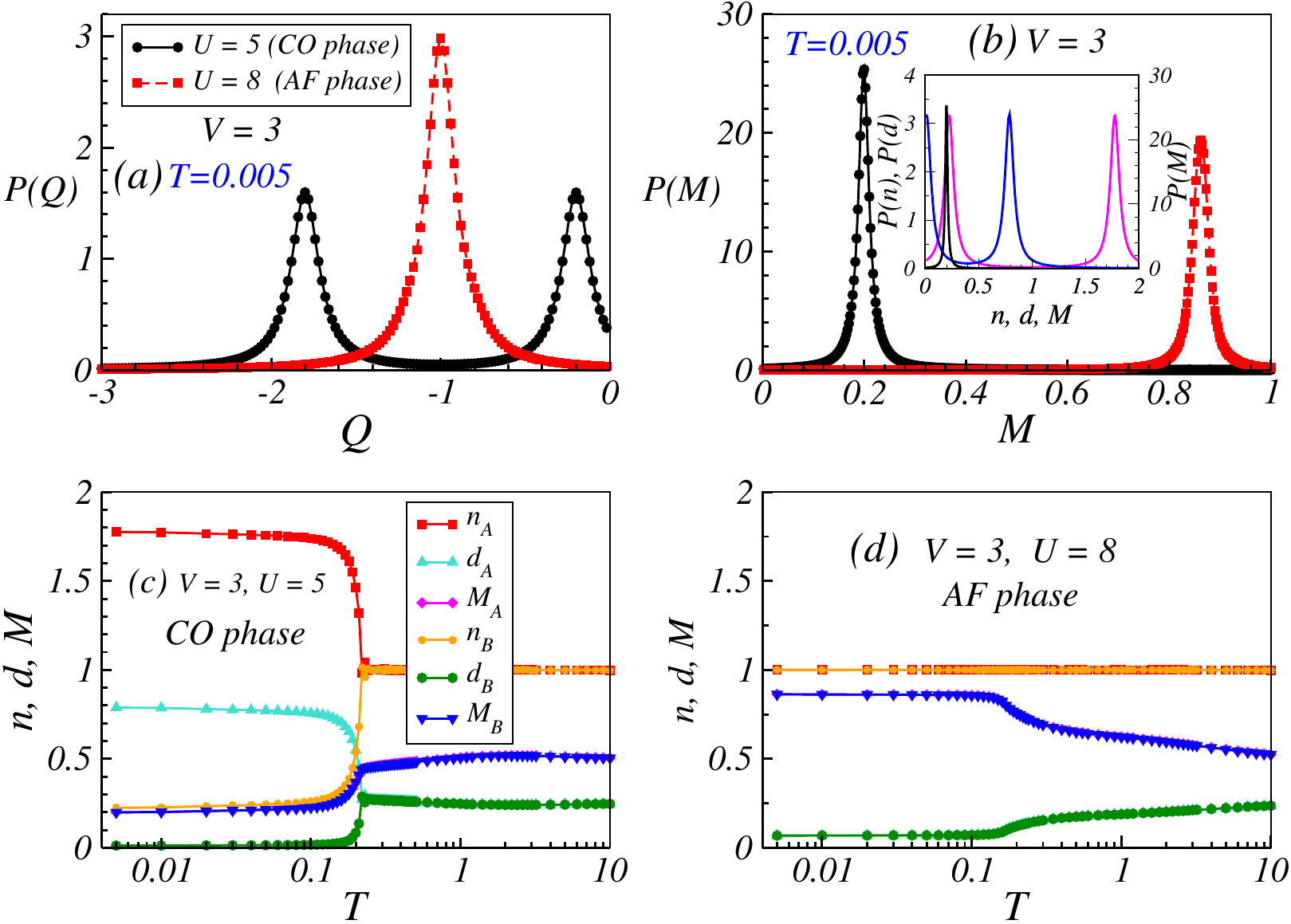}}
\caption{
(a) Phonon distribution $P(Q)$ at $T = 0.005$ for fixed $V = 3$ in the charge-ordered ($U = 5$) and
antiferromagnetic ($U = 8$) phases. $P(Q)$ exhibits a bimodal distribution in the CO phase and a
unimodal distribution in the AF phase.
(b) Corresponding local moment distributions $P(M)$, showing unimodal behavior in both phases.
Inset: In the CO phase, the density $P(n)$ (magenta) and double occupancy $P(d = n_{\uparrow}n_{\downarrow})$
(blue) are bimodal, whereas $P(M)$ (black) remains unimodal. Legends in (a) and (b) are identical.
Temperature evolution of sublattice-resolved densities ($n_{A}, n_{B}$), double occupancies ($d_{A}, d_{B}$),
and local moments ($M_{A}, M_{B}$) for (c) $U = 5, V = 3$ (CO phase) and (d) $U = 8, V = 3$ (AF phase).  
} 
\label{mtp_02}
\end{figure}
%*****************************************************************************************

To further characterize the ground-state phases beyond transport and spectral properties, we examine the phonon
distribution function P(Q) and the local moment distribution P(M) at low temperature $T = 0.005$, shown in
Fig.~\ref{mtp_02}(a) and (b). In the CO phase ($V = 3, U = 5$), P(Q) exhibits a bimodal distribution~\cite{Ciuchi,Pai},
indicating a staggered pattern of elongated and contracted phonon modes, consistent
with experimental observations~\cite{Haule,Munoz,Alonso}. In contrast, the antiferromagnetic (AF) phase
($V =3, U = 8$) shows a unimodal phonon distribution, reflecting uniform lattice distortion and nearly uniform
site occupancy.
The local moment distribution P(M), however, remains unimodal in both phases (Fig.~\ref{mtp_02}(b)). In the AF phase,
the peak occurs at a larger moment due to reduced double occupancy at larger $U$. In the CO phase, the peak appears
at a smaller moment. Interestingly, despite the bimodal distributions of density $n$  and double occupancy
$d = n_{\uparrow}n_{\downarrow}$ in the CO phase (inset of Fig.~\ref{mtp_02}(b)), P(M) remains unimodal.
To clarify this behavior, we compute sublattice-resolved quantities $n_{A,B}$, $d_{A,B}$, and $M_{A,B}$ (Fig.~\ref{mtp_02}(c)).
Below the charge-ordering temperature $T_{CO}$, the A-sublattice shows an increase in density and double occupancy,
while the B-sublattice shows a corresponding decrease. This results in equivalent local moments on both sublattices 
($M_{A}=n_{A}-2d_{A}\equiv M_{B}=n_{B}-2d_{B}$), leading to a uniform local moment distribution in the CO phase.
The sublattice splitting of double occupancy is consistent with earlier studies~\cite{Bauer}.
For completeness, the corresponding sublattice-resolved quantities for the AF phase are shown in Fig.~\ref{mtp_02}(d).
Here $n_{A}\sim n_{B}$  and $d_{A}\sim d_{B}$, with very small double occupancy, producing a larger average
local moment compared to the CO phase, while still satisfying $M_{A}\sim M_{B}$.

%*****************************************************************************************
\begin{figure}[!t]
\centerline{
\includegraphics[width=8.5cm,height=6.30cm,clip=true]{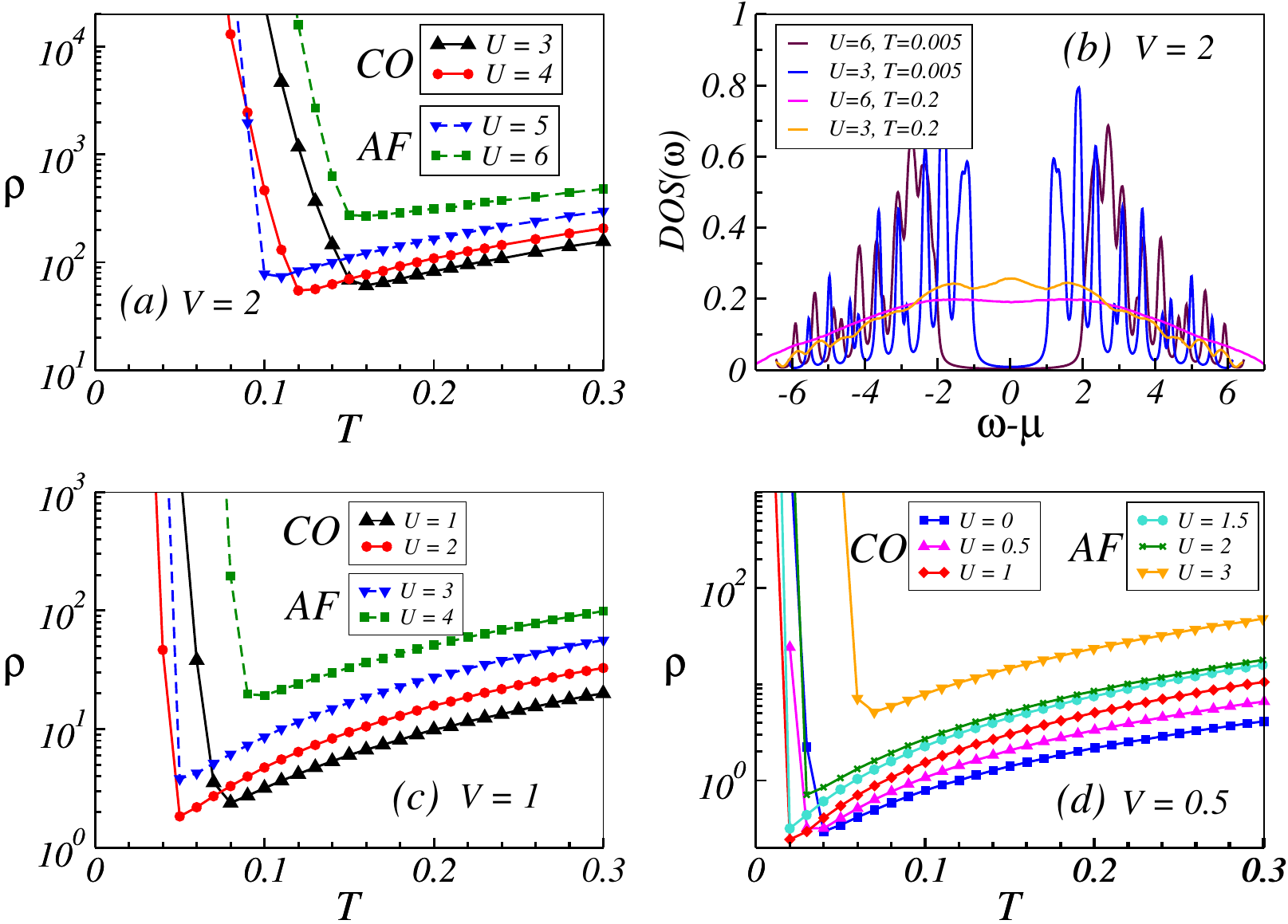}}
\caption{
(a) Resistivity $\rho$ vs temperature $T$ for fixed $V = 2$ and $U = 3$--$6$ across the phase
boundary in Fig.~\ref{uv_01}(a). MITs for $U = 3, 4$ correspond to charge ordering (solid) and
for $U = 5, 6$ to spin ordering (dashed).
(b) DOS for CO ($V = 2, U = 3$) and AF ($V = 2, U = 6$) phases at $T = 0.005$ and $0.2$. At low $T$,
both phases show a clear gap; at $T = 0.2$ (above ordering), the DOS at the Fermi level is finite.
$\rho$ vs $T$ for (c) $V = 1$, $U = 1$--$4$ and (d) $V = 0.5$, $U = 0$--$3$. All cases exhibit MITs,
with insulating ground states at low temperatures.
} 
\label{mtp_03}
\end{figure}
%*****************************************************************************************

Most previous studies of the half-filled Hubbard-Holstein model in lower dimensions ($1D$ and $2D$) report
a metallic phase at small $U$ and $V$ near the boundary between the AF and CO phases. In contrast, our $U$-$V$
phase diagram in Fig.~\ref{uv_01}(a) shows no evidence of a metallic phase in this regime. To substantiate this
result, we examine the magnetic, transport, and electronic properties for small $U$ and $V$.
We first analyze the magnetotransport properties at $V = 2$ by varying $U = 3, 4, 5$, and $6$ across the phase
boundary in Fig.~\ref{uv_01}(a). For $U = 3, 4$ the system is in the CO phase, while for $U = 5, 6$ it is in the
AF phase, as confirmed by $S(\pi, \pi, \pi)$ and $CO(\pi, \pi, \pi)$ calculations (not shown). As shown in Fig.~\ref{mtp_03}(a),
the system exhibits a metal-insulator transition for all values of $U$. Specifically, the transitions for $U = 3, 4$ ($U = 5, 6$)
correspond to charge (spin) ordering. In this weak-coupling regime, CO arises from Peierls instability~\cite{Peierls,Hwang}, while AF order
originates from Slater instability~\cite{Slater} driven by Fermi-surface nesting~\cite{Costa,Lee}. Consequently, $T_{MIT}$ coincides with the
corresponding ordering temperature ($T_{CO}$ or $T_{N}$). Moreover, $T_{MIT}$ increases away from the phase
boundary--decreasing $U$ in the CO phase and increasing $U$ in the AF phase. 
The DOS at low temperature ($T = 0.005$) exhibits a clear gap at the Fermi level for both CO ($U = 3$)
and AF ($U = 6$) phases, confirming insulating ground states (Fig.~\ref{mtp_03}(b)). At higher temperature
($T = 0.2$), above the ordering temperatures, a finite DOS at the Fermi level indicates metallic behavior,
consistent with the resistivity results.
We further compute resistivity for smaller values $V = 1, 0.5$, and below $0.5$ while varying $U$ across
the phase boundary. In all cases, the system undergoes a metal-insulator transition and remains insulating
at low temperatures, as shown in Fig.~\ref{mtp_03}(c) and (d). These results collectively support the absence
of a metallic phase in the $U$-$V$ phase diagram of Fig.~\ref{uv_01}(a). Overall, the observed AF-I and CO-I
phases are more robust in three dimensions than in lower dimensions even for small $U$ and $V$ combinations.

%*****************************************************************************************
\begin{figure}[!t]
\centerline{
\includegraphics[width=8.5cm,height=6.30cm,clip=true]{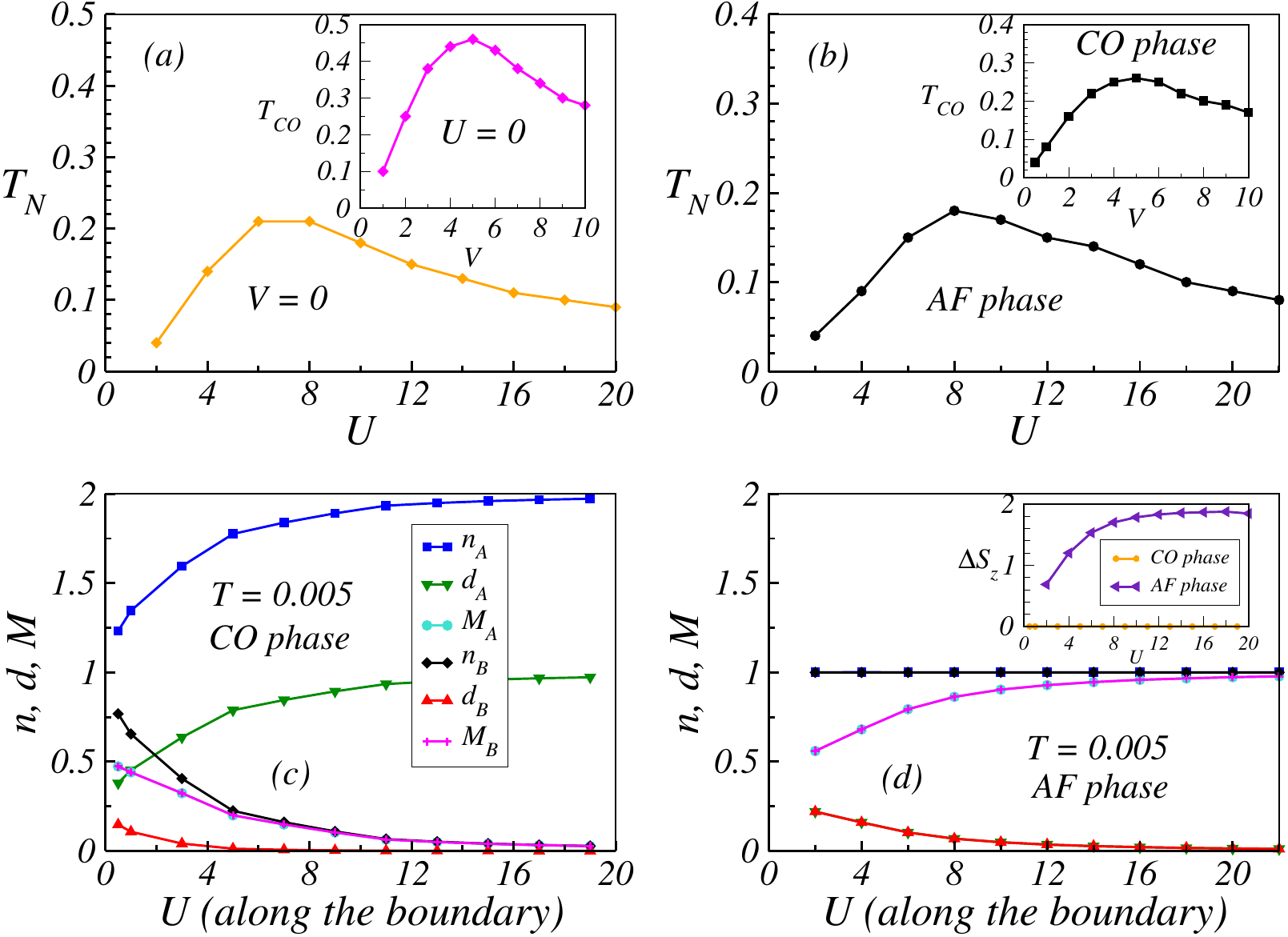}}
\caption{
(a) The Neel temperature $T_{N}$ for the pure Hubbard model ($V = 0$) shows a nonmonotonic dependence on $U$,
reaching a maximum around $U \sim 8$. The inset displays the charge-ordering temperature $T_{CO}$ for the
pure Holstein model ($U = 0$), which also shows nonmonotonic behavior with an optimum near $V \sim 5$.
(b) $T_{N}$ is plotted for a series of ($U, V$) points along the AF boundary (blue line) in Fig.~\ref{uv_01}(a):
($2, 0.5$), ($4, 1$), ($6, 2$), ($8, 3$), ($10, 4$), ($12, 5$), ($14, 6$), ($16, 7$), ($18, 8$), ($20, 10$),
($22, 10$) (set-I). For clarity, the plot shows only the variation with $U$. The inset shows $T_{CO}$ along
the CO boundary (red line) in Fig.~\ref{uv_01}(a) for the points ($0.5, 0.5$), ($1, 1$), ($3, 2$), ($5, 3$), ($7, 4$), ($9, 5$),
($11, 6$), ($13, 7$), ($15, 8$), ($17, 9$), ($19, 10$) (set-II), displayed as a function of $V$.
Sublattice-resolved densities ($n_{A}, n_{B}$), double occupancies ($d_{A}, d_{B}$), and local moments ($M_{A}, M_{B}$)
are shown along the boundary line in the (c) CO phase (set-II) and (d) AF phase (set-I). The inset of (d) presents the
variation of $\Delta S_{z} = |S_{z}^{A}-S_{z}^{B}|$ along the CO and AF boundaries.
} 
\label{diagonal}
\end{figure}
%*****************************************************************************************

\section{Features along the boundary of the phase diagram}
In the previous sections, we characterized the phases across the $U$-$V$ phase diagram by varying $U$ at fixed $V$
and vice versa. Here, we investigate the properties along the phase boundary, indicated by the blue (AF) and red (CO)
lines in Fig.~\ref{uv_01}(a). Since phases near the boundary are highly sensitive to perturbations, factors such
as doping, strain, frustration, or long-range hopping may drive transitions or stabilize new phases. Therefore,
understanding the physics along the boundary is important.
To examine this, we consider several ($U, V$) points along the CO boundary (red line in Fig~\ref{uv_01}(a)):
($0.5, 0.5$), ($1, 1$), ($3, 2$), ($5, 3$), ($7, 4$), ($9, 5$), ($11, 6$), ($13, 7$),
($15, 8$), ($17, 9$), ($19, 10$). Similarly, along the AF boundary (blue line in Fig.~\ref{uv_01}(a)) we consider:
($2, 0.5$), ($4, 1$), ($6, 2$), ($8, 3$), ($10, 4$), ($12, 5$), ($14, 6$), ($16, 7$), ($18, 8$), ($20, 10$), ($22, 10$).

In order to compare our results obtained along the phase boundary in presence of both $U$ and $V$ with 
pure Hubbard ($U = 0$) or Holstein ($V = 0$) limits, we first describe the properties in these pure limits 
and then discuss the case varying both $U$ and $V$ along the boundary.
For the pure Hubbard limit ($V = 0$), the system remains antiferromagnetic for any finite $U$, and the Neel
temperature $T_{N}$ varies nonmonotonically~\cite{Staudt,Hirsch,Scalettar,Jarrell,Ulmke} with $U$ (Fig.~\ref{diagonal}(a)). This nonmonotonic
behavior with an optimum $T_{N}$ around $U \sim 8$ is consistent with the previous QMC study~\cite{Song}. The maximum
value arises at intermediate $U$ due to the competition between quantum and thermal fluctuations. However, the QMC study
is unable to capture low-temperature physics in the regime $T < 0.1$ due to the severe sign problem, whereas our method
can probe very low temperatures down to $T = 0.005$. In the pure Holstein limit ($U = 0$), the charge-ordering temperature
$T_{CO}$ also shows a nonmonotonic dependence on $V$, as shown in the inset of Fig.~\ref{diagonal}(a). Here, the observed
charge order structure factor peak is found at the wave vector $\mathbf{q} = (\pi, \pi, \pi)$, which is also observed in
the Langevin QMC study~\cite{Stead}. A direct one-to-one comparison is not possible since our study is in the adiabatic
limit ($\omega = 0$) in contrast to the Langevin QMC study at a finite phonon frequency ($\omega \ne 0$ ). Although no
direct results exist for the 3D Holstein model in the adiabatic phonon limit, similar behavior has been reported in two
dimensions~\cite{Pai}, despite debates about long-range order due to the Mermin-Wagner theorem~\cite{Mermin,Hohenberg}.
A DMFT study on a bipartite lattice~\cite{Ciuchi} also reported nonmonotonic $T_{CO}$ with
unimodal and bimodal phonon distributions depending on the electron-phonon coupling. In contrast,
our study (in the adiabatic limit) shows only bimodal phonon distributions.
Along the AF boundary, the nonmonotonic behavior of $T_{N}$ persists even with finite $V$ (Fig.~\ref{diagonal}(b)),
reaching a maximum near $(U, V) = (8, 3)$. However, the magnitude of $T_{N}$ is slightly reduced compared to the
pure Hubbard case, indicating that the physics in the AF phase is primarily governed by $U$, with $V$ providing a
secondary effect. Similarly, along the CO boundary, $T_{CO}$ retains its nonmonotonic behavior (inset of Fig.~\ref{diagonal}(b))
and reaches a maximum near $(U, V) = (9, 5)$. The magnitude of $T_{CO}$, however, is significantly reduced in the
presence of finite $U$ compared to the pure Holstein model.

At low temperature $T = 0.005$ in the CO regime, the A-sublattice density $n_{A}$ increases with $U$, while $n_{B}$
decreases along the boundary line (Fig.~\ref{diagonal}(c)). Although both $U$ and $V$ vary, the plot is shown versus
$U$ for clarity. The densities saturate at $n_{A}\sim 2$ and $n_{B}\sim 0$ for $U \ge 11$ ($V \ge 5$). Thus, densities
split along the boundary even for small $(U, V)$, with a similar splitting in double occupancies: $d_{A}$ increases
and $d_{B}$ decreases, saturating at $d_{A}\sim 1$ and $d_{B}\sim 0$. This corresponds to an alternating pattern of
doubly occupied and empty sites in the CO phase. Despite this, the local moments remain equal in both sublattices
($M_{A} = M_{B}$), decreasing from the paramagnetic limit ($0.5$) toward the bipolaronic limit ($0$) with increasing $U$.
In the AF regime, the sublattice densities remain equal ($n_{A} = n_{B} = 1$) along the boundary (Fig.~\ref{diagonal}(d)).
The double occupancies decrease with increasing $U$, leading to larger local moments, with $M_{A} = M_{B}$. To further
distinguish the phases, we evaluate $\Delta S_{z} = |S_{z}^{A}-S_{z}^{B}|$ (inset of Fig.~\ref{diagonal}(d)).
$\Delta S_{z}$ vanishes in the CO phase but is finite in the AF phase and increases with $U$. These results provide a
comprehensive description of the behavior of the AF and CO phases in the presence of both $U$ and $V$ along the phase boundary,
and are therefore important for understanding and designing quantum materials that host both AF and CO phases, as well as
for studying the proximity effects between them.

\section{Conclusions}
Motivated by the lack of theoretical studies in three dimensions and the experimental relevance of coexisting electronic
correlations and electron-phonon interactions, we investigate the magnetotransport properties of the half-filled one-band
Hubbard-Holstein model using an exact diagonalization based semiclassical Monte Carlo (s-MC) approach.
We first establish the ground-state $U$-$V$ phase diagram, where the antiferromagnetic (AF) and charge-ordered
(CO) phases are separated by a first-order transition line. Throughout the phase diagram, the system remains
insulating, as confirmed by resistivity and density-of-states (DOS) calculations.
In particular, no metallic phase is found even for small $U$ and $V$, suggesting that the observed
AF-I and CO-I phases are more robust in three dimensions than in lower dimensions.
For an optimal correlation strength $U = 8$, the $V$-$T$ phase diagram reveals a rich set of phases,
including AF-I, CO-I, MH-I, BP-I, BP-M*, and BP-M. Robust first-order metal-to-metal and insulator-to-insulator
transitions occur near $V\sim 3.75$. The distinction between BP-M and BP-M* is identified from the
temperature dependence of the bipolaronic order parameter. Notably, the DOS collapses near the
transition line above the ordering temperatures, indicating a universal high-temperature behavior
dominated by electronic contributions.
The coupled charge-lattice inhomogeneities may contribute to
pseudogap formation, while preformed bipolaron pairing remains a possibility. These aspects are systematically
investigated through calculations of the susceptibilities, DOS, and bipolaronic order parameter, as well as
real-space snapshots of charge density and lattice distortions.

We also examine magnetic and charge-order profiles along the phase boundaries of the $U$-$V$ diagram, where
both $U$ and $V$ vary. Along these boundaries, the nonmonotonic behavior of $T_{N}$ ($T_{CO}$) with $U$ ($V$)
persists in the AF (CO) regimes. However, $T_{N}$ is slightly reduced by finite $V$, while $T_{CO}$ is strongly
suppressed by $U$. Understanding the behavior of physical properties along these boundaries may aid in designing
heterostructures combining AF and CO components, where interface effects and the competition or coexistence of
these phases could lead to new phenomena. We leave such studies for future work.

\begin{center}
\textbf{ACKNOWLEDGMENT}
\end{center}

S. H. acknowledges Kreitman School of Advanced Graduate Studies at BGU for providing the Kreitman Postdoctoral Fellowship. M. S. acknowledges support by the Israel Science Foundation Grant No. 3679/24.

\bibliographystyle{unsrt}
\bibliography{hub_holstein_3D_revised.bib}

@article{Dagotto,
author = {Elbio Dagotto },
title = {Complexity in Strongly Correlated Electronic Systems},
journal = {Science},
volume = {309},
number = {5732},
pages = {257-262},
year = {2005},
doi = {10.1126/science.1107559},
URL = {https://www.science.org/doi/abs/10.1126/science.1107559},
}

@article{Imada,
  title = {Metal-insulator transitions},
  author = {Imada, Masatoshi and Fujimori, Atsushi and Tokura, Yoshinori},
  journal = {Rev. Mod. Phys.},
  volume = {70},
  issue = {4},
  pages = {1039--1263},
  numpages = {0},
  year = {1998},
  month = {Oct},
  publisher = {American Physical Society},
  doi = {10.1103/RevModPhys.70.1039},
  url = {https://link.aps.org/doi/10.1103/RevModPhys.70.1039},
}

@article{Damascelli,
  title = {Angle-resolved photoemission studies of the cuprate superconductors},
  author = {Damascelli, Andrea and Hussain, Zahid and Shen, Zhi-Xun},
  journal = {Rev. Mod. Phys.},
  volume = {75},
  issue = {2},
  pages = {473--541},
  numpages = {0},
  year = {2003},
  month = {Apr},
  publisher = {American Physical Society},
  doi = {10.1103/RevModPhys.75.473},
  url = {https://link.aps.org/doi/10.1103/RevModPhys.75.473},
}

@article{Kulic,
title = {Interplay of electron–phonon interaction and strong correlations: the possible way to high-temperature superconductivity},
journal = {Physics Reports},
volume = {338},
number = {1},
pages = {1-264},
year = {2000},
issn = {0370-1573},
doi = {https://doi.org/10.1016/S0370-1573(00)00008-9},
url = {https://www.sciencedirect.com/science/article/pii/S0370157300000089},
author = {Miodrag L. Kulić},
}

@article{Capone1,
author = {Capone, M. and Castellani, C. and Grilli, M.},
title = {Electron-Phonon Interaction in Strongly Correlated Systems},
journal = {Advances in Condensed Matter Physics},
volume = {2010},
number = {1},
pages = {920860},
doi = {https://doi.org/10.1155/2010/920860},
url = {https://onlinelibrary.wiley.com/doi/abs/10.1155/2010/920860},
year = {2010},
}

@article{Mannhart,
author = {J. Mannhart  and D. G. Schlom },
title = {Oxide Interfaces—An Opportunity for Electronics},
journal = {Science},
volume = {327},
number = {5973},
pages = {1607-1611},
year = {2010},
doi = {10.1126/science.1181862},
URL = {https://www.science.org/doi/abs/10.1126/science.1181862},
eprint = {https://www.science.org/doi/pdf/10.1126/science.1181862}}

@article{Dagotto1,
title = {Colossal magnetoresistant materials: the key role of phase separation},
journal = {Physics Reports},
volume = {344},
number = {1},
pages = {1-153},
year = {2001},
issn = {0370-1573},
doi = {https://doi.org/10.1016/S0370-1573(00)00121-6},
url = {https://www.sciencedirect.com/science/article/pii/S0370157300001216},
author = {Elbio Dagotto and Takashi Hotta and Adriana Moreo},
}

@Article{Choi,
author="Choi, Byoung Ki
and Moreschini, Luca
and Bostwick, Aaron
and Williams, R. Stanley
and Chang, Young Jun
and Rotenberg, Eli",
title="Correlation control of the Mott transition in LaTiO3/SrTiO3 heterostructures",
journal="Communications Materials",
year="2025",
volume="7",
number="1",
pages="18",
issn="2662-4443",
doi="10.1038/s43246-025-01032-6",
url="https://doi.org/10.1038/s43246-025-01032-6",
}

@Article{Ahn,
author="Ahn, C. H.
and Triscone, J.-M.
and Mannhart, J.",
title="Electric field effect in correlated oxide systems",
journal="Nature",
year="2003",
volume="424",
number="6952",
pages="1015--1018",
issn="1476-4687",
doi="10.1038/nature01878",
url="https://doi.org/10.1038/nature01878",
}

@article{Scheiderer,
author = {Scheiderer, Philipp and Schmitt, Matthias and Gabel, Judith and Zapf, Michael and Stübinger, Martin and Schütz, Philipp and Dudy, Lenart and Schlueter, Christoph and Lee, Tien-Lin and Sing, Michael and Claessen, Ralph},
title = {Tailoring Materials for Mottronics: Excess Oxygen Doping of a Prototypical Mott Insulator},
journal = {Advanced Materials},
volume = {30},
number = {25},
pages = {1706708},
doi = {https://doi.org/10.1002/adma.201706708},
url = {https://advanced.onlinelibrary.wiley.com/doi/abs/10.1002/adma.201706708},
year = {2018},
}

@ARTICLE{YZhou,
  author={Zhou, You and Ramanathan, Shriram},
  journal={Proceedings of the IEEE}, 
  title={Mott Memory and Neuromorphic Devices}, 
  year={2015},
  volume={103},
  number={8},
  pages={1289-1310},
  doi={10.1109/JPROC.2015.2431914},
}

@article{Chang,
title = {Resistance random access memory},
journal = {Materials Today},
volume = {19},
number = {5},
pages = {254-264},
year = {2016},
issn = {1369-7021},
doi = {https://doi.org/10.1016/j.mattod.2015.11.009},
url = {https://www.sciencedirect.com/science/article/pii/S1369702115003843},
author = {Ting-Chang Chang and Kuan-Chang Chang and Tsung-Ming Tsai and Tian-Jian Chu and Simon M. Sze},
}

@article{Goteti,
author = {Uday S. Goteti  and Ivan A. Zaluzhnyy  and Shriram Ramanathan  and Robert C. Dynes  and Alex Frano },
title = {Low-temperature emergent neuromorphic networks with correlated oxide devices},
journal = {Proceedings of the National Academy of Sciences},
volume = {118},
number = {35},
pages = {e2103934118},
year = {2021},
doi = {10.1073/pnas.2103934118},
URL = {https://www.pnas.org/doi/abs/10.1073/pnas.2103934118},
}

@article{Wadley,
author = {P. Wadley  and B. Howells  and J. Železný  and C. Andrews  and V. Hills  and R. P. Campion  and V. Novák  and K. Olejník  and F. Maccherozzi  and S. S. Dhesi  and S. Y. Martin  and T. Wagner  and J. Wunderlich  and F. Freimuth  and Y. Mokrousov  and J. Kuneš  and J. S. Chauhan  and M. J. Grzybowski  and A. W. Rushforth  and K. W. Edmonds  and B. L. Gallagher  and T. Jungwirth },
title = {Electrical switching of an antiferromagnet},
journal = {Science},
volume = {351},
number = {6273},
pages = {587-590},
year = {2016},
doi = {10.1126/science.aab1031},
URL = {https://www.science.org/doi/abs/10.1126/science.aab1031},
}

@Article{Marti,
author="Marti, X.
and Fina, I.
and Frontera, C.
and Liu, Jian
and Wadley, P.
and He, Q.
and Paull, R. J.
and Clarkson, J. D.
and Kudrnovsk{\~A}{\oe}, J.
and Turek, I.
and Kune{\AA}{\textexclamdown}, J.
and Yi, D.
and Chu, J.-H.
and Nelson, C. T.
and You, L.
and Arenholz, E.
and Salahuddin, S.
and Fontcuberta, J.
and Jungwirth, T.
and Ramesh, R.",
title="Room-temperature antiferromagnetic memory resistor",
journal="Nature Materials",
year="2014",
volume="13",
number="4",
pages="367--374",
issn="1476-4660",
doi="10.1038/nmat3861",
url="https://doi.org/10.1038/nmat3861",
}

@article{Yang,
   author = "Yang, Zheng and Ko, Changhyun and Ramanathan, Shriram",
   title = "Oxide Electronics Utilizing Ultrafast Metal-Insulator Transitions", 
   journal= "Annual Review of Materials Research",
   year = "2011",
   volume = "41",
   number = "Volume 41, 2011",
   pages = "337-367",
   doi = "https://doi.org/10.1146/annurev-matsci-062910-100347",
   url = "https://www.annualreviews.org/content/journals/10.1146/annurev-matsci-062910-100347",
   publisher = "Annual Reviews",
   issn = "1545-4118",
   type = "Journal Article",
}

@article{Zhou,
author = {Zhou, Guangdong and Wang, Zhongrui and Sun, Bai and Zhou, Feichi and Sun, Linfeng and Zhao, Hongbin and Hu, Xiaofang and Peng, Xiaoyan and Yan, Jia and Wang, Huamin and Wang, Wenhua and Li, Jie and Yan, Bingtao and Kuang, Dalong and Wang, Yuchen and Wang, Lidan and Duan, Shukai},
title = {Volatile and Nonvolatile Memristive Devices for Neuromorphic Computing},
journal = {Advanced Electronic Materials},
volume = {8},
number = {7},
pages = {2101127},
doi = {https://doi.org/10.1002/aelm.202101127},
url = {https://advanced.onlinelibrary.wiley.com/doi/abs/10.1002/aelm.202101127},
year = {2022},
}

@Article{Kwon,
author="Kwon, Ki Chang
and Baek, Ji Hyun
and Hong, Kootak
and Kim, Soo Young
and Jang, Ho Won",
title="Memristive Devices Based on Two-Dimensional Transition Metal Chalcogenides for Neuromorphic Computing",
journal="Nano-Micro Letters",
year="2022",
volume="14",
number="1",
pages="58",
issn="2150-5551",
doi="10.1007/s40820-021-00784-3",
url="https://doi.org/10.1007/s40820-021-00784-3",
}

@article{YLi,
  title = {Simulation of memristive synapses and neuromorphic computing on a quantum computer},
  author = {Li, Ying},
  journal = {Phys. Rev. Res.},
  volume = {3},
  issue = {2},
  pages = {023146},
  numpages = {11},
  year = {2021},
  month = {May},
  publisher = {American Physical Society},
  doi = {10.1103/PhysRevResearch.3.023146},
  url = {https://link.aps.org/doi/10.1103/PhysRevResearch.3.023146}
}

@article{SDHa,
  title = {Neuromimetic Circuits with Synaptic Devices Based on Strongly Correlated Electron Systems},
  author = {Ha, Sieu D. and Shi, Jian and Meroz, Yasmine and Mahadevan, L. and Ramanathan, Shriram},
  journal = {Phys. Rev. Appl.},
  volume = {2},
  issue = {6},
  pages = {064003},
  numpages = {11},
  year = {2014},
  month = {Dec},
  publisher = {American Physical Society},
  doi = {10.1103/PhysRevApplied.2.064003},
  url = {https://link.aps.org/doi/10.1103/PhysRevApplied.2.064003},
}

@Article{Boybat,
author="Boybat, Irem
and Le Gallo, Manuel
and Nandakumar, S. R.
and Moraitis, Timoleon
and Parnell, Thomas
and Tuma, Tomas
and Rajendran, Bipin
and Leblebici, Yusuf
and Sebastian, Abu
and Eleftheriou, Evangelos",
title="Neuromorphic computing with multi-memristive synapses",
journal="Nature Communications",
year="2018",
volume="9",
number="1",
pages="2514",
issn="2041-1723",
doi="10.1038/s41467-018-04933-y",
url="https://doi.org/10.1038/s41467-018-04933-y",
}

@Article{Franchini,
author="Franchini, Cesare
and Reticcioli, Michele
and Setvin, Martin
and Diebold, Ulrike",
title="Polarons in materials",
journal="Nature Reviews Materials",
year="2021",
volume="6",
number="7",
pages="560--586",
issn="2058-8437",
doi="10.1038/s41578-021-00289-w",
url="https://doi.org/10.1038/s41578-021-00289-w",
}

@article{Peng,
  title = {Enhanced Electron-Phonon Coupling for Charge-Density-Wave Formation in 
  ${\mathrm{La}}_{1.8\ensuremath{-}x}{\mathrm{Eu}}_{0.2}{\mathrm{Sr}}_{x}{\mathrm{CuO}}_{4+\ensuremath{\delta}}$},
  author = {Peng, Y. Y. and Husain, A. A. and Mitrano, M. and Sun, S. X.-L. and Johnson, T. A. and Zakrzewski, A. V. and MacDougall, G. J. and Barbour, A. and Jarrige, I. and Bisogni, V. and Abbamonte, P.},
  journal = {Phys. Rev. Lett.},
  volume = {125},
  issue = {9},
  pages = {097002},
  numpages = {6},
  year = {2020},
  month = {Aug},
  publisher = {American Physical Society},
  doi = {10.1103/PhysRevLett.125.097002},
  url = {https://link.aps.org/doi/10.1103/PhysRevLett.125.097002},
}

@Book{Peierls,
  title = { Quantum Theory of Solids},
  year = {1955},
  publisher = {Clarendon Oxford},
  editor = {R. E. Peierls},
  isbn = {978-1-4613-1469-1},
  doi = {10.1007/978-1-4613-1469-1},
  url = {https://doi.org/10.1007/978-1-4613-1469-1},
}

@article{Alonso,
  title = {Charge Disproportionation in 
  $\mathit{R}{\mathrm{NiO}}_{3}$ Perovskites: Simultaneous Metal-Insulator and Structural Transition in ${\mathrm{YNiO}}_{3}$},
  author = {Alonso, J. A. and Garc\'{\i}a-Mu\~noz, J. L. and Fern\'andez-D\'{\i}az, M. T. and Aranda, M. A. G. and Mart\'{\i}nez-Lope, M. J. and Casais, M. T.},
  journal = {Phys. Rev. Lett.},
  volume = {82},
  issue = {19},
  pages = {3871--3874},
  numpages = {0},
  year = {1999},
  month = {May},
  publisher = {American Physical Society},
  doi = {10.1103/PhysRevLett.82.3871},
  url = {https://link.aps.org/doi/10.1103/PhysRevLett.82.3871},
}

@article{Kim,
  title = {Signatures of spin-orbital states of 
  ${{t}_{2g}}^{2}$ system in optical conductivity: $R{\mathrm{VO}}_{3}$ ($R=\text{Y}$ and La)},
  author = {Kim, Minjae},
  journal = {Phys. Rev. B},
  volume = {97},
  issue = {15},
  pages = {155141},
  numpages = {9},
  year = {2018},
  month = {Apr},
  publisher = {American Physical Society},
  doi = {10.1103/PhysRevB.97.155141},
  url = {https://link.aps.org/doi/10.1103/PhysRevB.97.155141},
}

@article{Prokofev,
  title = {Polaron Problem by Diagrammatic Quantum Monte Carlo},
  author = {Prokof'ev, Nikolai V. and Svistunov, Boris V.},
  journal = {Phys. Rev. Lett.},
  volume = {81},
  issue = {12},
  pages = {2514--2517},
  numpages = {0},
  year = {1998},
  month = {Sep},
  publisher = {American Physical Society},
  doi = {10.1103/PhysRevLett.81.2514},
  url = {https://link.aps.org/doi/10.1103/PhysRevLett.81.2514},
}

@article{Kornilovitch,
  title = {Continuous-Time Quantum Monte Carlo Algorithm for the Lattice Polaron},
  author = {Kornilovitch, P. E.},
  journal = {Phys. Rev. Lett.},
  volume = {81},
  issue = {24},
  pages = {5382--5385},
  numpages = {0},
  year = {1998},
  month = {Dec},
  publisher = {American Physical Society},
  doi = {10.1103/PhysRevLett.81.5382},
  url = {https://link.aps.org/doi/10.1103/PhysRevLett.81.5382},
}

@article{Bonca,
  title = {Mobile Bipolaron},
  author = {Bonc\ifmmode \breve{}\else \u{}\fi{}a, J. and Katras\ifmmode \breve{}\else \u{}\fi{}nik, T. and Trugman, S. A.},
  journal = {Phys. Rev. Lett.},
  volume = {84},
  issue = {14},
  pages = {3153--3156},
  numpages = {0},
  year = {2000},
  month = {Apr},
  publisher = {American Physical Society},
  doi = {10.1103/PhysRevLett.84.3153},
  url = {https://link.aps.org/doi/10.1103/PhysRevLett.84.3153},
}

@article{Bai,
author = {Bai, Zhiying and He, Dawei and Fu, Shaohua and Miao, Qing and Liu, Shuangyan and Huang, Mohan and Zhao, Kun and Wang, Yongsheng and Zhang, Xiaoxian},
title = {Recent progress in electron–phonon interaction of two-dimensional materials},
journal = {Nano Select},
volume = {3},
number = {7},
pages = {1112-1122},
doi = {https://doi.org/10.1002/nano.202100367},
url = {https://onlinelibrary.wiley.com/doi/abs/10.1002/nano.202100367},
year = {2022},
}

@article{Chen,
    author = {Chen, J. K. and Latham, W. P. and Beraun, J. E.},
    title = {The role of electron–phonon coupling in ultrafast laser heating},
    journal = {Journal of Laser Applications},
    volume = {17},
    number = {1},
    pages = {63-68},
    year = {2005},
    month = {02},   
    issn = {1042-346X},
    doi = {10.2351/1.1848522},
    url = {https://doi.org/10.2351/1.1848522},
}

@article{Ulbricht,
  title = {Carrier dynamics in semiconductors studied with time-resolved terahertz spectroscopy},
  author = {Ulbricht, Ronald and Hendry, Euan and Shan, Jie and Heinz, Tony F. and Bonn, Mischa},
  journal = {Rev. Mod. Phys.},
  volume = {83},
  issue = {2},
  pages = {543--586},
  numpages = {0},
  year = {2011},
  month = {Jun},
  publisher = {American Physical Society},
  doi = {10.1103/RevModPhys.83.543},
  url = {https://link.aps.org/doi/10.1103/RevModPhys.83.543},
}

@Article{Sayers,
author="Sayers, Charles J.
and Genco, Armando
and Trovatello, Chiara
and Conte, Stefano Dal
and Khaustov, Vladislav O.
and Cervantes-Villanueva, Jorge
and Sangalli, Davide
and Molina-Sanchez, Alejandro
and Coletti, Camilla
and Gadermaier, Christoph
and Cerullo, Giulio",
title="Strong Coupling of Coherent Phonons to Excitons in Semiconducting Monolayer MoTe2",
journal="Nano Letters",
year="2023",
month="Oct",
day="25",
publisher="American Chemical Society",
volume="23",
number="20",
pages="9235--9242",
issn="1530-6984",
doi="10.1021/acs.nanolett.3c01936",
url="https://doi.org/10.1021/acs.nanolett.3c01936",
}

@Article{Benyamini,
author="Benyamini, A.
and Hamo, A.
and Kusminskiy, S. Viola
and von Oppen, F.
and Ilani, S.",
title="Real-space tailoring of the electron{\^a}??phonon coupling in ultraclean nanotube mechanical resonators",
journal="Nature Physics",
year="2014",
volume="10",
number="2",
pages="151--156",
issn="1745-2481",
doi="10.1038/nphys2842",
url="https://doi.org/10.1038/nphys2842",
}

@Article{Yan,
author="Yan, Jun
and Kim, M.-H.
and Elle, J. A.
and Sushkov, A. B.
and Jenkins, G. S.
and Milchberg, H. M.
and Fuhrer, M. S.
and Drew, H. D.",
title="Dual-gated bilayer graphene hot-electron bolometer",
journal="Nature Nanotechnology",
year="2012",
volume="7",
number="7",
pages="472--478",
issn="1748-3395",
doi="10.1038/nnano.2012.88",
url="https://doi.org/10.1038/nnano.2012.88",
}

@Article{Niehues,
author="Niehues, Iris
and Schmidt, Robert
and Dr{\~A}{\OE}ppel, Matthias
and Marauhn, Philipp
and Christiansen, Dominik
and Selig, Malte
and Bergh{\~A}{\texteuro}user, Gunnar
and Wigger, Daniel
and Schneider, Robert
and Braasch, Lisa
and Koch, Rouven
and Castellanos-Gomez, Andres
and Kuhn, Tilmann
and Knorr, Andreas
and Malic, Ermin
and Rohlfing, Michael
and Michaelis de Vasconcellos, Steffen
and Bratschitsch, Rudolf",
title="Strain Control of Exciton{\^a}??Phonon Coupling in Atomically Thin Semiconductors",
journal="Nano Letters",
year="2018",
month="Mar",
day="14",
publisher="American Chemical Society",
volume="18",
number="3",
pages="1751--1757",
issn="1530-6984",
doi="10.1021/acs.nanolett.7b04868",
url="https://doi.org/10.1021/acs.nanolett.7b04868",
}

@article{Tang,
    author = {Tang, Dao-Sheng and Qin, Guang-Zhao and Hu, Ming and Cao, Bing-Yang},
    title = {Thermal transport properties of GaN with biaxial strain and electron-phonon coupling},
    journal = {Journal of Applied Physics},
    volume = {127},
    number = {3},
    pages = {035102},
    year = {2020},
    month = {01},
    issn = {0021-8979},
    doi = {10.1063/1.5133105},
    url = {https://doi.org/10.1063/1.5133105},
    eprint = {https://pubs.aip.org/aip/jap/article-pdf/doi/10.1063/1.5133105/15241007/035102_1_online.pdf},
}

@Article{Lanzara,
author="Lanzara, A.
and Bogdanov, P. V.
and Zhou, X. J.
and Kellar, S. A.
and Feng, D. L.
and Lu, E. D.
and Yoshida, T.
and Eisaki, H.
and Fujimori, A.
and Kishio, K.
and Shimoyama, J.-I.
and Noda, T.
and Uchida, S.
and Hussain, Z.
and Shen, Z.-X.",
title="Evidence for ubiquitous strong electron{\^a}??phonon coupling in high-temperature superconductors",
journal="Nature",
year="2001",
volume="412",
number="6846",
pages="510--514",
issn="1476-4687",
doi="10.1038/35087518",
url="https://doi.org/10.1038/35087518",
}

@article{Gunnarsson,
  title = {Superconductivity in fullerides},
  author = {Gunnarsson, O.},
  journal = {Rev. Mod. Phys.},
  volume = {69},
  issue = {2},
  pages = {575--606},
  numpages = {0},
  year = {1997},
  month = {Apr},
  publisher = {American Physical Society},
  doi = {10.1103/RevModPhys.69.575},
  url = {https://link.aps.org/doi/10.1103/RevModPhys.69.575},
}

@Article{Haule,
author="Haule, Kristjan
and Pascut, Gheorghe L.",
title="Mott Transition and Magnetism in Rare Earth Nickelates and its Fingerprint on the X-ray Scattering",
journal="Scientific Reports",
year="2017",
volume="7",
number="1",
pages="10375",
issn="2045-2322",
doi="10.1038/s41598-017-10374-2",
url="https://doi.org/10.1038/s41598-017-10374-2",
}

@Article{Stepanov,
author="Stepanov, Evgeny A.
and Vandelli, Matteo
and Lichtenstein, Alexander I.
and Lechermann, Frank",
title="Charge density wave ordering in NdNiO2: effects of multiorbital nonlocal correlations",
journal="npj Computational Materials",
year="2024",
volume="10",
number="1",
pages="108",
issn="2057-3960",
doi="10.1038/s41524-024-01298-3",
url="https://doi.org/10.1038/s41524-024-01298-3",
}

@article{Tranquada,
  title = {Cooperative ordering of holes and spins in 
  ${\mathrm{La}}_{2}$${\mathrm{NiO}}_{4.125}$},
  author = {Tranquada, J. M. and Lorenzo, J. E. and Buttrey, D. J. and Sachan, V.},
  journal = {Phys. Rev. B},
  volume = {52},
  issue = {5},
  pages = {3581--3595},
  numpages = {0},
  year = {1995},
  month = {Aug},
  publisher = {American Physical Society},
  doi = {10.1103/PhysRevB.52.3581},
  url = {https://link.aps.org/doi/10.1103/PhysRevB.52.3581},
}

@article{Millis,
  title = {Double Exchange Alone Does Not Explain the Resistivity of 
  ${{\mathrm{La}}_{1}}_{\ensuremath{-}\mathit{x}}{\mathrm{Sr}}_{\mathit{x}}{\mathrm{MnO}}_{3}$},
  author = {Millis, A. J. and Littlewood, P. B. and Shraiman, B. I.},
  journal = {Phys. Rev. Lett.},
  volume = {74},
  issue = {25},
  pages = {5144--5147},
  numpages = {0},
  year = {1995},
  month = {Jun},
  publisher = {American Physical Society},
  doi = {10.1103/PhysRevLett.74.5144},
  url = {https://link.aps.org/doi/10.1103/PhysRevLett.74.5144},
}

@Article{Millis1,
author="Millis, A. J.",
title="Lattice effects in magnetoresistive manganese perovskites",
journal="Nature",
year="1998",
volume="392",
number="6672",
pages="147--150",
issn="1476-4687",
doi="10.1038/32348",
url="https://doi.org/10.1038/32348",
}

@article{Powell,
doi = {10.1088/0953-8984/18/45/R03},
url = {https://doi.org/10.1088/0953-8984/18/45/R03},
year = {2006},
month = {oct},
publisher = {},
volume = {18},
number = {45},
pages = {R827},
author = {Powell, B J and McKenzie, Ross H},
title = {Strong electronic correlations in superconducting organic charge transfer salts},
journal = {Journal of Physics: Condensed Matter},
}

@article{Zhan,
  title = {Cooperation between Electron-Phonon Coupling and Electronic Interaction in Bilayer Nickelates 
  ${\mathrm{La}}_{3}{\mathrm{Ni}}_{2}{\mathrm{O}}_{7}$},
  author = {Zhan, Jun and Gu, Yuhao and Wu, Xianxin and Hu, Jiangping},
  journal = {Phys. Rev. Lett.},
  volume = {134},
  issue = {13},
  pages = {136002},
  numpages = {7},
  year = {2025},
  month = {Mar},
  publisher = {American Physical Society},
  doi = {10.1103/PhysRevLett.134.136002},
  url = {https://link.aps.org/doi/10.1103/PhysRevLett.134.136002},
}

@article{RWang,
title = {Role of electronic correlation effect on charge ordering in $\beta$-V2OPO4},
journal = {Computational Materials Science},
volume = {173},
pages = {109433},
year = {2020},
issn = {0927-0256},
doi = {https://doi.org/10.1016/j.commatsci.2019.109433},
url = {https://www.sciencedirect.com/science/article/pii/S0927025619307323},
author = {Runxue Wang and Xipeng Zhang and Hengyang Xie and Yuanhui Xu and Keju Sun and Yanlong Yu and Yongshan Liu and Xianfeng Hao},
}

@article{QWang,
author = {Qisi Wang  and Karin von Arx  and Masafumi Horio  and Deepak John Mukkattukavil  and Julia Küspert  and Yasmine Sassa  and Thorsten Schmitt  and Abhishek Nag  and Sunseng Pyon  and Tomohiro Takayama  and Hidenori Takagi  and Mirian Garcia-Fernandez  and Ke-Jin Zhou  and Johan Chang },
title = {Charge order lock-in by electron-phonon coupling in 
${\mathrm{La}}_{1.675}{\mathrm{Eu}}_{0.2}{\mathrm{Sr}}_{0.125}{\mathrm{CuO}}_{4}$},
journal = {Science Advances},
volume = {7},
number = {27},
pages = {eabg7394},
year = {2021},
doi = {10.1126/sciadv.abg7394},
URL = {https://www.science.org/doi/abs/10.1126/sciadv.abg7394},
}

@Article{Georgescu,
author="Georgescu, Alexandru B.
and Millis, Andrew J.",
title="Quantifying the role of the lattice in metal{\^a}??insulator phase transitions",
journal="Communications Physics",
year="2022",
volume="5",
number="1",
pages="135",
issn="2399-3650",
doi="10.1038/s42005-022-00909-z",
url="https://doi.org/10.1038/s42005-022-00909-z",
}

@article{Ge,
  title = {Lattice-driven magnetoresistivity and metal-insulator transition in single-layered iridates},
  author = {Ge, M. and Qi, T. F. and Korneta, O. B. and De Long, D. E. and Schlottmann, P. and Crummett, W. P. and Cao, G.},
  journal = {Phys. Rev. B},
  volume = {84},
  issue = {10},
  pages = {100402},
  numpages = {5},
  year = {2011},
  month = {Sep},
  publisher = {American Physical Society},
  doi = {10.1103/PhysRevB.84.100402},
  url = {https://link.aps.org/doi/10.1103/PhysRevB.84.100402},
}

@article{Duan,
doi = {10.1088/0256-307X/42/9/090712},
url = {https://doi.org/10.1088/0256-307X/42/9/090712},
year = {2025},
month = {sep},
publisher = {Chinese Physical Society and IOP Publishing Ltd},
volume = {42},
number = {9},
pages = {090712},
author = {Duan, Qingzhuo and Jia, Zixuan and Fan, Zenghui and Ma, Runyu and Meng, Jingyao and Huang, Bing and Ma, Tianxing},
title = {Breathing-Driven Metal-Insulator Transition in Correlated Kagome Systems},
journal = {Chinese Physics Letters},
}

@article{Ngai,
   author = "Ngai, J.H. and Walker, F.J. and Ahn, C.H.",
   title = "Correlated Oxide Physics and Electronics", 
   journal= "Annual Review of Materials Research",
   year = "2014",
   volume = "44",
   number = "Volume 44, 2014",
   pages = "1-17",
   doi = "https://doi.org/10.1146/annurev-matsci-070813-113248",
   url = "https://www.annualreviews.org/content/journals/10.1146/annurev-matsci-070813-113248",
   publisher = "Annual Reviews",
   issn = "1545-4118",
   type = "Journal Article",
}

@article{Bark,
author = {C. W. Bark  and D. A. Felker  and Y. Wang  and Y. Zhang  and H. W. Jang  and C. M. Folkman  and J. W. Park  and S. H. Baek  and H. Zhou  and D. D. Fong  and X. Q. Pan  and E. Y. Tsymbal  and M. S. Rzchowski  and C. B. Eom },
title = {Tailoring a two-dimensional electron gas at the LaAlO<sub>3</sub>/SrTiO<sub>3</sub> (001) interface by epitaxial strain},
journal = {Proceedings of the National Academy of Sciences},
volume = {108},
number = {12},
pages = {4720-4724},
year = {2011},
doi = {10.1073/pnas.1014849108},
URL = {https://www.pnas.org/doi/abs/10.1073/pnas.1014849108},
}

@article{Watanabe,
    title={DFT
    ${+U}$ accurate for strain effect and overall properties of perovskite oxide ferroelectronics and polaron},
    author={Watanabe, Yukio},
    volume={135},
    url={http://dx.doi.org/10.1063/5.0213487},
    doi={10.1063/5.0213487},
    number={22},
    journal={Journal of Applied Physics},
    publisher={AIP Publishing},
    year={2024},
    month={Jun},
    language={en},
}

@article{Mirjolet,
author = {Mirjolet, Mathieu and Rivadulla, Francisco and Marsik, Premysl and Borisov, Vladislav and Valentí, Roser and Fontcuberta, Josep},
title = {Electron–Phonon Coupling and Electron–Phonon Scattering in SrVO3},
journal = {Advanced Science},
volume = {8},
number = {15},
pages = {2004207},
doi = {https://doi.org/10.1002/advs.202004207},
url = {https://advanced.onlinelibrary.wiley.com/doi/abs/10.1002/advs.202004207},
year = {2021},
}

@article{ZOu,
author = {Ou, Zhenwei and Peng, Bin and Chu, Weibin and Li, Zhe and Wang, Cheng and Zeng, Yan and Chen, Hongyi and Wang, Qiuyu and Dong, Guohua and Wu, Yongyi and Qiu, Ruibin and Ma, Li and Zhang, Lili and Liu, Xiaoze and Li, Tao and Yu, Ting and Hu, Zhongqiang and Wang, Ti and Liu, Ming and Xu, Hongxing},
title = {Strong Electron-Phonon Coupling Mediates Carrier Transport in BiFeO3},
journal = {Advanced Science},
volume = {10},
number = {22},
pages = {2301057},
doi = {https://doi.org/10.1002/advs.202301057},
url = {https://advanced.onlinelibrary.wiley.com/doi/abs/10.1002/advs.202301057},
year = {2023},
}

@article{Baldini,
author = {Edoardo Baldini  and Michael A. Sentef  and Swagata Acharya  and Thomas Brumme  and Evgeniia Sheveleva  and Fryderyk Lyzwa  and Ekaterina Pomjakushina  and Christian Bernhard  and Mark van Schilfgaarde  and Fabrizio Carbone  and Angel Rubio  and Cédric Weber },
title = {Electron–phonon-driven three-dimensional metallicity in an insulating cuprate},
journal = {Proceedings of the National Academy of Sciences},
volume = {117},
number = {12},
pages = {6409-6416},
year = {2020},
doi = {10.1073/pnas.1919451117},
URL = {https://www.pnas.org/doi/abs/10.1073/pnas.1919451117},
}

@article{Singh,
title = {Electron-phonon interactions and Helmholtz free energy in confined systems: Advancing low-dimensional superconductors for quantum technologies},
journal = {Results in Physics},
volume = {73},
pages = {108270},
year = {2025},
issn = {2211-3797},
doi = {https://doi.org/10.1016/j.rinp.2025.108270},
url = {https://www.sciencedirect.com/science/article/pii/S2211379725001640},
author = {Nitin Pratap Singh and Kusum Lata and Linga Reddy Cenkeramaddi},
}

@article{Karakuzu,
  title = {Superconductivity, charge-density waves, antiferromagnetism, and phase separation in the Hubbard-Holstein model},
  author = {Karakuzu, Seher and Tocchio, Luca F. and Sorella, Sandro and Becca, Federico},
  journal = {Phys. Rev. B},
  volume = {96},
  issue = {20},
  pages = {205145},
  numpages = {10},
  year = {2017},
  month = {Nov},
  publisher = {American Physical Society},
  doi = {10.1103/PhysRevB.96.205145},
  url = {https://link.aps.org/doi/10.1103/PhysRevB.96.205145},
}

@article{Han,
  title = {Strong Coupling Limit of the Holstein-Hubbard Model},
  author = {Han, Zhaoyu and Kivelson, Steven A. and Yao, Hong},
  journal = {Phys. Rev. Lett.},
  volume = {125},
  issue = {16},
  pages = {167001},
  numpages = {6},
  year = {2020},
  month = {Oct},
  publisher = {American Physical Society},
  doi = {10.1103/PhysRevLett.125.167001},
  url = {https://link.aps.org/doi/10.1103/PhysRevLett.125.167001},
}

@article{Koller,
doi = {10.1209/epl/i2003-10228-6},
url = {https://doi.org/10.1209/epl/i2003-10228-6},
year = {2004},
month = {may},
publisher = {},
volume = {66},
number = {4},
pages = {559},
author = {W. Koller and D. Meyer and Y. Ōno and A. C. Hewson},
title = {First- and second-order phase transitions  in the
Holstein-Hubbard model},
journal = {Europhysics Letters},
}

@article{Koller1,
  title = {Dynamic response functions for the Holstein-Hubbard model},
  author = {Koller, W. and Meyer, D. and Hewson, A. C.},
  journal = {Phys. Rev. B},
  volume = {70},
  issue = {15},
  pages = {155103},
  numpages = {12},
  year = {2004},
  month = {Oct},
  publisher = {American Physical Society},
  doi = {10.1103/PhysRevB.70.155103},
  url = {https://link.aps.org/doi/10.1103/PhysRevB.70.155103},
}

@article{Koller2,
  title = {Polaronic Quasiparticles in a Strongly Correlated Electron Band},
  author = {Koller, W. and Hewson, A. C. and Edwards, D. M.},
  journal = {Phys. Rev. Lett.},
  volume = {95},
  issue = {25},
  pages = {256401},
  numpages = {4},
  year = {2005},
  month = {Dec},
  publisher = {American Physical Society},
  doi = {10.1103/PhysRevLett.95.256401},
  url = {https://link.aps.org/doi/10.1103/PhysRevLett.95.256401},
}

@Article{Costa,
author="Costa, Natanael C.
and Seki, Kazuhiro
and Yunoki, Seiji
and Sorella, Sandro",
title="Phase diagram of the two-dimensional Hubbard-Holstein model",
journal="Communications Physics",
year="2020",
volume="3",
number="1",
pages="80",
issn="2399-3650",
doi="10.1038/s42005-020-0342-2",
url="https://doi.org/10.1038/s42005-020-0342-2",
}

@article{Nowadnick,
  title = {Competition Between Antiferromagnetic and Charge-Density-Wave Order in the Half-Filled Hubbard-Holstein Model},
  author = {Nowadnick, E. A. and Johnston, S. and Moritz, B. and Scalettar, R. T. and Devereaux, T. P.},
  journal = {Phys. Rev. Lett.},
  volume = {109},
  issue = {24},
  pages = {246404},
  numpages = {5},
  year = {2012},
  month = {Dec},
  publisher = {American Physical Society},
  doi = {10.1103/PhysRevLett.109.246404},
  url = {https://link.aps.org/doi/10.1103/PhysRevLett.109.246404},
}

@article{Johnston,
  title = {Determinant quantum Monte Carlo study of the two-dimensional single-band Hubbard-Holstein model},
  author = {Johnston, S. and Nowadnick, E. A. and Kung, Y. F. and Moritz, B. and Scalettar, R. T. and Devereaux, T. P.},
  journal = {Phys. Rev. B},
  volume = {87},
  issue = {23},
  pages = {235133},
  numpages = {13},
  year = {2013},
  month = {Jun},
  publisher = {American Physical Society},
  doi = {10.1103/PhysRevB.87.235133},
  url = {https://link.aps.org/doi/10.1103/PhysRevB.87.235133},
}

@article{Mendl,
  title = {Doping dependence of ordered phases and emergent quasiparticles in the doped Hubbard-Holstein model},
  author = {Mendl, C. B. and Nowadnick, E. A. and Huang, E. W. and Johnston, S. and Moritz, B. and Devereaux, T. P.},
  journal = {Phys. Rev. B},
  volume = {96},
  issue = {20},
  pages = {205141},
  numpages = {7},
  year = {2017},
  month = {Nov},
  publisher = {American Physical Society},
  doi = {10.1103/PhysRevB.96.205141},
  url = {https://link.aps.org/doi/10.1103/PhysRevB.96.205141},
}

@article{Ohgoe,
  title = {Competition among Superconducting, Antiferromagnetic, and Charge Orders with Intervention by Phase Separation in the 2D Holstein-Hubbard Model},
  author = {Ohgoe, Takahiro and Imada, Masatoshi},
  journal = {Phys. Rev. Lett.},
  volume = {119},
  issue = {19},
  pages = {197001},
  numpages = {5},
  year = {2017},
  month = {Nov},
  publisher = {American Physical Society},
  doi = {10.1103/PhysRevLett.119.197001},
  url = {https://link.aps.org/doi/10.1103/PhysRevLett.119.197001},
}

@article{Huang,
  title = {Electron-phonon vertex in the two-dimensional one-band Hubbard model},
  author = {Huang, Z. B. and Hanke, W. and Arrigoni, E. and Scalapino, D. J.},
  journal = {Phys. Rev. B},
  volume = {68},
  issue = {22},
  pages = {220507},
  numpages = {4},
  year = {2003},
  month = {Dec},
  publisher = {American Physical Society},
  doi = {10.1103/PhysRevB.68.220507},
  url = {https://link.aps.org/doi/10.1103/PhysRevB.68.220507},
}

@article{Nocera,
  title = {Interplay of charge, spin, and lattice degrees of freedom in the spectral properties of the one-dimensional Hubbard-Holstein model},
  author = {Nocera, A. and Soltanieh-ha, M. and Perroni, C. A. and Cataudella, V. and Feiguin, A. E.},
  journal = {Phys. Rev. B},
  volume = {90},
  issue = {19},
  pages = {195134},
  numpages = {9},
  year = {2014},
  month = {Nov},
  publisher = {American Physical Society},
  doi = {10.1103/PhysRevB.90.195134},
  url = {https://link.aps.org/doi/10.1103/PhysRevB.90.195134},
}

@article{SLi,
  title = {Phase competition in a one-dimensional three-orbital Hubbard-Holstein model},
  author = {Li, Shaozhi and Tang, Yanfei and Maier, Thomas A. and Johnston, Steven},
  journal = {Phys. Rev. B},
  volume = {97},
  issue = {19},
  pages = {195116},
  numpages = {11},
  year = {2018},
  month = {May},
  publisher = {American Physical Society},
  doi = {10.1103/PhysRevB.97.195116},
  url = {https://link.aps.org/doi/10.1103/PhysRevB.97.195116},
}

@article{Xiao,
  title = {Competition between phase separation and spin density wave or charge density wave order: Role of long-range interactions},
  author = {Xiao, Bo and H\'ebert, F. and Batrouni, G. and Scalettar, R. T.},
  journal = {Phys. Rev. B},
  volume = {99},
  issue = {20},
  pages = {205145},
  numpages = {9},
  year = {2019},
  month = {May},
  publisher = {American Physical Society},
  doi = {10.1103/PhysRevB.99.205145},
  url = {https://link.aps.org/doi/10.1103/PhysRevB.99.205145},
}

@article{Hebert,
  title = {One-dimensional Hubbard-Holstein model with finite-range electron-phonon coupling},
  author = {H\'ebert, F. and Xiao, Bo and Rousseau, V. G. and Scalettar, R. T. and Batrouni, G. G.},
  journal = {Phys. Rev. B},
  volume = {99},
  issue = {7},
  pages = {075108},
  numpages = {11},
  year = {2019},
  month = {Feb},
  publisher = {American Physical Society},
  doi = {10.1103/PhysRevB.99.075108},
  url = {https://link.aps.org/doi/10.1103/PhysRevB.99.075108},
}

@Article{Lavanya,
author="Lavanya, Ch Uma
and Sankar, I. V.
and Chatterjee, Ashok",
title="Metallicity in a Holstein-Hubbard Chain at Half Filling with Gaussian Anharmonicity",
journal="Scientific Reports",
year="2017",
volume="7",
number="1",
pages="3774",
issn="2045-2322",
doi="10.1038/s41598-017-03985-2",
url="https://doi.org/10.1038/s41598-017-03985-2",
}

@article{Hohenadler1,
  title = {Excitation spectra and spin gap of the half-filled Holstein-Hubbard model},
  author = {Hohenadler, Martin and Assaad, Fakher F.},
  journal = {Phys. Rev. B},
  volume = {87},
  issue = {7},
  pages = {075149},
  numpages = {10},
  year = {2013},
  month = {Feb},
  publisher = {American Physical Society},
  doi = {10.1103/PhysRevB.87.075149},
  url = {https://link.aps.org/doi/10.1103/PhysRevB.87.075149},
}

@article{Matsueda,
  title = {Electron-phonon coupling and spin-charge separation in one-dimensional Mott insulators},
  author = {Matsueda, H. and Tohyama, T. and Maekawa, S.},
  journal = {Phys. Rev. B},
  volume = {74},
  issue = {24},
  pages = {241103},
  numpages = {4},
  year = {2006},
  month = {Dec},
  publisher = {American Physical Society},
  doi = {10.1103/PhysRevB.74.241103},
  url = {https://link.aps.org/doi/10.1103/PhysRevB.74.241103},
}

@article{Ning,
  title = {Phonon Effects on Spin-Charge Separation in One Dimension},
  author = {Ning, Wen-Qiang and Zhao, Hui and Wu, Chang-Qin and Lin, Hai-Qing},
  journal = {Phys. Rev. Lett.},
  volume = {96},
  issue = {15},
  pages = {156402},
  numpages = {4},
  year = {2006},
  month = {Apr},
  publisher = {American Physical Society},
  doi = {10.1103/PhysRevLett.96.156402},
  url = {https://link.aps.org/doi/10.1103/PhysRevLett.96.156402},
}

@article{Tezuka,
  title = {Density-Matrix Renormalization Group Study of Pairing when Electron-Electron and Electron-Phonon Interactions Coexist: Effect of the Electronic Band Structure},
  author = {Tezuka, Masaki and Arita, Ryotaro and Aoki, Hideo},
  journal = {Phys. Rev. Lett.},
  volume = {95},
  issue = {22},
  pages = {226401},
  numpages = {4},
  year = {2005},
  month = {Nov},
  publisher = {American Physical Society},
  doi = {10.1103/PhysRevLett.95.226401},
  url = {https://link.aps.org/doi/10.1103/PhysRevLett.95.226401},
}

@article{Tezuka1,
  title = {Phase diagram for the one-dimensional Hubbard-Holstein model: A density-matrix renormalization group study},
  author = {Tezuka, Masaki and Arita, Ryotaro and Aoki, Hideo},
  journal = {Phys. Rev. B},
  volume = {76},
  issue = {15},
  pages = {155114},
  numpages = {11},
  year = {2007},
  month = {Oct},
  publisher = {American Physical Society},
  doi = {10.1103/PhysRevB.76.155114},
  url = {https://link.aps.org/doi/10.1103/PhysRevB.76.155114},
}

@article{Clay,
  title = {Intermediate Phase of the One Dimensional Half-Filled Hubbard-Holstein Model},
  author = {Clay, R. T. and Hardikar, R. P.},
  journal = {Phys. Rev. Lett.},
  volume = {95},
  issue = {9},
  pages = {096401},
  numpages = {4},
  year = {2005},
  month = {Aug},
  publisher = {American Physical Society},
  doi = {10.1103/PhysRevLett.95.096401},
  url = {https://link.aps.org/doi/10.1103/PhysRevLett.95.096401},
}

@article{Fehske,
doi = {10.1209/0295-5075/84/57001},
url = {https://doi.org/10.1209/0295-5075/84/57001},
year = {2008},
month = {nov},
publisher = {},
volume = {84},
number = {5},
pages = {57001},
author = {Fehske, H. and Hager, G. and Jeckelmann, E.},
title = {Metallicity in the half-filled Holstein-Hubbard model},
journal = {Europhysics Letters},
}

@article{Fradkin,
  title = {Phase diagram of one-dimensional electron-phonon systems. I. The Su-Schrieffer-Heeger model},
  author = {Fradkin, Eduardo and Hirsch, Jorge E.},
  journal = {Phys. Rev. B},
  volume = {27},
  issue = {3},
  pages = {1680--1697},
  numpages = {0},
  year = {1983},
  month = {Feb},
  publisher = {American Physical Society},
  doi = {10.1103/PhysRevB.27.1680},
  url = {https://link.aps.org/doi/10.1103/PhysRevB.27.1680},
}

@article{Brink,
  title = {Charge ordering and magnetism in quarter-filled Hubbard-Holstein model},
  author = {Kumar, Sanjeev and van den Brink, Jeroen},
  journal = {Phys. Rev. B},
  volume = {78},
  issue = {15},
  pages = {155123},
  numpages = {5},
  year = {2008},
  month = {Oct},
  publisher = {American Physical Society},
  doi = {10.1103/PhysRevB.78.155123},
  url = {https://link.aps.org/doi/10.1103/PhysRevB.78.155123},
}

@article{Berger,
  title = {Two-dimensional Hubbard-Holstein model},
  author = {Berger, E. and Val\'a\ifmmode \check{s}\else \v{s}\fi{}ek, P. and von der Linden, W.},
  journal = {Phys. Rev. B},
  volume = {52},
  issue = {7},
  pages = {4806--4814},
  numpages = {0},
  year = {1995},
  month = {Aug},
  publisher = {American Physical Society},
  doi = {10.1103/PhysRevB.52.4806},
  url = {https://link.aps.org/doi/10.1103/PhysRevB.52.4806},
}

@article{Hotta,
  title = {Effect of electron correlation on phonons in a strongly coupled electron-phonon system},
  author = {Hotta, Takashi and Takada, Yasutami},
  journal = {Phys. Rev. B},
  volume = {56},
  issue = {21},
  pages = {13916--13926},
  numpages = {0},
  year = {1997},
  month = {Dec},
  publisher = {American Physical Society},
  doi = {10.1103/PhysRevB.56.13916},
  url = {https://link.aps.org/doi/10.1103/PhysRevB.56.13916},
}

@article{Nowadnick1,
  title = {Renormalization of spectra by phase competition in the half-filled Hubbard-Holstein model},
  author = {Nowadnick, E. A. and Johnston, S. and Moritz, B. and Devereaux, T. P.},
  journal = {Phys. Rev. B},
  volume = {91},
  issue = {16},
  pages = {165127},
  numpages = {11},
  year = {2015},
  month = {Apr},
  publisher = {American Physical Society},
  doi = {10.1103/PhysRevB.91.165127},
  url = {https://link.aps.org/doi/10.1103/PhysRevB.91.165127},
}

@article{Dupuis,
  title = {Spin fluctuations and pseudogap in the two-dimensional half-filled Hubbard model at weak coupling},
  author = {Dupuis, N.},
  journal = {Phys. Rev. B},
  volume = {65},
  issue = {24},
  pages = {245118},
  numpages = {5},
  year = {2002},
  month = {Jun},
  publisher = {American Physical Society},
  doi = {10.1103/PhysRevB.65.245118},
  url = {https://link.aps.org/doi/10.1103/PhysRevB.65.245118},
}

@article{Pai,
  title = {Holstein-Hubbard model at half filling: A static auxiliary field study},
  author = {Pradhan, Saurabh and Pai, G. Venketeswara},
  journal = {Phys. Rev. B},
  volume = {92},
  issue = {16},
  pages = {165124},
  numpages = {10},
  year = {2015},
  month = {Oct},
  publisher = {American Physical Society},
  doi = {10.1103/PhysRevB.92.165124},
  url = {https://link.aps.org/doi/10.1103/PhysRevB.92.165124},
}

@article{Wang,
  title = {Zero-temperature phases of the two-dimensional Hubbard-Holstein model: A non-Gaussian exact diagonalization study},
  author = {Wang, Yao and Esterlis, Ilya and Shi, Tao and Cirac, J. Ignacio and Demler, Eugene},
  journal = {Phys. Rev. Res.},
  volume = {2},
  issue = {4},
  pages = {043258},
  numpages = {19},
  year = {2020},
  month = {Nov},
  publisher = {American Physical Society},
  doi = {10.1103/PhysRevResearch.2.043258},
  url = {https://link.aps.org/doi/10.1103/PhysRevResearch.2.043258},
}

@article{Marsiglio,
title = {Pairing in the Holstein model in the dilute limit},
journal = {Physica C: Superconductivity},
volume = {244},
number = {1},
pages = {21-34},
year = {1995},
issn = {0921-4534},
doi = {https://doi.org/10.1016/0921-4534(95)00046-1},
url = {https://www.sciencedirect.com/science/article/pii/0921453495000461},
author = {F. Marsiglio},
}

@article{Weber,
  title = {Two-dimensional Holstein-Hubbard model: Critical temperature, Ising universality, and bipolaron liquid},
  author = {Weber, Manuel and Hohenadler, Martin},
  journal = {Phys. Rev. B},
  volume = {98},
  issue = {8},
  pages = {085405},
  numpages = {8},
  year = {2018},
  month = {Aug},
  publisher = {American Physical Society},
  doi = {10.1103/PhysRevB.98.085405},
  url = {https://link.aps.org/doi/10.1103/PhysRevB.98.085405},
}

@article{Murakami,
  title = {Ordered phases in the Holstein-Hubbard model: Interplay of strong Coulomb interaction and electron-phonon coupling},
  author = {Murakami, Yuta and Werner, Philipp and Tsuji, Naoto and Aoki, Hideo},
  journal = {Phys. Rev. B},
  volume = {88},
  issue = {12},
  pages = {125126},
  numpages = {14},
  year = {2013},
  month = {Sep},
  publisher = {American Physical Society},
  doi = {10.1103/PhysRevB.88.125126},
  url = {https://link.aps.org/doi/10.1103/PhysRevB.88.125126},
}

@article{Bauer,
  title = {Competition between antiferromagnetic and charge order in the Hubbard-Holstein model},
  author = {Bauer, Johannes and Hewson, Alex C.},
  journal = {Phys. Rev. B},
  volume = {81},
  issue = {23},
  pages = {235113},
  numpages = {17},
  year = {2010},
  month = {Jun},
  publisher = {American Physical Society},
  doi = {10.1103/PhysRevB.81.235113},
  url = {https://link.aps.org/doi/10.1103/PhysRevB.81.235113},
}

@article{Bauer1,
doi = {10.1209/0295-5075/90/27002},
url = {https://doi.org/10.1209/0295-5075/90/27002},
year = {2010},
month = {may},
publisher = {},
volume = {90},
number = {2},
pages = {27002},
author = {Bauer, J.},
title = {Competing interactions and symmetry breaking in the Hubbard-Holstein model},
journal = {Europhysics Letters},
}

@article{Hayashida,
  title = {Lattice dynamics of the infinite-layer nickelate 
  ${\mathrm{LaNiO}}_{2}$},
  author = {Hayashida, Shohei and Sundaramurthy, Vignesh and Wu, Wenfeng and Puphal, 
  Pascal and Keller, Thomas and F\aa{}k, Bj\"orn and Isobe, Masahiko and Keimer, Bernhard and Held, Karsten and Si, Liang and Hepting, Matthias},
  journal = {Phys. Rev. B},
  volume = {112},
  issue = {20},
  pages = {205104},
  numpages = {10},
  year = {2025},
  month = {Nov},
  publisher = {American Physical Society},
  doi = {10.1103/rk6x-2f3r},
  url = {https://link.aps.org/doi/10.1103/rk6x-2f3r},
}

@article{Talantsev,
author="E.F. Talantsev and V.V. Chistyakov",
title="Debye temperature, electron-phonon coupling constant, and three-dome shape of crystalline strain as a function of pressure in highly compressed La3Ni2O7-${\delta}$",
publisher="Letters on Materials",
year="2024",
journal="Letters on Materials",
volume="14",
number="3",
pages="262-268",
url="https://lettersonmaterials.com/en/Readers/Article.aspx?aid=44771",
doi="10.48612/letters/2024-3-262-268",
}

@Article{Ouyang,
author="Ouyang, Zhenfeng
and Gao, Miao
and Lu, Zhong-Yi",
title="Absence of electron-phonon coupling superconductivity in the bilayer phase of La3Ni2O7 under pressure",
journal="npj Quantum Materials",
year="2024",
volume="9",
number="1",
pages="80",
issn="2397-4648",
doi="10.1038/s41535-024-00689-5",
url="https://doi.org/10.1038/s41535-024-00689-5",
}

@Article{Zhang,
author="Zhang, Yang
and Lin, Ling-Fang
and Moreo, Adriana
and Maier, Thomas A.
and Dagotto, Elbio",
title="Structural phase transition, s{\^A}{\textpm}-wave pairing, and magnetic stripe order in bilayered superconductor La3Ni2O7 under pressure",
journal="Nature Communications",
year="2024",
volume="15",
number="1",
pages="2470",
issn="2041-1723",
doi="10.1038/s41467-024-46622-z",
url="https://doi.org/10.1038/s41467-024-46622-z",
}

@article{Munoz,
  title = {Structure and charge order in the antiferromagnetic band-insulating phase of 
  ${\mathrm{NdNiO}}_{3}$},
  author = {Garc\'{\i}a-Mu\~noz, J. L. and Aranda, M. A. G. and Alonso, J. A. and Mart\'{\i}nez-Lope, M. J.},
  journal = {Phys. Rev. B},
  volume = {79},
  issue = {13},
  pages = {134432},
  numpages = {5},
  year = {2009},
  month = {Apr},
  publisher = {American Physical Society},
  doi = {10.1103/PhysRevB.79.134432},
  url = {https://link.aps.org/doi/10.1103/PhysRevB.79.134432},
}

@article{Johnston1,
  title = {Charge Disproportionation without Charge Transfer in the Rare-Earth-Element Nickelates as a Possible Mechanism for the Metal-Insulator Transition},
  author = {Johnston, Steve and Mukherjee, Anamitra and Elfimov, Ilya and Berciu, Mona and Sawatzky, George A.},
  journal = {Phys. Rev. Lett.},
  volume = {112},
  issue = {10},
  pages = {106404},
  numpages = {5},
  year = {2014},
  month = {Mar},
  publisher = {American Physical Society},
  doi = {10.1103/PhysRevLett.112.106404},
  url = {https://link.aps.org/doi/10.1103/PhysRevLett.112.106404},
}

@article{Alonso1,
  title = {Room-temperature monoclinic distortion due to charge disproportionation in 
  $R{\mathrm{NiO}}_{3}$ perovskites with small rare-earth cations $(R=\mathrm{Ho},$ Y, Er, Tm, Yb, and Lu): A neutron diffraction study},
  author = {Alonso, J. A. and Mart\'{\i}nez-Lope, M. J. and Casais, M. T. and Garc\'{\i}a-Mu\~noz, J. L. and Fern\'andez-D\'{\i}az, M. T.},
  journal = {Phys. Rev. B},
  volume = {61},
  issue = {3},
  pages = {1756--1763},
  numpages = {0},
  year = {2000},
  month = {Jan},
  publisher = {American Physical Society},
  doi = {10.1103/PhysRevB.61.1756},
  url = {https://link.aps.org/doi/10.1103/PhysRevB.61.1756},
}

@article{Staub,
  title = {Direct Observation of Charge Order in an Epitaxial 
  ${\mathrm{NdNiO}}_{3}$ Film},
  author = {Staub, U. and Meijer, G. I. and Fauth, F. and Allenspach, R. and Bednorz, J. G. and Karpinski, J. and Kazakov, S. M. and Paolasini, L. and d'Acapito, F.},
  journal = {Phys. Rev. Lett.},
  volume = {88},
  issue = {12},
  pages = {126402},
  numpages = {4},
  year = {2002},
  month = {Mar},
  publisher = {American Physical Society},
  doi = {10.1103/PhysRevLett.88.126402},
  url = {https://link.aps.org/doi/10.1103/PhysRevLett.88.126402},
}

@Article{Kumar,
author="Kumar, S.
and Majumdar, P.",
title="A travelling cluster approximation for lattice fermions strongly coupled to classical degrees of freedom",
journal="The European Physical Journal B - Condensed Matter and Complex Systems",
year="2006",
volume="50",
number="4",
pages="571--579",
issn="1434-6036",
doi="10.1140/epjb/e2006-00173-2",
url="https://doi.org/10.1140/epjb/e2006-00173-2",
}

@article{Hardikar,
  title = {Phase diagram of the one-dimensional Hubbard-Holstein model at half and quarter filling},
  author = {Hardikar, R. P. and Clay, R. T.},
  journal = {Phys. Rev. B},
  volume = {75},
  issue = {24},
  pages = {245103},
  numpages = {10},
  year = {2007},
  month = {Jun},
  publisher = {American Physical Society},
  doi = {10.1103/PhysRevB.75.245103},
  url = {https://link.aps.org/doi/10.1103/PhysRevB.75.245103},
}

@article{S1,
  title = {Layer-resolved magnetotransport properties in antiferromagnetic/paramagnetic superlattices: Proximity effect induced antiferromagnetism in a paramagnetic metal},
  author = {Halder, Sandip and Chakraborty, Sourav and Pradhan, Kalpataru},
  journal = {Phys. Rev. B},
  volume = {110},
  issue = {19},
  pages = {195147},
  numpages = {15},
  year = {2024},
  month = {Nov},
  publisher = {American Physical Society},
  doi = {10.1103/PhysRevB.110.195147},
  url = {https://link.aps.org/doi/10.1103/PhysRevB.110.195147},
}

@article{S2,
  title = {Microscopic study of interlayer magnetic coupling across the interface in antiferromagnetic bilayers},
  author = {Halder, Sandip and Mandal, Sudip and Pradhan, Kalpataru},
  journal = {Phys. Rev. B},
  volume = {111},
  issue = {12},
  pages = {125149},
  numpages = {15},
  year = {2025},
  month = {Mar},
  publisher = {American Physical Society},
  doi = {10.1103/PhysRevB.111.125149},
  url = {https://link.aps.org/doi/10.1103/PhysRevB.111.125149},
}

@article{Metropolis,
    author = {Metropolis, Nicholas and Rosenbluth, Arianna W. and Rosenbluth, Marshall N. and Teller, Augusta H. and Teller, Edward},
    title = {Equation of State Calculations by Fast Computing Machines},
    journal = {The Journal of Chemical Physics},
    volume = {21},
    number = {6},
    pages = {1087-1092},
    year = {1953},
    month = {06},
    issn = {0021-9606},
    doi = {10.1063/1.1699114},
    url = {https://doi.org/10.1063/1.1699114},
    eprint = {https://pubs.aip.org/aip/jcp/article-pdf/21/6/1087/18802390/1087_1_online.pdf},
}

@article{Halder,
  title = {Interfacial antiferromagnetic coupling driven magnetotransport properties in ferromagnetic superlattices},
  author = {Halder, Sandip and Das, Subrat K. and Pradhan, Kalpataru},
  journal = {Phys. Rev. B},
  volume = {108},
  issue = {23},
  pages = {235111},
  numpages = {13},
  year = {2023},
  month = {Dec},
  publisher = {American Physical Society},
  doi = {10.1103/PhysRevB.108.235111},
  url = {https://link.aps.org/doi/10.1103/PhysRevB.108.235111},
}

@article{S3,
  title = {Magnetization reversal in nanocrystalline 
  ${\mathrm{Gd}}_{0.5}{\mathrm{Sr}}_{0.5}{\mathrm{MnO}}_{3}$},
  author = {Chatterjee, Soma and Halder, Sandip and Das, Kalipada and Pradhan, Kalpataru and Das, I.},
  journal = {Phys. Rev. B},
  volume = {110},
  issue = {15},
  pages = {155138},
  numpages = {16},
  year = {2024},
  month = {Oct},
  publisher = {American Physical Society},
  doi = {10.1103/PhysRevB.110.155138},
  url = {https://link.aps.org/doi/10.1103/PhysRevB.110.155138},
}

@article{Chakraborty,
  title = {Itinerant ferromagnetism in a spin-fermion model for diluted spin systems},
  author = {Chakraborty, Sourav and Halder, Sandip and Pradhan, Kalpataru},
  journal = {Phys. Rev. B},
  volume = {108},
  issue = {16},
  pages = {165110},
  numpages = {9},
  year = {2023},
  month = {Oct},
  publisher = {American Physical Society},
  doi = {10.1103/PhysRevB.108.165110},
  url = {https://link.aps.org/doi/10.1103/PhysRevB.108.165110},
}

@Book{Mahan,
  title = {Many-Particle Physics},
  year = {2012},
  publisher = {Springer New York, NY},
  editor = {Gerald D. Mahan},
  isbn = {978-1-4613-1469-1},
  doi = {10.1007/978-1-4613-1469-1},
  url = {https://doi.org/10.1007/978-1-4613-1469-1},
}

@article{Kumar1,
doi = {10.1209/epl/i2003-10056-8},
url = {https://doi.org/10.1209/epl/i2003-10056-8},
year = {2004},
month = {jan},
publisher = {},
volume = {65},
number = {1},
pages = {75},
author = {S. Kumar and P. Majumdar},
title = {Anti-localisation to strong localisation:
The interplay  of magnetic scattering and structural disorder},
journal = {Europhysics Letters},
}

@article{Bulanchuk,
title = {On the delta function broadening in the Kubo–Greenwood equation},
journal = {Computer Physics Communications},
volume = {261},
pages = {107714},
year = {2021},
issn = {0010-4655},
doi = {https://doi.org/10.1016/j.cpc.2020.107714},
url = {https://www.sciencedirect.com/science/article/pii/S0010465520303532},
author = {Pavlo Bulanchuk},
}

@article{White,
  title = {Numerical study of the two-dimensional Hubbard model},
  author = {White, S. R. and Scalapino, D. J. and Sugar, R. L. and Loh, E. Y. and Gubernatis, J. E. and Scalettar, R. T.},
  journal = {Phys. Rev. B},
  volume = {40},
  issue = {1},
  pages = {506--516},
  numpages = {0},
  year = {1989},
  month = {Jul},
  publisher = {American Physical Society},
  doi = {10.1103/PhysRevB.40.506},
  url = {https://link.aps.org/doi/10.1103/PhysRevB.40.506}
}

@article{Mondaini,
  title = {Magnetism, transport, and thermodynamics in two-dimensional half-filled Hubbard superlattices},
  author = {Mondaini, Rubem and Paiva, Thereza},
  journal = {Phys. Rev. B},
  volume = {95},
  issue = {7},
  pages = {075142},
  numpages = {13},
  year = {2017},
  month = {Feb},
  publisher = {American Physical Society},
  doi = {10.1103/PhysRevB.95.075142},
  url = {https://link.aps.org/doi/10.1103/PhysRevB.95.075142}
}

@article{Araujo,
  title = {Two-dimensional 
  $t\ensuremath{-}{t}^{\ensuremath{'}}$ Holstein model},
  author = {Ara\'ujo, Maykon V. and de Lima, Jos\'e P. and Sorella, Sandro and Costa, Natanael C.},
  journal = {Phys. Rev. B},
  volume = {105},
  issue = {16},
  pages = {165103},
  numpages = {8},
  year = {2022},
  month = {Apr},
  publisher = {American Physical Society},
  doi = {10.1103/PhysRevB.105.165103},
  url = {https://link.aps.org/doi/10.1103/PhysRevB.105.165103}
}

@article{Slater,
  title = {Magnetic Effects and the Hartree-Fock Equation},
  author = {Slater, J. C.},
  journal = {Phys. Rev.},
  volume = {82},
  issue = {4},
  pages = {538--541},
  numpages = {0},
  year = {1951},
  month = {May},
  publisher = {American Physical Society},
  doi = {10.1103/PhysRev.82.538},
  url = {https://link.aps.org/doi/10.1103/PhysRev.82.538},
}

@Book{Gruner,
  title = {Density Waves in Solids},
  year = {2000},
  publisher = {Westview Press},
  editor = {G. Gruner},
  isbn = {978-1-4613-1469-1},
  doi = {10.1007/978-1-4613-1469-1},
  url = {https://doi.org/10.1007/978-1-4613-1469-1},
}

@article{Jeckelmann,
  title = {Metal-insulator transition in the one-dimensional Holstein model at half filling},
  author = {Jeckelmann, Eric and Zhang, Chunli and White, Steven R.},
  journal = {Phys. Rev. B},
  volume = {60},
  issue = {11},
  pages = {7950--7955},
  numpages = {0},
  year = {1999},
  month = {Sep},
  publisher = {American Physical Society},
  doi = {10.1103/PhysRevB.60.7950},
  url = {https://link.aps.org/doi/10.1103/PhysRevB.60.7950},
}

@Article{Hohenadler,
author="Hohenadler, Martin
and Fehske, Holger",
title="Density waves in strongly correlated quantum chains",
journal="The European Physical Journal B",
year="2018",
volume="91",
number="9",
pages="204",
issn="1434-6036",
doi="10.1140/epjb/e2018-90354-7",
url="https://doi.org/10.1140/epjb/e2018-90354-7",
}

@article{Prelov,
  title = {Self-Localization of Composite Spin-Lattice Polarons},
  author = {Prelov\ifmmode \check{s}\else \v{s}\fi{}ek, Peter and Zeyher, Roland and Horsch, Peter},
  journal = {Phys. Rev. Lett.},
  volume = {96},
  issue = {8},
  pages = {086402},
  numpages = {4},
  year = {2006},
  month = {Mar},
  publisher = {American Physical Society},
  doi = {10.1103/PhysRevLett.96.086402},
  url = {https://link.aps.org/doi/10.1103/PhysRevLett.96.086402},
}

@article{Greco,
title = {Stability of correlated electronic systems under the influence of the electron-phonon interaction},
journal = {Solid State Communications},
volume = {99},
number = {7},
pages = {473-477},
year = {1996},
issn = {0038-1098},
doi = {https://doi.org/10.1016/0038-1098(96)00307-9},
url = {https://www.sciencedirect.com/science/article/pii/0038109896003079},
author = {A. Greco and A. Dobry},
}

@article{Sangiovanni,
  title = {Electron-Phonon Interaction Close to a Mott Transition},
  author = {Sangiovanni, G. and Capone, M. and Castellani, C. and Grilli, M.},
  journal = {Phys. Rev. Lett.},
  volume = {94},
  issue = {2},
  pages = {026401},
  numpages = {4},
  year = {2005},
  month = {Jan},
  publisher = {American Physical Society},
  doi = {10.1103/PhysRevLett.94.026401},
  url = {https://link.aps.org/doi/10.1103/PhysRevLett.94.026401},
}

@article{Sangiovanni1,
  title = {Electron-Phonon Interaction and Antiferromagnetic Correlations},
  author = {Sangiovanni, G. and Gunnarsson, O. and Koch, E. and Castellani, C. and Capone, M.},
  journal = {Phys. Rev. Lett.},
  volume = {97},
  issue = {4},
  pages = {046404},
  numpages = {4},
  year = {2006},
  month = {Jul},
  publisher = {American Physical Society},
  doi = {10.1103/PhysRevLett.97.046404},
  url = {https://link.aps.org/doi/10.1103/PhysRevLett.97.046404},
}

@article{Borejsza,
  title = {Antiferromagnetism and single-particle properties in the two-dimensional half-filled Hubbard model: A nonlinear sigma model approach},
  author = {Borejsza, K. and Dupuis, N.},
  journal = {Phys. Rev. B},
  volume = {69},
  issue = {8},
  pages = {085119},
  numpages = {17},
  year = {2004},
  month = {Feb},
  publisher = {American Physical Society},
  doi = {10.1103/PhysRevB.69.085119},
  url = {https://link.aps.org/doi/10.1103/PhysRevB.69.085119},
}

@article{Borejsza1,
doi = {10.1209/epl/i2003-00584-7},
url = {https://doi.org/10.1209/epl/i2003-00584-7},
year = {2003},
month = {sep},
publisher = {},
volume = {63},
number = {5},
pages = {722},
author = {K. Borejsza and N. Dupuis},
title = {Antiferromagnetism and single-particle properties  in the
two-dimensional half-filled Hubbard model:  Slater vs.
Mott-Heisenberg},
journal = {Europhysics Letters},
}

@article{Jeon,
  title = {Dynamical mean-field theory of the Hubbard-Holstein model at half filling: Zero temperature metal-insulator and insulator-insulator transitions},
  author = {Jeon, Gun Sang and Park, Tae-Ho and Han, Jung Hoon and Lee, Hyun C. and Choi, Han-Yong},
  journal = {Phys. Rev. B},
  volume = {70},
  issue = {12},
  pages = {125114},
  numpages = {6},
  year = {2004},
  month = {Sep},
  publisher = {American Physical Society},
  doi = {10.1103/PhysRevB.70.125114},
  url = {https://link.aps.org/doi/10.1103/PhysRevB.70.125114},
}

@article{Esterlis,
  title = {Pseudogap crossover in the electron-phonon system},
  author = {Esterlis, I. and Kivelson, S. A. and Scalapino, D. J.},
  journal = {Phys. Rev. B},
  volume = {99},
  issue = {17},
  pages = {174516},
  numpages = {5},
  year = {2019},
  month = {May},
  publisher = {American Physical Society},
  doi = {10.1103/PhysRevB.99.174516},
  url = {https://link.aps.org/doi/10.1103/PhysRevB.99.174516},
}

@article{Ciuchi,
  title = {Charge-ordered state from weak to strong coupling},
  author = {Ciuchi, S. and de Pasquale, F.},
  journal = {Phys. Rev. B},
  volume = {59},
  issue = {8},
  pages = {5431--5440},
  numpages = {0},
  year = {1999},
  month = {Feb},
  publisher = {American Physical Society},
  doi = {10.1103/PhysRevB.59.5431},
  url = {https://link.aps.org/doi/10.1103/PhysRevB.59.5431},
}

@article{Hwang,
doi = {10.1088/1361-6633/ad36d3},
url = {https://doi.org/10.1088/1361-6633/ad36d3},
year = {2024},
month = {apr},
publisher = {IOP Publishing},
volume = {87},
number = {4},
pages = {044502},
author = {Hwang, Jinwoong and Ruan, Wei and Chen, Yi and Tang, Shujie and Crommie, Michael F and Shen, Zhi-Xun and Mo, Sung-Kwan},
title = {Charge density waves in two-dimensional transition metal dichalcogenides},
journal = {Reports on Progress in Physics},
}

@Article{Lee,
author="Lee, Yung-Ting
and Chen, Po-Tuan
and Li, Zheng-Hong
and Wu, Jyun-Yu
and Kuo, Chia-Nung
and Lue, Chin Shan
and Wu, Chien-Te
and Kuo, Chien-Cheng
and Chiang, Cheng-Tien
and Ozaki, Taisuke
and Lin, Chun-Liang
and Lee, Chi-Cheng
and Hsueh, Hung-Chung
and Chung, Ming-Chiang",
title="Revealing the Charge Density Wave Caused by Peierls Instability in Two-Dimensional NbSe2",
journal="ACS Materials Letters",
year="2024",
month="Jul",
day="01",
publisher="American Chemical Society",
volume="6",
number="7",
pages="2941--2947",
doi="10.1021/acsmaterialslett.4c00142",
url="https://doi.org/10.1021/acsmaterialslett.4c00142",
}

@article{Mermin,
  title = {Absence of Ferromagnetism or Antiferromagnetism in One- or Two-Dimensional Isotropic Heisenberg Models},
  author = {Mermin, N. D. and Wagner, H.},
  journal = {Phys. Rev. Lett.},
  volume = {17},
  issue = {22},
  pages = {1133--1136},
  numpages = {0},
  year = {1966},
  month = {Nov},
  publisher = {American Physical Society},
  doi = {10.1103/PhysRevLett.17.1133},
  url = {https://link.aps.org/doi/10.1103/PhysRevLett.17.1133},
}

@article{Hohenberg,
  title = {Existence of Long-Range Order in One and Two Dimensions},
  author = {Hohenberg, P. C.},
  journal = {Phys. Rev.},
  volume = {158},
  issue = {2},
  pages = {383--386},
  numpages = {0},
  year = {1967},
  month = {Jun},
  publisher = {American Physical Society},
  doi = {10.1103/PhysRev.158.383},
  url = {https://link.aps.org/doi/10.1103/PhysRev.158.383},
}

@article{Schulz,
  title = {Effective action for strongly correlated fermions from functional integrals},
  author = {Schulz, H. J.},
  journal = {Phys. Rev. Lett.},
  volume = {65},
  issue = {19},
  pages = {2462--2465},
  numpages = {0},
  year = {1990},
  month = {Nov},
  publisher = {American Physical Society},
  doi = {10.1103/PhysRevLett.65.2462},
  url = {https://link.aps.org/doi/10.1103/PhysRevLett.65.2462}
}

@article{Mukherjee,
  title = {Testing the Monte Carlo--mean field approximation in the one-band Hubbard model},
  author = {Mukherjee, Anamitra and Patel, Niravkumar D. and Dong, Shuai and Johnston, Steve and Moreo, Adriana and Dagotto, Elbio},
  journal = {Phys. Rev. B},
  volume = {90},
  issue = {20},
  pages = {205133},
  numpages = {13},
  year = {2014},
  month = {Nov},
  publisher = {American Physical Society},
  doi = {10.1103/PhysRevB.90.205133},
  url = {https://link.aps.org/doi/10.1103/PhysRevB.90.205133},
}

@article{Santos,
  author = {Santos, R. R. d.},
  title = {Introduction to quantum Monte Carlo simulations for fermionic systems},
  journal = {Brazilian Journal of Physics},
  year = {2003},
  volume = {33},
  issue = {1},
  pages = {36-54},
  doi = {10.1590/s0103-97332003000100003},
}

@Article{Staudt,
author="Staudt, R.
and Dzierzawa, M.
and Muramatsu, A.",
title="Phase diagram of the three-dimensional Hubbard model at half filling",
journal="The European Physical Journal B - Condensed Matter and Complex Systems",
year="2000",
volume="17",
number="3",
pages="411--415",
issn="1434-6036",
doi="10.1007/s100510070120",
url="https://doi.org/10.1007/s100510070120",
}

@article{Hirsch,
  title = {Antiferromagnetism, localization, and pairing in a two-dimensional model for 
  ${\mathrm{CuO}}_{2}$},
  author = {Hirsch, J. E.},
  journal = {Phys. Rev. Lett.},
  volume = {59},
  issue = {2},
  pages = {228--231},
  numpages = {0},
  year = {1987},
  month = {Jul},
  publisher = {American Physical Society},
  doi = {10.1103/PhysRevLett.59.228},
  url = {https://link.aps.org/doi/10.1103/PhysRevLett.59.228},
}

@article{Scalettar,
  title = {Phase diagram of the half-filled 3D Hubbard model},
  author = {Scalettar, R. T. and Scalapino, D. J. and Sugar, R. L. and Toussaint, D.},
  journal = {Phys. Rev. B},
  volume = {39},
  issue = {7},
  pages = {4711--4714},
  numpages = {0},
  year = {1989},
  month = {Mar},
  publisher = {American Physical Society},
  doi = {10.1103/PhysRevB.39.4711},
  url = {https://link.aps.org/doi/10.1103/PhysRevB.39.4711},
}

@article{Jarrell,
  title = {Hubbard model in infinite dimensions: A quantum Monte Carlo study},
  author = {Jarrell, M.},
  journal = {Phys. Rev. Lett.},
  volume = {69},
  issue = {1},
  pages = {168--171},
  numpages = {0},
  year = {1992},
  month = {Jul},
  publisher = {American Physical Society},
  doi = {10.1103/PhysRevLett.69.168},
  url = {https://link.aps.org/doi/10.1103/PhysRevLett.69.168},
}

@article{Ulmke,
  title = {Anderson-Hubbard model in infinite dimensions},
  author = {Ulmke, M. and Jani{\ifmmode \check{s}\else \v{s}\fi{}}, V. and Vollhardt, D.},
  journal = {Phys. Rev. B},
  volume = {51},
  issue = {16},
  pages = {10411--10426},
  numpages = {0},
  year = {1995},
  month = {Apr},
  publisher = {American Physical Society},
  doi = {10.1103/PhysRevB.51.10411},
  url = {https://link.aps.org/doi/10.1103/PhysRevB.51.10411},
}

@article{Song,
  title = {Magnetic, thermodynamic, and dynamical properties of the three-dimensional fermionic Hubbard model: A comprehensive Monte Carlo study},
  author = {Song, Yu-Feng and Deng, Youjin and He, Yuan-Yao},
  journal = {Phys. Rev. B},
  volume = {111},
  issue = {3},
  pages = {035123},
  numpages = {28},
  year = {2025},
  month = {Jan},
  publisher = {American Physical Society},
  doi = {10.1103/PhysRevB.111.035123},
  url = {https://link.aps.org/doi/10.1103/PhysRevB.111.035123},
}

@article{Stead,
  title = {Langevin simulations of the half-filled cubic Holstein model},
  author = {Cohen-Stead, B. and Barros, Kipton and Meng, ZY and Chen, Chuang and Scalettar, R. T. and Batrouni, G. G.},
  journal = {Phys. Rev. B},
  volume = {102},
  issue = {16},
  pages = {161108(R)},
  numpages = {5},
  year = {2020},
  month = {Oct},
  publisher = {American Physical Society},
  doi = {10.1103/PhysRevB.102.161108},
  url = {https://link.aps.org/doi/10.1103/PhysRevB.102.161108},
}

@article{Kovac,
  title = {Signature of Preformed Pairs in Angle-Resolved Photoemission Spectroscopy},
  author = {Kova{\ifmmode \check{c}\else \v{c}\fi{}}, Klemen and Nocera, Alberto and Damascelli, Andrea and Bon{\ifmmode \check{c}\else \v{c}\fi{}}a, Janez and Berciu, Mona},
  journal = {Phys. Rev. Lett.},
  volume = {134},
  issue = {9},
  pages = {096502},
  numpages = {8},
  year = {2025},
  month = {Mar},
  publisher = {American Physical Society},
  doi = {10.1103/PhysRevLett.134.096502},
  url = {https://link.aps.org/doi/10.1103/PhysRevLett.134.096502},
}

\end{document}